\documentclass[12pt]{article}
\pdfoutput=1

\newlength{\abstractwidth}
\usepackage{amsmath, nccmath}
\usepackage{amsfonts}
\usepackage{amssymb}
\usepackage{latexsym}

\usepackage{color}
\usepackage{graphicx}
\usepackage{tikz}
\usepackage{dsfont}

\usetikzlibrary{arrows,shapes,positioning}
\usetikzlibrary{decorations.markings}
\usepackage[rightcaption]{sidecap}
\tikzstyle arrowstyle=[scale=1]
\tikzstyle directed=[postaction={decorate,decoration={markings,
    mark=at position .65 with {\arrow[arrowstyle]{stealth}}}}]
\tikzstyle reverse directed=[postaction={decorate,decoration={markings,
    mark=at position .65 with {\arrowreversed[arrowstyle]{stealth};}}}]
\usetikzlibrary{positioning}

\usepackage{hyperref}
\definecolor{darkred}{rgb}{0.8,0.1,0.1}
\hypersetup{colorlinks=true, linkcolor=darkred, citecolor=blue, linktoc=page}

\usepackage{cite}

\renewcommand{\thefootnote}{\fnsymbol{footnote}}
\renewcommand{\thanks}[1]{\footnote{#1}}
\newcommand{\starttext}{
\setcounter{footnote}{0}
\renewcommand{\thefootnote}{\arabic{footnote}}}

\newcommand{\bea}{\begin{eqnarray}}
\newcommand{\eea}{\end{eqnarray}}
\newcommand{\be}{\begin{eqnarray}}
\newcommand{\ee}{\end{eqnarray}}
\newcommand{\bma}{\begin{matrix}}
\newcommand{\ema}{\cr\end{matrix}}
\newcommand{\<}{\langle}
\renewcommand{\>}{\rangle}

\newtheorem{thm}{Theorem}[section]
\newtheorem{lem}[thm]{Lemma}
\newtheorem{prop}[thm]{Proposition}
\newtheorem{cor}[thm]{Corollary}

\def\cA{{\cal A}}
\def\cB{{\cal B}}
\def\cC{{\cal C}}

\def\cE{{\cal E}}
\def\cF{{\cal F}}
\def\cG{{\cal G}}
\def\cH{{\cal H}}

\def\cL{{\cal L}}
\def\cM{{\cal M}}

\def\cO{{\cal O}}
\def\cP{{\cal P}}
\def\cQ{{\cal Q}}
\def\cR{{\cal R}}
\def\cS{{\cal S}}
\def\cT{{\cal T}}
\def\cU{{\cal U}}
\def\cV{{\cal V}}
\def\cW{{\cal W}}

\def\mA{\mathfrak{A}}

\def\mC{\mathfrak{C}}

\def\mF{\mathfrak{F}}

\def\mI{\mathfrak{I}}

\def\mS{\mathfrak{S}}

\def\mc{\mathfrak{c}}

\def\ZZ{{\mathbb Z}}
\def\RR{{\mathbb R}}
\def\NN{{\mathbb N}}
\def\CC{{\mathbb C}}

\def\QQ{{\mathbb Q}}

\def\Re{{\rm Re \,}}
\def\Im{{\rm Im \,}}

\def\half{{1\over 2}}
\def\thalf{{\tfrac{1}{2}}}
\def\p{\partial}
\def\DD{\nabla }
\def\DDb{\overline{ \nabla}}

\def\a{\alpha}
\def\b{\beta}
\def\g{\gamma}
\def\G{\Gamma}
\def\eps{\epsilon}
\def\f{\varphi}

\def\ep{\varepsilon}

\def\zsv{{\zeta_{{\rm sv}}}}

\def\no{\nonumber}
\def\sm{\smallskip}

\begin{document}
\starttext
\setcounter{footnote}{0}

\begin{flushright}
2019 February 11  \\
revised 2021 February 5\,\,
\end{flushright}

\vskip 0.3in

\begin{center}

{\Large \bf Modular graph functions  and odd cuspidal functions}

\vskip 0.1in

{\large \bf  Fourier and Poincar\'e series\footnote{This research  is supported in part by the National Science Foundation under grant PHY-16-19926.}}

\vskip 0.3in

{\large  Eric D'Hoker and Justin Kaidi} 

\vskip 0.1in

 { \sl Mani L. Bhaumik Institute for Theoretical Physics}\\
{\sl Department of Physics and Astronomy }\\
{\sl University of California, Los Angeles, CA 90095, USA} 

\vskip 0.05in

{\tt \small dhoker@physics.ucla.edu, jkaidi@physics.ucla.edu}

\end{center}

\begin{abstract}

Modular graph functions are $SL(2,\ZZ)$-invariant functions associated with Feynman graphs of a two-dimensional conformal field theory on a torus of modulus $\tau$. For one-loop graphs they reduce to real analytic Eisenstein series. We obtain the Fourier series, including the constant and non-constant Fourier modes, of all two-loop modular graph functions, as well as their Poincar\'e series with respect to $\Gamma _\infty \backslash PSL(2,\ZZ)$. 
The Fourier and Poincar\'e series  provide the tools to compute the Petersson inner product of two-loop modular graph functions using Rankin-Selberg-Zagier methods. Modular graph functions which are odd under $\tau \to - \bar \tau$ are cuspidal functions, with exponential decay near the cusp, and exist starting at two loops. Holomorphic subgraph reduction and the sieve algorithm, developed in earlier work, are used to give a lower bound on the dimension of the space $\mA_w$ of odd two-loop modular graph functions  of weight $w$. For $w \leq 11$ the bound is saturated and we exhibit a basis for $\mA_w$. 

\end{abstract}

\newpage

\baselineskip=15pt
\setcounter{equation}{0}
\setcounter{footnote}{0}

\newpage
\setcounter{tocdepth}{2}
\tableofcontents
\newpage

\section{Introduction}
\setcounter{equation}{0}
\label{sec:1}

A modular graph function is a function associated to a certain Feynman graph for a two-dimensional conformal field theory on a torus with complex structure $\tau$. It is invariant under transformations of $\tau$ by the modular group $SL(2,\ZZ)$. Modular graph functions may be decomposed into even and odd pieces depending on their parity under $\tau \to - \bar \tau$, and their {\sl loop order} is given by the number of loops of the corresponding Feynman graph. The {\sl weight} $w$ of a modular graph function is  half of the degree of homogeneity in the momenta of the graph, and provides a grading on the ring of such functions. One-loop modular graph functions are always even and, for weight $w$, reduce to real analytic Eisenstein series $E_w$.

\sm

The origin of modular graph functions at higher loop order may be traced back to string theory. The analytic structure of the genus-one superstring amplitude for four strings was analyzed in \cite{DHoker:1994gnm}, and the low-energy expansion was organized in terms of modular functions in \cite{Green:1999pv,Green:2008uj}. A systematic study of modular graph functions was initiated in \cite{DHoker:2015gmr,DHoker:2015sve} where they were  shown to obey a system of differential and algebraic identities and in \cite{DHoker:2015wxz} where they were shown to be intimately related to single-valued elliptic polylogarithms. Further identities were derived in  \cite{Basu:2015ayg,Basu:2016kli,Basu:2016mmk,Kleinschmidt:2017ege}. A key ingredient in the construction of these identities is the Laurent polynomial in $\Im(\tau)$ of the modular graph function. It governs the leading  behavior at the cusp $\tau \to i \infty$ and was obtained in closed form  for an infinite class of even two-loop modular graph functions  in \cite{DHoker:2017zhq}, and for all graphs with four or fewer vertices in \cite{Zerbini}.

\sm

Modular graph \textit{forms}, which associate an $SL(2,\ZZ)$ non-holomorphic modular form to a certain type of (decorated) graph, were introduced in \cite{DHoker:2016mwo,DHoker:2016quv,Gerken:2018zcy}, where a systematic method was developed to construct all the identities between them. An alternative construction in terms of Eichler type integrals was developed in \cite{Broedel:2018izr}. The underlying nature of these identities remains to be fully uncovered, but it is already  clear that  they generalize to modular graph forms some of the algebraic identities which exist between multiple zeta values (see for example \cite{ZagierMZV,Hoffmann,W,Zudilin,Blumlein:2009cf,Brown:2013gia}). Direct connections between modular graph functions and single-valued elliptic polylogarithms were obtained in \cite{DHoker:2015wxz,Brown:2017qwo,Brown2,Vanhove:2018elu}. The role of multiple zeta values   in string amplitudes has been investigated extensively  in \cite{SCHLotterer:2012ny,Broedel:2013tta,Stieberger:2013wea,Broedel:2014vla,Broedel:2015hia}. Generalizations of modular graph functions for Heterotic string amplitudes were studied  in \cite{Basu:2017zvt,Basu:2017nhs,Gerken:2018jrq}.  Higher genus modular graph functions were introduced and analyzed in   \cite{DHoker:2017pvk, DHoker:2018mys}, building on earlier studies of special cases in  \cite{DHoker:2013fcx,DHoker:2014oxd,Pioline:2015qha}, and further developed in \cite{Basu:2018eep,Pioline:2018pso,Basu:2018bde}.

\sm

In the present paper, we shall obtain the Fourier series, including constant and non-constant Fourier modes,  of arbitrary two-loop modular graph functions, thereby generalizing the results of \cite{DHoker:2017zhq}. The coefficients of the Fourier modes involve odd zeta values as well as a novel generalization of divisor sums. We also construct the Poincar\'e series of all two-loop modular graph functions  in terms of a sum over  $\Gamma _\infty \backslash PSL(2,\ZZ)$, with $\Gamma_\infty$ the Borel subgroup of $PSL(2,\ZZ)$,  compute the corresponding seed functions, prove absolute convergence of the series, and outline the generalization to  arbitrary weight and loop order. 

\sm

The Fourier and Poincar\'e series expansions provide all the tools needed to evaluate integrals of two-loop modular graph functions over the fundamental domain of $PSL(2,\ZZ)$, as well as the Petersson inner product between modular graph functions, using the Rankin-Selberg-Zagier methods.  Some of these integrals are needed to  evaluate the contributions to the genus-one string amplitudes which are analytic in the external momenta \cite{Green:1999pv,DHoker:2015gmr}. 

\sm
Finally, we investigate the space $\mA_w$ of odd two-loop modular graph functions of weight $w$, and show that $\mA_w$ consists entirely of cuspidal functions, namely modular functions which have exponential decay near the cusp.  Using the techniques of holomorphic subgraph reduction and the sieve algorithm developed in \cite{DHoker:2016mwo,DHoker:2016quv,Gerken:2018zcy} we produce a lower bound on the dimension of $\mA_w$ by exhibiting families of linearly independent odd modular graph functions at each weight. For $w \leq 4$ the space $\mA_w$ is empty, while for $5 \leq w \leq 11$ the lower bound is saturated and we exhibit an explicit  basis for $\mA_w$. 

\subsection*{Organization} 

In Section~\ref{sec:2}, we  review the definition and basic properties of general modular graph functions and forms, define their parity under $\tau \to - \bar \tau$, discuss the one-loop  and two-loop cases, and prove some basic decomposition formulas for two-loop modular graph functions. In Section~\ref{sec:3}, we obtain the Poincar\'e series for general modular graph functions, compute the seed functions for the two-loop case and prove absolute convergence of the series. In Section~\ref{sec:4} we carry out the calculation of the Fourier series expansion for arbitrary two-loop modular graph functions. In Section~\ref{sec:5} we simplify the result for the constant Fourier mode, and extract the expansion for the non-constant Fourier modes. In Section~\ref{sec:6}, we review the holomorphic subgraph reduction procedure and sieve algorithm,  construct infinite families of linearly independent two-loop odd modular graph functions, and produce a lower bound for the dimension of the space $\mA_w$ of odd modular graph functions of weight $w$. Finally, in Section~\ref{sec:7} we discuss the action of the Laplace operator on two-loop odd modular graph functions and study Petersson inner products between them. Technical details of the calculations and some explicit formulas for complicated expansion coefficients are relegated to the Appendices.

\subsection*{Acknowledgments}

We are happy to thank Bill Duke, Michael Green, Boris Pioline, and Oliver Schlotterer for their interest in this project and for several stimulating discussions. The research of ED  is supported in part by the National Science Foundation under research grant PHY-16-19926, and was supported by a Fellowship from the Simons Foundation in 2017-2018.  The research of JK is supported in part by funding from the Bhaumik Institute. 

\newpage

\section{Modular graph functions and forms}
\setcounter{equation}{0}
\label{sec:2}

In this section, we shall review the definitions and basic properties of modular graph functions and forms, define even and odd modular graph functions, exhibit the action of differential operators, and illustrate the cases of one and two loops.

\subsection{Definitions and basic properties}

A decorated connected graph $(\Gamma, A,B)$ with $V$ vertices and $R$ edges is defined as follows. The connectivity matrix $\Gamma$ of the graph  has components $\Gamma _{v \, r}$ where the index $v=1,\cdots, V$ labels the vertices of the graph and the index $r=1,\cdots, R$ labels its edges. No edge is allowed to begin and end on the same vertex. When edge $r$ contains vertex $v$ we have $\Gamma _{v \, r}=\pm 1 $,  while otherwise we have $\Gamma _{v\, r} = 0$. The  decoration $(A,B)$ of the graph is defined as follows,
\bea
\label{ab}
A = [a_1, \cdots, , a_R] & \hskip 1in & a=a_1+ \cdots + a_R
\no \\
B = [b_1, \cdots, , b_R] && b=b_1+ \cdots + b_R
\eea
where $a_r, b_r \in \CC$ with $a_r-b_r \in \ZZ$ for all $r = 1, \cdots , R$.  The pair $(a_r,b_r)$ is associated with edge $r$, while $a,b$ are associated with the full decorated graph.

\sm

To a decorated graph  $(\Gamma, A, B)$ we associate a complex-valued function  on the Poincar\'e upper half plane $\cH$, defined by the following Kronecker-Eisenstein sum, 
\bea
\label{2a1}
\cC_\Gamma \left [ \begin{matrix} A \cr B \cr \end{matrix} \right ]  (\tau) 
=   \left ( { \tau_2 \over \pi} \right ) ^{ \half a+ \half b} \sum_{p_1,\dots,p_R \in \Lambda '}  ~ \prod_{r =1}^R
  { 1 \over  (p_r) ^{a_r} ~ (\bar p _r) ^{b_r} }\, \prod_{v =1}^V
  \delta \left ( \sum_{s =1}^R \Gamma _{v \, s} \, p_s \right )
\eea 
whenever this sum is absolutely convergent. The momentum lattice  is given by $\Lambda= \ZZ + \tau \ZZ$ for $\tau \in \cH$ with $\Lambda ' = \Lambda \setminus \{ 0 \}$. The momenta~$p_r$ depend  holomorphically  on $\cH$. We  refer to $a_r$ and $b_r$ as the exponents of respectively holomorphic and anti-holomorphic momenta. The Kronecker $\delta$-symbol equals 1 when its argument vanishes and  0  otherwise.  The number of loops $L$  is the number of independent momenta, given by  $L=R-V+1$.  The domain of absolute convergence of the sums in (\ref{2a1}) is given by a system of inequalities on the combinations $\Re (a_r+b_r)$, beyond which the functions $\cC$ of (\ref{2a1}) may be defined by analytic continuation in the variables $a_r+ b_r$.

\sm

The function $\cC_\Gamma$ in (\ref{2a1}) vanishes whenever the integer $a-b$ is odd, or whenever the graph $\Gamma$ becomes disconnected upon severing a single edge. A function $\cC_\Gamma$ associated with a connected graph $\Gamma$ which is the union of two graphs $\Gamma = \Gamma _1 \cup \Gamma_2$ whose intersection $\Gamma _ 1 \cap \Gamma_2$ consists of a single vertex factorizes into the product $\cC_{\Gamma_1} \cC_{\Gamma_2}$ with the corresponding partitions of the exponents.  Henceforth, we shall assume that $a-b$ is even, and that the graph $\Gamma$ remains connected upon the removal of any single edge or vertex.

\sm

Under $SL(2,\ZZ)$, the  functions defined in (\ref{2a1})  transform as follows, 
\bea
\label{mod}
\cC_\Gamma  \left [ \begin{matrix} A \cr B \cr \end{matrix}\right ] \left  ({ \alpha \tau + \beta \over \gamma \tau + \delta} \right ) 
= \left ( { \gamma \tau+\delta \over \gamma  \bar \tau +\delta } \right ) ^{\half a - \half b}
\cC_\Gamma  \left [ \begin{matrix} A \cr B \cr \end{matrix} \right ] (\tau) 
\eea
where $\alpha, \beta , \gamma, \delta \in \ZZ$ and $\alpha \delta - \beta \gamma =1$. The  {\sl modular weight} of  $\cC_\Gamma$ is given by the pair $({a-b \over 2}  , {b-a \over 2})$ which has integer entries since $a-b$ is even.  
For $a \not =b$, the function $\cC_\Gamma$ transforms as a non-holomorphic  {\sl modular graph form}. For $a=b$ the function $\cC_\Gamma$ is $SL(2,\ZZ)$-invariant and referred to as a  {\sl modular graph function} of {\sl weight}\footnote{This {\sl weight} is not to be confused with the {\sl modular weight}, which vanishes for $a=b$.}  $w=a=b$.  

\sm

A modular graph form is  invariant under permutations of its vertices and its pairs of exponents $(a_r,b_r)$. It obeys a momentum conservation identity at each vertex $v=1,\cdots, V$, 
\bea
\label{3d3}
\sum _{r=1}^R \Gamma _{v\, r} \, \cC_\Gamma \left [ \begin{matrix} A -S_r \cr B \cr \end{matrix} \right ]  =
\sum _{r=1}^R \Gamma _{v\, r} \, \cC_\Gamma \left [ \begin{matrix} A \cr B - S_r \cr \end{matrix} \right ]  =0
\eea
The $R$-dimensional array $S_r$ is defined to have zeroes in all slots except for the $r$-th, 
\bea
\label{Sr}
S_r = [ 0_{r-1}, 1, 0_{R-r}]
\eea 
where $0_\ell$ stands for an array of $\ell$ zeros. The momentum conservation identities  provide linear algebraic relations between modular graph forms of the same modular weight.

\subsection{Graphical representation}
Though it will not be used much in the present work, we now briefly review the graphical representation of modular graph functions and forms. The generating function of modular graph functions with $N$ vertices is given as follows, 
\bea
\cB_{N}(s_{ij}|\tau) = \prod_{k=1}^N \int_\Sigma {d^2 z_k \over \tau_2} \,\mathrm{exp}\left(\sum_{1 \leq i < j \leq N} s_{ij} \, G(z_i - z_j | \tau) \right)
\eea
where $\Sigma$ is a torus with complex structure modulus $\tau$ and $G(z_i - z_j | \tau)$ is the scalar Green function on $\Sigma$, given as a Fourier sum by 
\bea
G(z|\tau) = \sum_{p \in \Lambda}' {\tau_2 \over \pi |p|^2}e^{2 \pi i (n \a - m \b)}
\eea
for $z = \a + \b \tau$. Modular graph functions may be represented by Feynman graphs on $\Sigma$ as follows. We begin by representing a Green function graphically by an edge in a Feynman diagram,  
\begin{align}
\begin{tikzpicture}[baseline=-0.5ex,scale=1.7]
%
\draw (1,0) -- (2.5,0) ;
\draw (1,0) [fill=white] circle(0.05cm) ;
\draw (2.5,0) [fill=white] circle(0.05cm) ;
%
\draw (1,-0.25) node{$z_i$};
\draw (2.5,-0.25) node{$z_j$};
\end{tikzpicture}
  =~ G(z_i-z_j|\tau)
\label{fig1}
\end{align}
The integration over the position of a vertex $z$ on which $r$  Green functions end is denoted by an unmarked  filled black dot, in contrast with an unintegrated vertex $z_i$ which is represented by a marked unfilled white dot. The basic ingredients in the graphical notation are depicted in the graph below,
\begin{align}
\begin{tikzpicture}[baseline=-0.5ex,scale=1.7]
%
\draw (2,0.7) -- (1,0) ;
\draw (2,0.7) -- (1.5,0) ;
\draw (2,0.7) -- (2.5,0) ;
\draw (2,0.7) -- (3,0) ;
\draw (2,0.2) node{$\cdots$};
\draw [fill=black]  (2,0.68)  circle [radius=.05] ;
\draw (1,0)    [fill=white] circle(0.05cm) ;
\draw (1.5,0) [fill=white] circle(0.05cm) ;
\draw (2.5,0)    [fill=white] circle(0.05cm) ;
\draw (3,0) [fill=white] circle(0.05cm) ;
%
\draw (1,-0.25) node{$z_1$};
\draw (1.5,-0.25) node{$z_2$};
\draw (2.57,-0.25) node{$z_{r-1}$};
\draw (3.1,-0.25) node{$z_r$};
\end{tikzpicture}
\label{fig2}
= \ \int _\Sigma {d^2 z \over  \tau_2} \, \prod _{i=1}^r G(z-z_i|\tau)
\end{align}
For our purposes, we will be interested only in those cases in which all positions on the torus have been integrated over, and hence all nodes in the diagrams are filled and unmarked.

More generally, modular graph forms may be obtained by considering derivatives of the scalar Green functions, which given rise to differing exponents $(a_r, b_r)$ for the holomorphic and anti-holomorphic momenta $p_r, \bar p_r$. These exponents are represented graphically by decorated edges as follows, 
\begin{align}
\begin{tikzpicture}[baseline=-0.5ex,scale=1.7]
%
\draw (0.7,0) --(1.4,0);
\draw (1.4+0.7,0)--(2.8,0) ;
\draw (0.7,0) [fill=white] circle(0.05cm) ;
\draw (2.8,0) [fill=white] circle(0.05cm) ;
\draw (1.4, -0.15) rectangle ++(0.7,0.3);
\draw(1.75,0) node{$a_r, b_r$};
\end{tikzpicture}
\,\,\,  \approx~ (p_r)^{-a_r} (\bar p_r)^{-b_r}
\label{fig3}
\end{align}
As a simple example, for $\Gamma_{1r}=1$ and $\Gamma_{2r}=-1$, $r=1,2,3$, one has 
\begin{align}
  \begin{tikzpicture}[scale=1.2]
  \begin{scope}[xshift=0cm,yshift=-1cm]
    \draw[directed,thick] (-1.2,-1) node{$\bullet$}..controls(-0.5,0) and (0.5,0).. (1.2,-1) node{$\bullet$};  
    \draw (-0.45, -0.45) [fill=white]rectangle ++(0.9,0.4);
    \draw[directed, thick] (0.75,-0.52) -- (0.85,-0.6) ; 
    \draw (0.02, -0.27) node{$a_1,b_1$};
    \draw (-3,-1) node{ $\cC_{\Gamma}\left[\begin{matrix} a_1 & a_2 & a_3 \\ b_1 & b_2& b_3 \end{matrix} \right] \,\,\,\,\,\, =\,\,\,\,\,\,$};
    \draw[directed,thick] (-1.2,-1) node{$\bullet$}-- (1.2,-1) node{$\bullet$};
    \draw (-0.45, -1.2) [fill=white]rectangle ++(0.9,0.4);
      \draw (0.02, -1) node{$a_2,b_2$};
       \draw[directed, thick] (0.75,-1) -- (0.85,-1) ; 
    \draw[directed, thick] (-1.2,-1) node{$\bullet$}..controls(-0.5,-2) and (0.5,-2).. node{$a_3,b_3$}(1.2,-1) node{$\bullet$};  
    \draw (-0.45, -1.95) [fill=white]rectangle ++(0.9,0.4);
    \draw (0.02, -1.75) node{$a_3,b_3$};
     \draw[directed, thick] (0.75,-1.48) -- (0.85,-1.395) ; 
     \end{scope}
  \end{tikzpicture}
  \label{fig4}
\end{align}

\subsection{Modular graph functions of even and odd parity}

Parity is an anti-holomorphic automorphism of the upper half plane $\tau \to - \bar \tau$ under which an arbitrary modular graph form transforms by swapping the arrays $A$ and $B$,  
\bea
\label{3d2}
 \cC_\Gamma \left [ \begin{matrix} A \cr B \cr \end{matrix} \right ]  (- \bar \tau)
= \cC_\Gamma \left [ \begin{matrix} B \cr A \cr \end{matrix} \right ]  (\tau)
\eea
thereby swapping the modular weights $({a-b \over 2}  , {b-a \over 2})$ and $({b-a \over 2}, {a-b \over 2})$. For $a=b$ it is consistent to decompose an arbitrary modular graph function into {\sl even and odd modular graph functions} $\cC_\Gamma = \cS_\Gamma+ \cA_\Gamma$ which satisfy,  
\bea
 \cS _\Gamma \left [ \begin{matrix} A \cr B \cr \end{matrix} \right ]  (- \bar \tau) = 
 \cS _\Gamma \left [ \begin{matrix} A \cr B \cr \end{matrix} \right ]  (\tau) 
\hskip 1in
\cA _\Gamma \left [ \begin{matrix} A \cr B \cr \end{matrix} \right ]  (- \bar \tau) = 
- \, \cA _\Gamma \left [ \begin{matrix} A \cr B \cr \end{matrix} \right ]  (\tau) 
\eea 
We note that whenever $A=B$ we have $\cA =0$. When the exponents $a_r, b_r$ are all integers, the parity operation on $\cC_\Gamma$ is equivalent to complex conjugation of $\cC_\Gamma$.

\subsection{Action of differential operators}

The choice made for the exponent of the $\tau_2$ prefactor in the definition of (\ref{2a1}) ensures the canonical normalization in which the modular weight vanishes for $a=b$.  For $a\not= b$, however,  there is no such canonical normalization available, and the action of differential operators is made more convenient by changing normalization to modular forms $\cC ^+_\Gamma$ and $\cC ^-_\Gamma$ of respective modular weights $(0,b-a)$ and  $(a-b,0)$ defined as follows,
\bea
\label{Cplus}
\cC_\Gamma  ^\pm\left [ \begin{matrix} A \cr B \cr \end{matrix} \right ] (\tau) 
= (\tau_2)^{\pm {a-b \over 2}} \cC_\Gamma \left [ \begin{matrix} A \cr B \cr \end{matrix} \right ] (\tau) 
\eea
The action of the first order operator $\nabla = 2 i \tau_2^2 \p_{\tau}$ on modular graph forms $\cC^+_\Gamma$
is simple since no connection is required, and the same is true for the action of $\DDb = -2 i \tau_2^2 \p_{\bar \tau}$ on modular graph forms $\cC^-_\Gamma$. The action of the operators $\nabla$ and $\DDb$ may be expressed as follows, 
\bea
\label{nab}
\nabla \cC^+_\Gamma  \left [  \begin{matrix} A \cr B \cr \end{matrix} \right] 
= \sum_{r=1}^R a_r \, \cC^+_\Gamma  \left [  \begin{matrix} A+S_r \cr B-S_r  \cr \end{matrix} \right] 
\hskip 0.8in 
\DDb \cC^-_\Gamma  \left [  \begin{matrix} A \cr B \cr \end{matrix} \right] 
= \sum_{r=1}^R b_r \, \cC^-_\Gamma \left [  \begin{matrix} A-S_r \cr B+S_r  \cr \end{matrix} \right] 
\eea
where the row matrix $S_r$ was defined in (\ref{Sr}).  The operator $\nabla$ maps modular graph forms of modular weight $(0,b-a)$ to those of modular weight $(0,b-a-2)$, and similarly for $\DDb$.

\sm

The Laplace-Beltrami operator on modular graph functions is the Laplace operator on $\cH$ given by $ \Delta =  4 \tau_2^2 \p_{\bar \tau} \p_\tau$ and maps the space of modular graph functions into itself, 
\bea
\label{Lap1}
(\Delta + a ) \, \cC_\Gamma \left [  \begin{matrix} A \cr B \cr \end{matrix} \right ]
= 
\sum _{r,s=1}^R a_r b_s \, \cC_\Gamma \left [ \begin{matrix} A+S_r - S_s \cr B-S_r+S_s  \cr \end{matrix}  \right ]
\eea
The action of the Laplace-Beltrami differentials on modular graph forms may be defined analogously, but will not be needed in this paper. Since the operator $\Delta $ commutes with parity $\tau \to - \bar \tau$, it maps even into even, and odd into odd modular graph functions. The explicit formula for these actions may be obtained by simply substituting $\cC$ on both sides of the equality in (\ref{Lap1}) by either $\cS_\Gamma$ or $\cA_\Gamma$.

\subsection{One-loop modular graph forms and Eisenstein series}

The combinatorial complexity of modular graphs forms increases rapidly with the number of loops. At one loop, the number of vertices equals the number of edges, so that all vertices are bivalent. The case $V=R=1$ is excluded since the edge begins and ends on the same vertex. For $V=R\geq 2$ all one-loop modular graphs forms are given for $a_r - b_r \in \ZZ$ by, 
\bea
\label{2a4}
\cC \left [ \begin{matrix} a_1 & a_2 \cr  b_1 & b_2 \cr \end{matrix} \right ] 
= (-)^{a_2+b_2} \, \cC \left [ \begin{matrix} a & 0 \cr  b & 0 \cr \end{matrix} \right ] 
\eea
where $a,b$ are given by (\ref{ab}). The non-holomorphic Eisenstein series $E_s$ is defined by, 
\bea
\label{Es}
E_s(\tau) = \sum _{ p \in \Lambda '}  { \tau_2^s \over \pi^s \, |p| ^{2s}}
= \sum _{{ m,n \in \ZZ \atop (m,n) \not= (0,0)}}   { \tau _2^s \over \pi^s  |m\tau  + n |^{2s}} 
\eea
The Fourier series of $E_s$ for $s \in \CC$ is given by,
\bea
\label{EF}
E_s(\tau) & = & { 2 \zeta(2s) \over \pi^s} \tau_2^ s + {2 \Gamma (s-\half ) \zeta(2s-1) \over \Gamma (s) \pi^{s-\half} \tau_2^{s-1}}
\no \\ &&
+ { 4 \sqrt{\tau_2} \over \Gamma (s)} \sum _{N=1}^\infty N^{s-\half} \sigma _{1-2s} (N) K_{s-\half} (2 \pi N \tau_2) 
\Big ( e^{2 \pi i N \tau_1} + e^{-2 \pi i N \tau_1} \Big )
\eea
where $K$ is the modified Bessel function and $\sigma_z(N)= \sum _{0<d|N} d^z$ is the divisor sum. The Poincar{\'e} series representation of $E_s$ is given by 
\bea
E_s(\tau) = {2 \zeta(2 s) \over \Gamma(s)} \sum_{\g \in \G_\infty \backslash SL(2, \mathbb{Z})}\left(\mathrm{Im}\g(\tau) \right)^s
\eea
\sm

The one-loop modular graph forms of (\ref{2a4}) may all be expressed in terms of $E_s$. There are no odd modular graph functions to one-loop order, and the Eisenstein series are the only even modular graph functions to one-loop order.   Successive application of the derivatives $\DD$ and $ \DDb$ to $E_s$ produces all one-loop  modular graph forms, 
\bea
\label{delE}
\DD ^k \, E_s
= { \Gamma (s+k) \over \Gamma (s)} \, \cC^+ \left [ \begin{matrix} s+k & 0 \cr  s-k & 0 \cr \end{matrix} \right ] 
\hskip 0.8in 
\DDb ^k \, E_s
= { \Gamma (s+k) \over \Gamma (s)} \, \cC^- \left [ \begin{matrix} s-k & 0 \cr  s+k & 0 \cr \end{matrix} \right ] 
\eea
For integer $s=k \geq 2$ the derivative is proportional to a holomorphic Eisenstein series $G_{2k}$,
\bea
\label{Gk}
\DD^k E_k (\tau)={ \Gamma (2k) \over \Gamma (k)} \,  (\tau_2)^{2k} G_{2k}(\tau)
\hskip 1in
G_{2k} (\tau) = \sum _{p \in \Lambda '}  { 1 \over \pi ^k \, p^{2k}}
\eea
For integer $s$ the Fourier series of $E_s$ simplifies and may be recast as follows, 
\bea
\label{Eas}
E_k(\tau) &=& - { B_{2k} \over (2k)!} (-4 \pi \tau_2)^k 
+ { 4 \, (2k-3)! \, \zeta (2k-1) \over (k-2)! \, (k-1)! \, (4 \pi \tau_2)^{k-1}}
\no\\
&\vphantom{.}&\hspace{0.1in}+{2\over (k-1)! }\sum_{N=1}^\infty N^{k-1}\sigma_{1-2k}(N)P_k(4N\pi \tau_2) \left(q^N + \bar q^N \right) 
\eea
where $q=e^{2 \pi i \tau}$, $B_{2k}$ are the Bernoulli numbers,  and $P_k(x)$ is a polynomial in $1/x$ given by, 
\bea
P_k(x) = \sum_{m=0}^{k-1}{(k+m-1)! \over m! \, (k-m-1)! \, x^m}
\eea

\subsection{Two-loop modular graph forms}

Two-loop modular graph forms are given by the following Kronecker-Eisenstein sum,
\bea
\label{4a1}
\cC \left [ \begin{matrix} a_1 & a_2 & a_3 \cr b_1 & b_2 & b_3 \cr \end{matrix} \right ]
= 
\sum _{p_1, \, p_2, \, p_3 \in \Lambda '} \,
{ \tau_2^w \, \delta _{p_1+p_2+p_3,0} \over  \pi^w \, p_1^{a_1} \, p_2^{a_2} \, p_3^{a_3} \, 
\bar p_1^{b_1} \, \bar p_2^{b_2} \, \bar p_3^{b_3} }
\eea 
where the weight $w$ is given by,
\bea
2w = a+b=
a_1+a_2+a_3+b_1+b_2+b_3
\eea
For $a_r, b_r \in \ZZ$ with $r=1,2,3$, the  sum is absolutely convergent provided $a_r+b_r+a_s+b_s \geq 3$ for any pair  $r \not=s$, which requires $w\geq 3$. Furthermore, we shall restrict attention to the case where all exponents are non-negative, as is natural from the point of view of string theory. 

\sm

The space of modular graph forms thus obtained contains forms that are effectively one-loop. Forms with $b_r=0$ and $a_r \geq 2$ are proportional to holomorphic modular forms $G_{2w}$. When both exponents of a given momentum vanish, we may use the {\sl algebraic reduction relation} of \cite{DHoker:2016mwo,DHoker:2016quv} to relate the form to  one-loop forms, 
\bea
\label{algred}
\cC  \left [ \begin{matrix} a_1 & a_2 & 0 \cr b_1 & b_2 & 0 \cr \end{matrix} \right ] 
=
\cC  \left [ \begin{matrix} a_1  & 0 \cr b_1& 0 \cr \end{matrix} \right ]
\cC  \left [ \begin{matrix}  a_2 & 0 \cr  b_2 & 0 \cr \end{matrix} \right ]
-(-)^{a_2+b_2} \cC  \left [ \begin{matrix} a & 0  \cr b & 0 \cr \end{matrix} \right ]
\eea
where $a=a_1+a_2$ and $b=b_1 + b_2$. Finally, when two anti-holomorphic exponents vanish, {\sl holomorphic subgraph reduction} may be used to relate the form to a one-loop form, as will be reviewed in more detail in Section \ref{sec:6}.

 The following proposition provides a systematic decomposition of two-loop modular graph forms into a reduced set of such forms,

{\prop
\label{prop31}
{\sl Two-loop modular graph forms with $a_r+a_s+b_r+b_s \geq 3$ for all $r \not=s$ and $a_1,a_2,b_1,b_2\geq 1$  admit the following decomposition, 
\bea
\label{thm1}
\cC \left [ \begin{matrix} a_1 & a_2 & a_3 \cr b_1 & b_2 & b_3 \cr \end{matrix} \right ]
& = & 
 \sum _{k=1}^{a_1} \Lambda_k(a_1,a_2)   \sum _{\ell=1}^{b_1} \Lambda_\ell(b_1,b_3) \, 
 \cC  \left [ \begin{matrix} a-k & 0 & k \cr 0 & b- \ell & \ell \cr \end{matrix} \right ] + (1 \leftrightarrow 2)
\\ && 
+  \sum _{k=1}^{a_1} \Lambda_k(a_1,a_2)   \sum _{\ell=1}^{b_3} \Lambda_\ell(b_3,b_1) \, 
 \cC \left [ \begin{matrix} k & 0 & a-k \cr 0 & b-\ell & \ell \cr \end{matrix} \right ] + (1 \leftrightarrow 2)
 \quad
\no \eea
 where the term $(1 \leftrightarrow 2)$ is obtained from the preceding term by swapping the pairs $(a_1, b_1)$ with $(a_2, b_2)$ leaving $(a_3, b_3)$ unchanged, and $\Lambda$ is proportional to a binomial coefficient,
\bea
\label{AA}
\Lambda_k (a_1, a_2) = (-)^{a_1+a_2+k} \mbinom{a_1+a_2-k-1}{a_2-1}
\eea
Odd (resp. even) two-loop modular graph functions admit the same decomposition with the symbol $\cC$ replaced by $\cA$ (resp. $\cS$) on both sides of the equality.}}

 To prove the proposition, we  perform a partial fraction decomposition of the holomorphic  factor of the summand in terms of the momentum variable $p_3=-p_1-p_2$, and obtain,  
\bea
\label{4a2}
{1 \over p_1^{a_1} p_2^{a_2} p_3^{a_3}} =
 \sum _{k=1}^{a_1} {\Lambda_k(a_1,a_2)  \over p_1^k \, p_3^{a-k} }
+ \sum _{k=1}^{a_2} {\Lambda_k (a_2,a_1) \over p_2^k \, p_3^{a-k} }
\eea
Using the restriction $b_r \geq 0$, the anti-holomorphic factor of the summand multiplying the first sum in (\ref{4a2}) may be decomposed as follows,
\bea
\label{4a3}
{1 \over \bar p_1^{b_1} \bar p_2^{b_2} \bar p_3^{b_3}} =
 \sum _{\ell=1}^{b_1} {\Lambda_\ell(b_1,b_3)  \over \bar p_1^\ell \, \bar p_2^{b-\ell} }
+ \sum _{\ell=1}^{b_3} {\Lambda_\ell (b_3,b_1) \over \bar p_3^\ell \, \bar p_2^{b-\ell} }
\eea
while the factor multiplying the second term in (\ref{4a2}) is given by (\ref{4a3}) with the indices $1$ and $2$ swapped. The resulting decomposition is then given by (\ref{thm1}). Symmetrization under $\tau \to - \bar \tau$ then immediately gives the final statement.

{\cor
\label{prop34}
{\sl For  $a_1+b_1, a_2+b_2  \geq 3$, we have the decomposition, 
\bea
\label{3b31}
 \cC  \left [ \begin{matrix} a_1 & a_2 & 0 \cr b_1 & b_2 & 0 \cr \end{matrix} \right ]
  =  \sum_{k=1}^{a_1} \Lambda_k(a_1,a_2) \, \cC  \left [ \begin{matrix} a-k & k & 0  \cr 0 & b_1 & b_2  \cr \end{matrix} \right ]
+ (1 \leftrightarrow 2)
 \eea
where $a=a_1+a_2 $.}}

The proof is an immediate consequence of Proposition~\ref{prop31} obtained by setting $a_3=b_3=0$.

{\cor
\label{prop33}
{\sl For  $k, \ell, a-k, b - \ell  \geq 1$, and $a+b-k-\ell \geq 3$, the reduced modular graph forms are related by the reflection formula,}}
\bea
\label{3b32}
 \cC  \left [ \begin{matrix} a-k & k & 0 \cr 0 & \ell & b-\ell \cr \end{matrix} \right ]
 & = & \sum_{m=1}^{a-k} \Lambda_m(a-k,k) \, \cC  \left [ \begin{matrix} m & a-m & 0 \cr 0 & b- \ell & \ell \cr \end{matrix} \right ]
 \no \\ &&
 + \sum_{m=1}^k \Lambda_m(k,a-k) \, \cC  \left [ \begin{matrix} m & a-m & 0 \cr \ell & b- \ell & 0 \cr \end{matrix} \right ]
 \eea
The proof proceeds by partial fraction decomposition of the holomorphic momenta onto~$p_3$. The modular graph functions appearing on the last line of Corollary \ref{prop33} may be expressed in terms of one-loop  functions using the algebraic reduction formula of (\ref{algred}) and in turn may be expressed in terms of Eisenstein series and their derivatives using (\ref{delE}).

\newpage

\section{Poincar\'e series for two-loop modular graph functions}
\setcounter{equation}{0}
\label{sec:3}

In this section, we shall construct the Poincar\'e series for two-loop modular graph functions  with respect to the coset $\Gamma _\infty \backslash \Gamma$ with $\Gamma = PSL(2,\ZZ)$, and outline the generalization of the construction to higher loop orders.

\subsection{Poincar\'e series for two loops}

For integer exponents we use Proposition \ref{prop31} to recast an arbitrary  modular graph function of weight $w=a=b$  as a linear combination of modular graph functions of the form, 
\bea
\label{Cuvw}
\cC_{u,v;w} (\tau)=  \cC  \left [ \begin{matrix} u & 0 & w-u \cr 0 & v & w-v \cr \end{matrix} \right ](\tau)
\eea
where $u,v,w$ are integers satisfying  $1 \leq u,v \leq w-1$. We furthermore restrict to $u+v>2$, lest the functions be divergent. These functions satisfy the complex conjugation condition  
$\cC_{u,v;w} = (\cC_{v,u;w})^*$.  Their Kronecker-Eisenstein sum is given as follows, 
\bea
\label{Adef2}
\cC_{u,v;w} (\tau)
=  \sum _{{m_r, n_r \in \ZZ \atop (m_r,n_r) \not = (0,0) }}  { \tau_2^w \over \pi^w} 
\left ( { m_2 \tau + n_2  \over m_1 \tau + n _1 } \right )^u
\left ( { m_2 \bar \tau  + n_2  \over m_3 \bar \tau  + n _3} \right )^v
{ \delta _{m,0} \delta _{n,0}  \over |m_2 \tau +n_2 |^{2w}}
\eea
where  $m=m_1+m_2+m_3$,  $n=n_1+n_2+n_3$ and the Kronecker $\delta_{m,0}\delta_{n,0}$ restricts the sum to contributions for which $m=n=0$.  

\sm

We shall obtain the Poincar\'e series for a generalization of the  functions $\cC_{u,v;w}$ where $u,v$ are positive integers but $w$ is analytically continued to $\CC$. This generalization will be useful when we evaluate Petersson inner products of modular graph functions.

{\thm
\label{thm52}
{\sl The Poincar\'e series for  $\cC_{u,v;w}(\tau) $ with respect to $\Gamma _\infty \backslash \Gamma $ for $u,v \in \NN$ with $u+v > 2$, and $w \in \CC$ with $\Re(w) >1+\thalf \max(u,v)$, is absolutely convergent and given by,}  
\bea
\label{CLambda}
\cC_{u,v;w}  (\tau)  = \sum _{g \in \Gamma _\infty \backslash \Gamma} \Lambda _{u,v;w} (g \tau)
\eea
{\sl where the seed function $\Lambda _{u,v;w} ( \tau)$ is given by,}
\bea
\label{Lambda}
\Lambda _{u,v;w} (\tau)  & = &   \ell_w (4 \pi \tau_2)^w  +  (-)^{u+v} \Lambda '_{u,v;w}(\tau) 
\no \\
\ell _w & = & \sum_{k=0}^{[u/2]} { 4 (-)^{u+v} \binom{u+v-2k-1}{v-1} \over (2\pi)^{2w} } \zeta (2k)\zeta(2w-2k)
+ (u \leftrightarrow v)
\no \\
\Lambda ' _{u,v;w} (\tau) & = &
 \sum _{m\not= 0} \, \sum _{\mu \in \ZZ} \, \sum _{n \not= 0} \, {\tau_2^w \over \pi^w} \,
{{(-)^{u} n^{u+v} } \over |n|^{2w} \, (m\tau+\mu)^u  \, (m \bar \tau + \mu +n)^v} 
\eea
}
We shall prove Theorem \ref{thm52} with the help of two lemmas, of which the first is standard  \cite{Apostol}.

{\lem 
\label{lem53}
{\sl For every pair $(m_2, n_2) \in \ZZ^2 \setminus \{(0,0) \}$ there exists a unique pair $(0,n)$ and a unique 
$g \in \Gamma _\infty \backslash \Gamma$, such that we have the following matrix relation,} 
\bea
(m_2~~ n_2) = (0~~ n) g 
\hskip 1in 
g =\left ( \begin{matrix} \a & \b \cr \gamma & \delta \cr \end{matrix} \right )
\hskip 0.5in \alpha, \beta, \gamma, \delta \in \ZZ
\eea
{\sl with $\alpha \delta - \beta \gamma=1$ and $n=\pm \gcd(m_2, n_2)$. The unique solution is given by  $\gamma = m_2/n$ and $\delta = n_2/n$ such that $\delta >0$ if $n_2 \not=0$ and $\gamma >0$ if $n_2=0$. }}

\sm

To prove Lemma \ref{lem53}, we show that the equations $m_2=n\gamma$, $n_2 = n \delta$, and $\alpha \delta - \beta \gamma=1$ have a unique solution given the conditions of the lemma. Since $(m_2, n_2 ) \not=0$ and $\gcd(\gamma , \delta)=1$, we have $n=\pm \gcd(m_2, n_2)$ where the greatest common divisor is defined to be positive. The sign may be fixed by  using $g \in \Gamma$ which allows us to choose $\delta >0$ if $n_2\not=0$ and $\gamma >0$ if $n_2 =0$. Thus $\gamma$ and $\delta$ are determined uniquely. The group $\Gamma _\infty$ is given by,
\bea
\Gamma _\infty = \left \{ \left ( \begin{matrix} 1 & \nu \cr 0 & 1 \cr \end{matrix} \right ) , ~ \nu \in \ZZ \right \}
\eea
Acting to the right on $(0,n)$ leaves $(0,n)$ invariant. Acting to the left on $g$ transforms $\alpha \to \alpha + \nu \gamma$ and $\beta \to \beta + \nu \delta$ for $\nu \in \ZZ$ mapping all solutions of the equation $\alpha \delta - \beta \gamma =1$ for given $\gamma, \delta$, into one another. Therefore, the coset element $g \in \Gamma _\infty \backslash \Gamma$ is unique. This completes the proof of Lemma \ref{lem53}.

\sm

{\lem 
\label{lem54}
{\sl The following Poincar\'e series representation is absolutely convergent\footnote{We thank Axel Kleinschmidt and Daniele Dorignoni for bringing their work on Poincar\'e series of modular graph functions in  \cite{Ahlen:2018wng,Dorigoni:2019yoq} to our attention, and for questions on convergence that led the authors to the proof of absolute convergence included here.} for $a_i \in \RR$  with $a_i+a_j > 1$
for all  $1\leq i \not= j \leq 3$, 
\bea
C_{a_1, a_2, a_3}(\tau)= \cC \left [ \begin{matrix} a_1 & a_2 & a_3 \cr a_1 & a_2 & a_3 \cr \end{matrix} \right ](\tau)
=  { (2\pi)^{2a} c_a \over 2 \zeta (2a)} E_a(\tau) + \sum _{g \in \Gamma _\infty \backslash \Gamma} 
\lambda _{a_1, a_2, a_3} (g\tau)
\eea
 $a=a_1+a_2+a_3$. The modular graph function $C_{a_1, a_2, a_3}$ was defined in (\ref{4a1}). The seed function $\lambda _{a_1, a_2, a_3} (\tau)$ is given by, 
\bea
\label{lambda}
\lambda _{a_1, a_2, a_3} (\tau)=  \sum _{m,n\not= 0} \, \sum _{\mu \in \ZZ}  
{ \tau_2^a \over  \pi^a n^{2a_2} \, |m\tau+\mu|^{2a_1}   \, |m  \tau + \mu +n|^{2a_3}} 
\eea
and $c_a$ is independent of $\tau$ and was given in equation (1.20) of \cite{DHoker:2017zhq}.}}

\sm

To prove Lemma \ref{lem54}, we use the results and the notations of Lemma \ref{lem53} to obtain,
\bea
\label{decomp}
{ |m_2 \tau+n_2|^2 \over \tau_2} = { n^2 \over \tau_2'} 
\hskip 0.6in 
{ m_2 \tau + n_2  \over m_r \tau + n _r } = { n \over m_r' \tau ' + n_r'}
\hskip 0.6in 
\tau ' = g \tau = {\alpha \tau + \b \over \gamma \tau + \delta}
\eea
where $(m_r, n_r) \not= (0,0)$ for $r=1,3$, $n \not=0$, and subject to the conditions $m_1'+m_3'=0$, and $n_1'+n+n_3'=0$. Parametrizing these variables by $n$, $m_1'=-m_3'=m$, $n_1'=\mu$, and $n_3'=-n-\mu$, and substituting these results into the Kronecker-Eisenstein sum which defines $C_{a_1, a_2, a_3}$ in  (\ref{4a1}), we prove the Poincar\'e series representation of Lemma \ref{lem54}. To prove that the Poincar\'e series is absolutely convergent, we use the fact that for $a_i \in \RR$, each term in the Kronecker-Eisenstein sum defining $C_{a_1, a_2, a_3}$ in  (\ref{4a1}) is real and positive. Recasting this sum in the form of a Poincar\'e series simply amounts to a rearrangement of the infinite series, which is always permitted by Tonelli's theorem since all terms are positive. 
This completes the proof of Lemma \ref{lem54}.  

\sm

To prove the Poincar\'e series representation of Theorem \ref{thm52}, we use Lemma \ref{lem53} and the decomposition formula (\ref{decomp}) of the Kronecker-Eisenstein sum into orbits under $\Gamma _\infty \backslash \Gamma$ to derive (\ref{Lambda}), in parallel with the derivation of $\lambda$ in Lemma \ref{lem54}. To prove absolute convergence of the Poincar\'e series (\ref{CLambda}),  we bound the series as follows, 
\bea
\label{neededineq}
\left | \cC_{u,v;w} - { (2\pi)^{2w} \ell_w \over 2 \zeta (2w)} E_w \right | \leq 
\sum _{g \in \Gamma _\infty \backslash \Gamma} \lambda _{a_1, a_2, a_3}(g \tau)
\eea
where $\lambda _{a_1, a_2, a_3}(\tau)$ was defined in (\ref{lambda}) and $a_1, a_2, a_3$ are given in terms of $u,v,w$ by $ a_1= \thalf u$, $ a_3 = \thalf v$, $ a_2 = \Re(w) -a_1-a_3$.  The left-hand side of (\ref{neededineq}) is the Poincar{\'e} series over $\Lambda'_{u,v;w}(\tau)$. Since the assumptions of Theorem \ref{thm52} include $u+v > 2$ and $\Re(w) > 1 + \thalf \max(u,v)$, it follows that $a_i+a_j >1$ and by Lemma \ref{lem54} the series in $\lambda _{a_1, a_2, a_3}(\tau)$ is absolutely convergent. This completes the proof of 
Theorem \ref{thm52}.

\subsection{Calculating the seed function for integer $w$}

Throughout this paper we shall use the following partial fraction decomposition formula, 
\bea
\label{parfrac}
{ 1 \over (\mu +x)^u (\mu +y)^v} = 
\sum _{k=1}^u {(-)^v \binom{u+v-k-1}{u-k}  \over (\mu +x)^k(x-y)^{u+v-k}} 
+ \sum _{k=1}^v { (-)^u \binom{u+v-k-1}{v-k}  \over (\mu +y)^k (y-x)^{u+v-k}}
\eea
for $u,v \in \NN$. We shall now derive an explicit formula for the seed function $\Lambda _{u,v;w}(\tau)$. 
Its first contribution involves the coefficient $\ell_w$ which is given by the $m=0$ term,
\bea
\label{cw}
\ell_w = {{(-)^{v}} \over (2 \pi)^{2w}} \sum _{n \not=0} \sum _{\mu \not= 0,-n} { 1 \over n^{2w-u-v} \mu^u (\mu+n)^v}
\eea
The summation over $\mu$ in (\ref{cw}) may be carried out using the partial fraction decomposition of the $\mu$-dependent factors with $x=0$ and $y=n$ in (\ref{parfrac}). For integer $w$, the remaining sum over $n$ gives even $\zeta$-values resulting in the formula for $\ell_w$ in (\ref{Lambda}). We may express the even zeta-values $\zeta(2k)$ in terms of Bernoulli numbers  $B_{2k}$ using the relation,
\bea
\zeta (2k) = \half (2 \pi)^{2k} (-)^{k+1} {B_{2k} \over (2k)!}
\eea 
and obtain the following equivalent formula,
\bea
\ell_w =  \sum_{k=0}^{[u/2]} \mbinom{u+v-2k-1}{v-1} { (-)^{u+v+w} B_{2k} B_{2w-2k} \over (2k)! (2w-2k)!} + ( u \leftrightarrow v)
\eea
This expression agrees with the result $(-)^w c_w$ found for even $u$ and $v $ in \cite{DHoker:2017zhq}, and generalizes that result for all $u,v$. The second contribution $\Lambda'$ is evaluated using the following lemma.

{\lem
\label{Lam}
{\sl For integer $w$, the  function $\Lambda'_{u,v;w}(\tau)$  is given by,} 
\bea
\label{Lam0}
\Lambda' _{u,v;w} (\tau) = 2 {(-\tau_2)^w \over \pi^w} 
\sum _{k=1}^u  \mbinom{u+v-k-1}{v-1} \sum _{m=1}^\infty
 \Phi_{2w-u-v;u+v-k} (2 m \tau_2) \,  \f_k (q^m) + (u \leftrightarrow v)^*
 \qquad
 \eea
{\sl where $q=e^{2 \pi i \tau}$ and the notation  $ (u \leftrightarrow v)^*$ stands for the complex conjugate of the preceding term in which $u$ and $v$ have been swapped.  The functions $\f_k$ and $\Phi_{a;b}$ are defined for integer $k,a,b$ with  $k \geq 2$ and $a+b \geq 2$ by the following infinite sums, }
\bea
\label{Lam1}
\f_k (e^{2 \pi i z} ) = \sum _{\mu \in \ZZ} { i^k \over (z+\mu )^k}
\hskip 1in
\Phi _{a;b} (z) = \sum _{n \not= 0} { (-i)^{a+b} \over n^a (n-iz)^b}
\eea
{\sl while for $k=1$ one has $ \f_1(q) =   \pi \, { 1+ q \over 1 - q}$. Explicit formulas for $\f_k$ will be given in (\ref{ff1}) and (\ref{ff2}) and for $\Phi_{a;b}$ in (\ref{ff3}).}}

\sm

The powers of $i$ are included  to give the functions  simple conjugation properties,
\begin{align}
\label{Lam2}
\overline{\f_k(e^{2 \pi i z}) } & = \f_k (e^{-2 \pi i \bar z}) &
\overline{\Phi_{a;b}(z)} & =  \Phi _{a;b} (\bar z) 
\no \\
\f_k(e^{-2 \pi i z})  & = (-)^k \f_k (e^{2 \pi i z}) &
\Phi_{a;b}(- z) & =  (-)^{a+b} \Phi_{a;b}(z)
\end{align}
In particular, the functions $\f_k$ and $\Phi_{a;b}$ are real functions of their respective argument.  The proof of Lemma \ref{Lam} proceeds by first computing the sum over $\mu \in \ZZ$ using the standard partial fraction decomposition formula of (\ref{parfrac}) with $u,v \in \NN$, $x=m \tau$, and $y= m \bar \tau +n$, and expressing the result of the summation in terms of $\f_k$, 
\bea
\sum _{\mu \in \ZZ}  {(-)^v  \over  (m \tau+\mu)^u  \, (m \bar \tau + \mu +n)^v} 
= \sum _{k=1} ^u { (-i)^k \binom{u+v-k-1}{v-1}  \over (2mi \tau_2-n)^{u+v-k}} \f_k (q^m) 
+ (u \leftrightarrow v)^*
\eea
Importantly, the factor $\f_k (q^m)$ is independent of $n$ which allows us to express the sum over $m$ in terms of the function $\Phi_{a;b}(2 m \tau_2)$ defined above, and this gives the expression in (\ref{Lam0}).

\subsection{Evaluating the functions $\f_k $ and $\Phi_{a;b}(z)$}

The functions $\f_k(q)$ for $k \geq 2$ may be expressed as derivatives of the function $\f_1(q)$, 
\bea
\label{ff1}
 \f_k (q ) = { (-i)^{k-1} \over \Gamma (k) } { \p^{k-1} \over \p z^{k-1}} \, \f_1( q )
 \hskip 1in 
 \f_1(q) =   \pi \, { 1+ q \over 1 - q}
\eea
Their Fourier series expansion for $|q|<1$ is obtained  as follows, 
\bea
\label{ff2}
\f_k(q) = \pi \delta _{k,1} + {(2 \pi)^k \over \Gamma (k)} \sum _{p=1}^\infty   p^{k-1}  \, q ^p
\eea

\sm

To evaluate the function $\Phi_{a;b}(z)$ for integer $a,b \geq 1$ we perform first the sum over $n$, using the partial fraction decomposition formula (\ref{parfrac}) for $\mu=n, x=0, y=-iz, u=a, v=b$, together with the definition of the function $\f_k$. We find,
\bea
\label{ff3}
\Phi _{a;b}(z)
=  \sum _{\alpha =0} ^{[a/2]}  
{(-)^{a+ \alpha} \, 2 \zeta (2 \alpha) \binom{a+b-2\a-1}{\b-1}  \over z^{a+b-2 \alpha} }
+ \sum _{\beta =1}^b 
{ (-)^a \, \binom{a+b-\beta -1}{a-1}  \f_\beta (e^{-2\pi z}) \over z^{a+b-\beta} }
\eea
The contribution from $\alpha=0$ in the sum over $\alpha$ arises from the subtraction required to exclude the  $m=0$ term when recasting the  sum over $n$ in terms of the function $\f_\beta$. Despite the appearance of poles at $z=0$  in the individual terms of the sum in (\ref{ff3}), the function $\Phi_{a;b}(z)$ is analytic near $z=0$ and,  for $a,b \geq 1$ with $z$ in the unit disc $|z|<1$, is given by the following Taylor expansion,
\bea
\label{Phiz}
\Phi_{a;b} (z) = - (2 \pi )^{a+b} \sum_{k=0}^\infty {\Gamma (k+b) B_{a+b+k} \over \Gamma (b) \Gamma (a+b+k+1) k!} (-2 \pi z)^k
\eea
The  $B_{a+b+k}$ vanish for $a+b+k$ odd since we have $a+b+k \geq 2$. The expression (\ref{Phiz})  confirms the earlier observation  that $\Phi_{a;b}(z) $ is a real function of $z$ with parity $(-)^{a+b}$.

\sm

We stress that both $\f_k$ and $\Phi_{a;b}$ are elementary functions of their argument.

\subsection{Analytic continuation in $w$}

The multiple summation for the seed function in Theorem \ref{thm52} is absolutely convergent for $u,v\in \NN$, $\Re(2w-u) >2$, and $\Re(2w-v)>2$, and the result may be analytically continued in $w$ to all of $\CC$ minus some poles. For $w \in \CC \setminus \NN$, the decomposition of Lemma \ref{Lam} in terms of elementary functions $\Phi_{a;b}$  is no longer available. Instead we have, 
\bea
\label{Ups}
\Lambda _{u,v;w}' & = & 2 { \tau_2^w \over \pi^w} \sum _{k=1}^u \mbinom{u+v-k-1}{v-1} 
\sum_{m=1 }^\infty \f_k(q^m) \, \Upsilon_{w,k,u+v}(2m\tau_2) 
\no \\ && 
+2 { \tau_2^w \over \pi^w} \sum _{k=1}^v \mbinom{u+v-k-1}{u-1} 
\sum_{m=1}^\infty \f_k(\bar q^m) \, \Upsilon_{w,k,u+v}(2m\tau_2)
\eea
where $\Upsilon$ is given by,
\bea
\Upsilon _{w, k, \ell} (z) = \sum_{n \not=0} { i^k n^\ell \over |n|^{2w} (n-iz)^{\ell-k}}
\eea
For integer $w$ we recover the expressions given in (\ref{Lambda}), but for arbitrary $w \in \CC$ we can no longer use partial fraction decomposition  to simplify $\Upsilon $. Its Taylor series expansion, 
\bea
\Upsilon_{w,k,\ell} = \sum_{m = [(k+1)/2]} ^\infty { 2 (-)^m \Gamma (\ell-2k+2m) \over \Gamma (\ell-k) \Gamma (2m+1-k) }  \zeta (2w+2m-2k) z^{2m-k}
\eea
is valid for all $w \in \CC$.

\subsection{Higher loops}
 
We shall generalize the construction of Poincar\'e series to higher loops and focus on the case of dihedral modular graphs functions whose Kronecker-Eisenstein sum is given by,  
 \bea
\label{4a11}
\cC \left [ \begin{matrix} a_1 & \cdots & a_R \cr b_1 & \cdots & b_R \cr \end{matrix} \right ]
= 
\sum _{p_1, \cdots, p_R \in \Lambda '} \,
{ \tau_2^w \, \delta _{p_1+\cdots+p_R,0} \over  \pi^w \, p_1^{a_1} \, \cdots \, p_R^{a_R} \, 
\bar p_1^{b_1} \cdots  \bar p_R^{b_R} }
\eea 
Here,  the number of loops is $R-1$ and we have $w=a=b$ with $a,b$ defined in (\ref{ab}). We single out any one of the  momentum edges, say $p_R=m_R\tau + n_R$, and apply Lemma \ref{lem53} to the pair $(m_R, n_R)$. In matrix notation, we thus recast the pair as follows, 
\bea
(m_R ~~ n_R)= (0~~ n) g 
\hskip 1in 
g =\left ( \begin{matrix} \a & \b \cr \gamma & \delta \cr \end{matrix} \right )
\hskip 0.5in \alpha, \beta, \gamma, \delta \in \ZZ
\eea
with $\alpha \delta - \beta \gamma=1$ and $n=\pm \gcd(m_R, n_R)$. The unique solution is given by  $\gamma = m_R/n$ and $\delta = n_R/n$ such that $\delta >0$ if $n_R \not=0$ and $\gamma >0$ if $n_R=0$. By generalizing the arguments used in the two-loop case, we thus have, 
\bea
\cC \left [ \begin{matrix} a_1 & \cdots & a_R \cr b_1 & \cdots & b_R \cr \end{matrix} \right ](\tau)
=
\sum _{g \in \Gamma _\infty \backslash PSL(2,\ZZ)} \Lambda \left [ \begin{matrix} a_1 & \cdots & a_R \cr b_1 & \cdots & b_R \cr \end{matrix} \right ] (g \tau)
\eea
The seed function $\Lambda$  is given by, 
\bea
 \Lambda \left [ \begin{matrix} a_1 & \cdots & a_R \cr b_1 & \cdots & b_R \cr \end{matrix} \right ] ( \tau)
=
\sum _{p_1, \cdots, p_{R-1} \in \Lambda '} \, \sum_{n \not=0} 
{ \tau_2^w \, \over \pi^w n^{a_R+b_R}} \, 
{ \delta _{p_1+\cdots+p_{R-1}+n,0} \over p_1^{a_1} \, \cdots \, p_{R-1}^{a_{R-1}} \, 
\bar p_1^{b_1} \cdots  \bar p_{R-1}^{b_{R-1} } }
\eea 
Effectively the sum over  $p_R$ in $\cC$ is replaced by a sum over $p_R=(0 ~~n)$ with $n \not=0$ in $\Lambda$.
 \newpage

\section{Fourier series for two-loop modular graph functions}
\setcounter{equation}{0}
\label{sec:4}

In this section we shall evaluate the Fourier series expansion of an arbitrary two-loop modular graph function with integer exponents.  As in the previous section we use Proposition \ref{prop31} to reduce this calculation to evaluation of the Fourier series of the reduced modular graph functions $\cC_{u,v;w} (\tau)$, defined in (\ref{Cuvw}),  for $u,v,w \in \NN$. We require $u+v \geq 3$, $2w-u \geq 3$, and $2w -v \geq 3$ to ensure convergence of the Kronecker-Eisenstein series which define $\cC_{u,v;w}$ in (\ref{Adef2}), and these conditions require $w \geq 3$. The Fourier decomposition of $\cC_{u,v;w}(\tau)$ is given by the following theorem.

{\thm 
\label{thm7}
{\sl The two-loop modular graph function $\cC_{u,v;w}$ with integer exponents $u,v,w\in\NN$ admits the following Fourier series expansion, 
\bea
\label{Four}
\cC_{u,v;w}(\tau) = \cL_{u,v;w}  (\tau_2) + \sum _{N=1}^\infty \cQ_{u,v;w} ^{(N)} (\tau_2) (q\bar q)^N  + \sum _{N=1}^\infty \left ( \cF_{u,v;w} ^{(N)} (\tau_2) q^N + \cF_{v,u;w} ^{(N)} (\tau_2) \bar q^N \right )
\eea
Here, the functions occurring in the constant Fourier mode, $\cL_{u,v;w}  (\tau_2)$ and $\cQ_{u,v;w} ^{(N)} (\tau_2)$, are real-valued Laurent polynomials in $\tau_2$. The functions $\cF^{(N)}_{u,v;w}(\tau_2)$ occurring in the non-constant Fourier modes  are real-valued functions which have an expansion in powers of $q\bar q$,
\bea
\label{Four2}
\cF_{u,v;w} ^{(N)} (\tau_2)  =  F_{u,v;w} ^{(N)} (\tau_2) 
+\sum _{L=1}^\infty G_{u,v;w}^{(N,L)} (\tau_2) (q \bar q)^L
\eea
where $F_{u,v;w} ^{(N)} (\tau_2)$ and $G_{u,v;w}^{(N,L)} (\tau_2)$ are real-valued Laurent polynomials in $\tau_2$.
The degrees of these Laurent polynomials are given as follows,
\bea
\deg \cL_{u,v;w}   & = & (w,\, 1-w)
\no \\
\deg \cQ_{u,v;w} ^{(N)} & = & (w-2,\, w-\b_+-1)
\no \\
\deg F_{u,v;w} ^{(N)}  & = & (w-1,\, w-\ell_+-\g_+)
\no \\
\deg G_{u,v;w}^{(N,L)} & = & (w-2,\, w-\ell_+-\b_+)
\eea
where $\b_+,\ell_+,$ and $\g_+$ are defined as
\bea
\beta_+ &=&  \mathrm{max} \left(w+|u-v|-1,\, u+v-1 \right)
\no\\
\ell_+ &=& \mathrm{max} \left(w-u, u \right)
\no\\
\g_+ &=& \mathrm{max} \left(u+v, \, 2w-u-v-\eps \right)
\eea Explicit formulas for these Laurent polynomials will be given in subsequent sections.}}

{\cor 
{\sl An immediate corollary of Theorem \ref{thm7}  is obtained for odd two-loop modular graph functions with integer exponents defined by $\cA_{u,v;w}(\tau) = \cC_{u,v;w}(\tau) - \cC_{v,u;w}(\tau)$. Their Fourier expansion is given by, 
\bea
\label{AF}
\cA_{u,v;w}(\tau) =  \sum _{N=1}^\infty \Big ( \cF_{u,v;w} ^{(N)} (\tau_2) - \cF_{v,u;w} ^{(N)} (\tau_2) \Big )  \left ( q^N - \bar q^N \right )
\eea
so that $\cA_{u,v;w}(\tau)$  is a cuspidal function with exponential decay at the cusp.}}

\subsection{Decomposition of the Kronecker-Eisenstein sum}

To prove Theorem \ref{thm7} and calculate the coefficient functions in the Fourier series (\ref{Four}), we start from  the Kronecker-Eisenstein sum representation of $\cC_{u,v;w}$, given by,
\bea
\cC_{u,v;w} = \sum _{(m_r,n_r)}' { \tau_2^w \, \delta _{m_1+m_2+m_3,0} \, \delta _{n_1+n_2+n_3,0} \over \pi^w \, (m_1\tau+n_1)^u (m_2 \bar \tau+n_2)^v (m_3 \tau+n_3)^{w-u} (m_3 \bar \tau+ n_3)^{w-v} }
\eea
and decompose the sum according to the vanishing of the summation variables $m_r$,
\bea
\cC_{u,v;w} = \sum_{k=0}^4 \cC_{u,v;w}^{(k)}  
\eea
The individual terms correspond to the following assignments of $m_r$ and $n_r$,
\begin{align}
\cC_{u,v;w}^{(0)} && m_i&= 0 &&   &&  n_i \not=0 && i=1,2,3
\no \\
\cC_{u,v;w}^{(1)} && m_1&=0 && m_2, m_3 \not=0 && n_1 \not=0 &&
\no \\
\cC_{u,v;w}^{(2)} && m_2&= 0 && m_3,m_1\not=0 && n_2 \not=0 &&
\no \\
\cC_{u,v;w}^{(3)} && m_3&=0 && m_1, m_2 \not=0 && n_3 \not=0 &&
\no \\
\cC_{u,v;w}^{(4)} && m_i&\not= 0 &&  &&  &&  i=1,2,3
\end{align}
We shall show that the Fourier series decompositions of the functions $\cC^{(k)} _{u,v;w}$ may be computed   in terms of the  functions $\Phi_{a;b}$ and $\f_k$ defined in (\ref{Lam1}). The calculation of $\cC^{(0)} _{u,v;w}$ was already carried out when we evaluated the coefficient $\ell_w$ in (\ref{cw}), and the result is given by,
\bea
\cC^{(0)} _{u,v;w} = \ell_w ( 4 \pi \tau_2)^w
\eea
with $\ell_w$ given explicitly in the second equation of (\ref{Lambda}). The remaining functions $\cC^{(k)} _{u,v;w}$ are considerably more involved and are the focus of the subsequent subsections.

\subsection{Calculation of $\cC_{u,v;w}^{(1)} $ and $\cC^{(2)}_{u,v;w}$}

The function $\cC^{(2)}_{u,v;w}$ is related to $\cC_{u,v;w}^{(1)} $ by swapping the summation variables $(m_1,n_1)$ with $(m_2,n_2)$ or equivalently by simultaneous complex conjugation and swapping $u$ and $v$, 
\bea
\cC_{u,v;w}^{(2)} (\tau) =  \cC_{v,u;w}^{(1)} (\tau)^*
\eea
an operation that will be denoted by $(u \leftrightarrow v)^*$ throughout. We begin by evaluating  $\cC_{u,v;w}^{(1)} $, and express the sum  in terms of the independent variables $m=m_3$, $n=n_3$, and $n_1$, 
\bea
\cC_{u,v;w}^{(1)} = {\tau_2^w \over \pi ^w} \sum _{m\not=0} \, \sum _{n_1 \not= 0} \sum _{n} 
{ (-)^v \over n_1^u (m \bar \tau+n+ n_1)^v \, (m \tau +n)^{w-u} \, (m \bar \tau +n)^{w-v} }
\eea
It is natural to perform first the summation over $n$. Since the denominator has three different factors involving $n$, we need to perform a double partial fraction decomposition. Performing first the partial fraction decomposition on the last two factors for the variable $n$ we find,
\bea
\cC_{u,v;w}^{(1)}  =
{\tau_2^w \over \pi ^w} (-)^{u+v} \sum _{m\not=0} 
\left ( \sum _{k=1}^{w-u} { \binom{2w-u-v-k-1}{w-u-k}   \over  (2m\tau_2)^{2w-u-v-k}} \, K_{u,v;k}
+ \sum _{k=1}^{w-v} {  \binom{2w-u-v-k-1}{w-v-k}  \over (2m\tau_2)^{2w-u-v-k}} \, L_{u,v;k} \right )
\eea
where the functions $K$ and $L$ are given by, 
\bea
K_{u,v;k} & = &  \sum _{n_1 \not= 0}  \sum _n { (-i)^{u+v-k} \over n_1^u \, (m \bar \tau+n+ n_1)^v (m\tau +n)^k } 
\no \\
L_{u,v;k} & = &  \sum _{n_1 \not= 0}  \sum _n { (-i)^{u+v+k} \over n_1^u \,  (m \bar \tau+n+ n_1)^v (m \bar \tau +n)^k }
\eea
Throughout, the dependence of $K_{u,v;k}$ and $L_{u,v;k}$ on $m$ and $\tau$ will be understood but not exhibited. The factors of $i$ have been inserted to make them real functions of their arguments.

\sm

To compute $K_{u,v;k}$, we carry out a partial fraction decomposition in $n$, and then sum over $n$ in terms of the function $\f_k$ defined in (\ref{Lam1}). The sum over $n_1$ may then be carried out in terms of the functions $\Phi_{a;b}$ defined in (\ref{Lam1}), and we find, 
\bea
\label{Kuvw}
K_{u,v;k} = 
 \sum _{\ell=1}^k   \mbinom{v+k-\ell-1}{k-\ell}   \Phi _{u;v+k-\ell}(2m\tau_2) \f_\ell (q^m) +(v \leftrightarrow k)^*
\eea
To compute $L_{u,v;k}$, we carry out a partial fraction decomposition in $n$ and express the sum over $n$ in terms of the functions $\f_k$,  
\bea
L_{u,v;k} =  
(-i)^{u+v-k} \sum _{n_1 \not= 0}  \left ( \sum _{\ell=1}^v { (-i)^{-\ell} \binom{v+k-\ell-1}{v-\ell} \over n_1^{u+v+k-\ell} } \f_\ell (\bar q^m ) 
+ \sum _{\ell=1}^k { (-i)^{\ell} \binom{v+k-\ell-1}{k-\ell} \over n_1^{u+v+k-\ell}  } \f_\ell ( \bar q^m ) \right ) 
\eea
The sum over $n_1$ vanishes unless $u+v+k-\ell$ is even. Thus we change summation variables from $\ell$ to $u+v+k-\ell= 2j$, so that the sum becomes, 
\bea
\label{Luvw}
(-)^kL_{u,v;k} =
\sum _{j=[{u+k+1\over 2} ]}^{[{u+v+k-1 \over 2}]}   2(-)^{j}  \mbinom{2j-u-1}{k-1}   \zeta (2j) \, \f_{u+v+k-2j} (\bar q^m) 
+ (-)^{u+v+k} (v \leftrightarrow k) 
\eea
Note that the $\ell=1$ contribution, which is given by the $j = [{u+v+k-1 \over 2}]$ contribution in the above sum, cancels exactly so that every term has a non-constant Fourier mode. 

\sm

Collecting both contributions $\cC_{u,v;w}^{(1)} + \cC_{u,v;w}^{(2)}$, we find, 
\bea
\label{C1C2}
\cC_{u,v;w}^{(1)} + \cC_{u,v;w}^{(2)}   = 
 \sum _{m\not=0}   \sum _{k=1}^{w-u} 
 { (-)^{u+v} \tau_2^w  \binom{2w-u-v-k-1}{w-u-k}  \over \pi^w (2m\tau_2)^{2w-u-v-k}} \, 
  \Big ( K_{u,v;k}+  L_{v,u;k}^* \Big ) 
+ ( u \leftrightarrow v)^*
\eea
The functions $K_{u,v;k}$ and $L_{u,v;k}$ involve $\f_k$, $\Phi_{a;b}$, and factors that are power behaved in $\tau_2$. 
As will be shown in the subsequent section, the infinite sum over $m$ may be performed for the terms that are purely power behaved in $\tau_2$. For the exponential terms the values of $m$ will label the powers in $q$ and $\bar q$ and, for a given assignment of powers of $q$ and $\bar q$, the sums over $m$  will reduce to finite sums.

\subsection{Calculation of $\cC_{u,v;w}^{(3)}$}

Expressing the sum in terms of  the independent summation variables $m=m_1$, $n=n_3$, and $n_1$, this contribution becomes,
\bea
\cC_{u,v;w}^{(3)} = {\tau_2^w \over \pi ^w} \sum _{m\not=0} \, \sum _{n \not= 0} \sum _{n_1} 
{ (-)^v \over (m\tau+n_1)^u \, (m\bar \tau +n_1+n)^v \, n^{2w-u-v}}
\eea
Carrying out the partial fraction decomposition in $n_1$, summing over $n_1$  in terms of the function $\f_k$, and expressing the sum over $n$ in terms of the function $\Phi_{a;b}$, we find,  
\bea
\label{C3}
\cC_{u,v;w}^{(3)} = {\tau_2^w \over \pi ^w} 
\sum_{k=1}^u  (-)^{u+v+w} \mbinom{u+v-k-1}{u-k}  \sum _{m\not=0} \,\Phi _{2w-u-v; u+v-k} (2 m \tau_2) \, \f_k(q^m)
+ ( u \leftrightarrow v)^*
\qquad
\eea

\subsection{Calculation of $\cC_{u,v;w}^{(4)}$}

For $\cC_{u,v;w}^{(4)}$ none of the summation variables $m_r$ vanish. We shall use independent continuous variables to abbreviate the $m$-dependence by letting $m_1 \tau \to \a$, $m_2 \bar \tau \to \b$, and $m_3 \tau \to \g$, and recast the sum over $m_r$ in terms of a function $\Omega_{u,v;w}$ of the continuous variables $\alpha, \beta, \gamma$, 
\bea
\cC_{u,v;w}^{(4)} = \sum _{m_1, m_2,m_3 \not=0}  { \tau_2^w \over \pi^w} \, \delta _{m_1+m_2+m_3,0} \,
\Omega_{u,v;w} \Big (m_1\tau, m_2 \bar \tau, m_3\tau, m_3 \bar \tau \Big ) 
\eea
where the function $\Omega$ is defined by, 
\bea
\Omega_{u,v;w} (\a, \b, \g, \bar \g) =
\sum _{n_1, n_2,n_3} {\delta_{n_1+n_2+n_3,0}   \over (\a+n_1)^u ( \b+n_2)^v
  (\g+n_3)^{w-u} (\bar \g+n_3)^{w-v} }
\eea
The partial fraction decomposition of the last two factors in the denominator with respect to $n_3$ may be expressed  in terms of a family of functions $\Psi _{u,v,k}$ defined by,
\bea
\Psi_{u,v,k}(\a, \b, \g) = \sum _{n_1, n_2,n_3}  { \delta_{n_1+n_2+n_3,0}  \over (\a+n_1)^u ( \b+n_2)^v (\g+n_3)^k  }
\eea
The function $\Psi_{u,v,k}(\a, \b, \g) $ is invariant under all six permutations of the pairs $(u,\a), (v,\b)$, and $(k,\g)$. In terms of $\Psi$, the function $\Omega$ may be expressed as follows, 
\bea
\Omega_{u,v;w} (\a, \b, \g, \bar \g) & = &
\sum_{k=1}^{w-u} { (-)^{w+v} \binom{2w-u-v-k-1}{w-u-k} \over  (\g-\bar \g)^{2w-u-v-k}} \, \Psi_{u,v,k}(\a, \b, \g)
\no \\ &&
+\sum_{k=1}^{w-v} { (-)^{w+v+k} \binom{2w-u-v-k-1}{w-v-k} \over  (\g- \bar \g)^{2w-u-v-k}} \, \Psi_{u,v,k}(\a, \b, \bar \g)
\eea
To calculate the function $\Psi$, we use the following  observation, 
\bea
\Psi _{u,v,k}(\a,\b,\g) = { (-)^{u+v+k-1} \over \Gamma (u) \Gamma (v) \Gamma (k)} 
\p_\a ^{u-1} \p_\b ^{v-1} \p_\g ^{k-1} \Psi _{1,1,1}(\a,\b,\g)
\eea
 $\Psi_{1,1,1}$ is readily evaluated by partial fraction decomposition and summation in the variables $n_1$ and $n_2$, and we find, 
\bea
\Psi_{1,1,1}(\alpha, \beta, \g) = 
- { \f_1(e^{2 \pi i \a}) \f_1(e^{2 \pi i \b}) + \f_1(e^{2 \pi i \b}) \f_1(e^{2 \pi i \g}) + \f_1(e^{2 \pi i \g}) \f_1(e^{2 \pi i \a}) + \pi^2  \over \a + \b + \g} 
\quad
\eea
The successive derivatives may be obtained using a trinomial expansion for Leibniz's rule, and we find the following expression,
\bea
\Psi _{u,v;k}(\a,\b,\g) & = & - \,  { \pi^2 \binom{u+v+k-3}{u-1,v-1,k-1} \over (\a+\b+\g)^{u+v+k-2}}
\no \\ &&
+ \sum _{a=1}^u \sum _{b=1}^v  { (-i)^{a+b} \binom{u+v+k-a-b-1}{u-a,v-b,k-1} \over (\a+\b + \g)^{u+v+k-a-b}} 
\f _a(e^{2 \pi i \a}) \f_b (e^{2 \pi i \b})
\no \\ &&
+ \sum _{b=1}^v \sum _{c=1}^k { (-i)^{b+c}  \binom{u+v+k-b-c-1}{u-1,v-b,k-c} \over (\a+\b + \g)^{u+v+k-b-c}} 
\f _b(e^{2 \pi i \b}) \f_c (e^{2 \pi i \g})
\no \\ &&
+ \sum _{c=1}^k \sum _{a=1}^u  { (-i)^{c+a} \binom{u+v+k-c-a-1}{u-a,v-1,k-c} \over (\a+\b + \g)^{u+v+k-c-a}} 
\f _a(e^{2 \pi i \a}) \f_c (e^{2 \pi i \g})
\eea
where we use the following notation for the trinomial coefficients, 
\bea
\mbinom{a+b+c}{a,\,b,\,c} = {(a+b+c)!\over a! \,b! \,c!}
\eea
Substituting the expression for $\Psi_{u,v,k}$  into $\Omega$ and substituting the original values for the continuous variables $\alpha=m_1\tau$, $ \beta=m_2\bar \tau$, and $ \gamma= m_3 \tau$, we find, 
\bea
\label{Omega}
\Omega_{u,v;w} & = & 
\sum_{k=1}^{w-u} \sum _{a=1}^u \sum _{b=1}^v 
{ (-)^{b+v} \, \binom{2w-u-v-k-1}{w-u-k}  \binom{u+v+k-a-b-1}{u-a,v-b,k-1}  \, \f _a(q^{m_1}) \f_b (\bar q^{m_2} ) \over 
(-m_2)^{u+v+k-a-b}  m_3^{2w-u-v-k} (2 \tau_2)^{2w-a-b}}  
\\ &&
+  \sum_{k=1}^{w-u} \sum _{b=1}^v \sum _{c=1}^k 
{ (-)^{b+v} \, \binom{2w-u-v-k-1}{w-u-k}  \binom{u+v+k-b-c-1}{u-1,v-b,k-c}  \, \f _b(\bar q^{m_2}) \f_c (q^{m_3}) \over
(- m_2)^{u+v+k-b-c}  m_3^{2w-u-v-k} (2 \tau_2)^{2w-b-c} } 
\no \\ &&
+  \sum_{k=1}^{w-u} \sum _{c=1}^k \sum _{a=1}^u 
{ (-)^v\binom{2w-u-v-k-1}{w-u-k}  \binom{u+v+k-c-a-1}{u-a,v-1,k-c}   \, \f _a(q^{m_1} ) \f_c (q^{m_3}) \over
 (- m_2)^{u+v+k-c-a}  m_3^{2w-u-v-k} (2 \tau_2)^{2w-c-a} } 
 \no \\ && 
+ \sum_{k=1}^{w-u} 
 { (-)^v \pi^2 \binom{2w-u-v-k-1}{w-u-k}   \binom{u+v+k-3}{u-1,v-1,k-1}  \over 
 (-m_2)^{u+v+k-2}  m_3^{2w-u-v-k} (2 \tau_2)^{2w-2}}
 \no \\ && 
+ ( [u,m_1,a] \leftrightarrow [v,m_2,b]) ^* 
\no
\eea
where the conjugation instruction $( [u,m_1,a] \leftrightarrow [v,m_2,b]) ^*$ applies to the entire expression and leaves $w,c,m_3$ unchanged. As was the case already for the functions $\cC^{(i)}_{u,v;w}$, and as we shall show explicitly in subsequent sections, the infinite sums over $m_r$ for $r=1,2,3$  may be performed for the terms that are purely power behaved in $\tau_2$. For the exponential terms the values of $m_r$ will control the powers in $q$ and $\bar q$ and, for a given assignment of powers of $q$ and $\bar q$, the sums over $m_r$  will reduce to finite sums.

 \newpage

\section{The Fourier coefficient functions}
\setcounter{equation}{0}
\label{sec:5}

In this section, we shall present the results for all the Fourier modes of $\cC_{u,v;w}$ in Theorem~\ref{thm7}.  The details of the calculations of the Laurent polynomial $\cL_{u,v;w}(\tau_2)$ will be relegated to Appendices~\ref{sec:A} and~\ref{sec:B}. Likewise, the details of the calculations of the exponential contributions, including the Laurent polynomials $\cQ_{u,v;w}^{(N)}$,  $F^{(N)} _{u,v;w}$, and $G^{(N,L)} _{u,v;w}$,  will be relegated to Appendix~\ref{sec:C}. Various coefficients in the Laurent polynomials $F^{(N)} _{u,v;w}$ and $G^{(N,L)} _{u,v;w}$ involve  rational numbers given by lengthy expressions of finite sums and will be presented in Appendix \ref{sec:CC}. In the present section, we shall summarize all of these results.

\subsection{Decomposition of the functions $\f_k$ and $\Phi _{a;b}$}

To extract the Fourier coefficient functions from the formulas for $C_{u,v;w}(\tau) $ computed in the previous section, we decompose  the elementary functions $\f_k$ and $\Phi_{a;b}$ into parts which are purely Laurent polynomial in $\tau_2$ and parts which are exponential, 
\bea
\label{decomp1}
\f_k (q^m) & = &  (2 \pi )^k \ep(m)^k \left (   \half \delta _{k,1} + \hat \f _k (q^{|m|})   \right )
\no \\ 
\Phi_{a;b} (2m \tau_2) & = & \ep(m)^{a+b} \Big ( \Phi_{a;b} ^{(0)} (2|m|\tau_2) + \Phi_{a;b} ^{(1)} (2|m|\tau_2) \Big ) 
\eea
where the sign function $\ep(m)$ is defined by,
\bea
\ep(m) = \left \{ \begin{matrix} 
+1 & \hbox{for} & m >0 \cr
0 & \hbox{for} & m =0 \cr  
-1 & \hbox{for} & m <0 \cr 
\end{matrix} \right .
\eea 
The functions appearing in (\ref{decomp1}) are given as follows, 
\bea
\label{decomp2}
\hat \f _k (q^{|m|}) & = & { 1 \over \Gamma (k) } \sum _{p=1}^\infty p^{k-1} q^{p |m|}
\no \\
\Phi _{a;b} ^{(0)} (2|m| \tau_2)
& = &
2 \sum _{\alpha =0} ^{[a/2]}  
{(-)^{a+ \alpha}   \zeta (2 \alpha) \, \binom{a+b-2\alpha-1}{b-1}  \over (2|m| \tau_2)^{a+b-2 \alpha} }
+    { (-)^a \pi  \binom{a+b-2}{a-1}  \over  (2|m| \tau_2)^{a+b-1}}
\no \\
\Phi _{a;b} ^{(1)} (2|m|\tau_2) & = &    
\sum _{\beta =1}^b  { (-)^a \, \Gamma( a+b-\beta) \over \Gamma (b-\beta +1) \Gamma (a)}
{ (2 \pi)^\beta \, \hat \f_\beta (e^{-4\pi |m|\tau_2}) \over (2 |m| \tau_2)^{a+b-\beta}} 
\eea
We have taken this opportunity to expose the dependence on the sign of the summation variable $m$ for all functions involved.

\subsection{The Laurent polynomial $\cL_{u,v;w}$}

The Laurent polynomial $\cL_{u,v;w}$ of $\cC_{u,v;w}$ is given by the following theorem.

{\thm \label{thm10}
{\sl The Laurent polynomial $\cL_{u,v;w}$ of $\cC_{u,v;w}$ is given as follows,
\bea
\label{Laurent}
\cL_{u,v;w}(\tau_2) = \ell_w (4 \pi \tau_2)^w  + \sum _{k=0}^{w-2} \ell_{2k-w+1} { \zeta (2w-2k-1) \over (4 \pi \tau_2)^{w-2k-1}} + { \ell_{2-w} \over ( 4 \pi \tau_2)^{w-2}}
\eea
The coefficient $\ell_w \in \QQ$ arises solely from $\cC^{(0)}_{u,v;w}$ and was given in (\ref{Lambda}). The  coefficients $\ell_{2k-w+1} \in \QQ$ arise from $\cC^{(i)} _{u,v;w}$ with $i=1,2,3$ and are given by, 
\bea
\label{L12f}
\ell_{2k-w+1}&=&
2 \theta \left ( \left [{ u \over 2} \right ]-k \right )  {  (-)^{v+1}  B_{2k}
\binom{2w-2k-2}{w-u-1}  \binom{u+v-1-2k}{u-2k} \over (2k)!  }
+(u \leftrightarrow v)
\no \\ && +
 2  \theta \left ( \left [{ 2w-u-v \over 2}  \right ] - k \right )
{  (-)^{w+1}B_{2k}  \binom{2w-2k-2}{2w-u-v-2k} \binom{u+v-2}{v-1} \over (2k)!  }
\eea
where $B_{2k}$ are the Bernoulli numbers and the symmetrization $u \leftrightarrow v$ applies only to the first line. The Heaviside $\theta$-function is defined such that $\theta(x)=1$ for $x \geq 0$ and zero otherwise. The coefficient $\ell_{2-w}$ arises from $\cC^{(i)} _{u,v;w}$ with $i=1,2,3,4$ and is given as follows,
 \bea
\label{L12g}
\ell_{2-w} &=&
\zeta(2w-2)  \mbinom{u+v-2}{u-1} \left [ (-)^v \mbinom{2w-3}{w-u-1}  + \half (-)^w \mbinom{2w-3}{u+v-2} \right ]
\no \\ &&
 + 2 (-)^v \sum_{k=u+v+1}^{w+v}  \mbinom{2w-k-1}{w+v-k} 
 { \Gamma (k-2) \,  \zeta (k-2, 2w-k)  \over \Gamma (u) \Gamma (v) \Gamma (k-u-v) }
  + ( u \leftrightarrow v) \quad
\eea
where the symmetrization $(u \leftrightarrow v)$ applies to the entire expression. The combination of multiple zeta values in the second term and even zeta values in the first may be simplified and gives $\ell_{2-w}$ as a bilinear combination of odd zeta values whose total weight is $2w-2$,
\bea
\label{L12h}
\ell_{2-w} = \sum_{k=1}^{w-3} \lambda _k \, \zeta (2k+1) \, \zeta (2w-2k-3) \hskip 1in \lambda _k \in \QQ
\eea
with the coefficients $\lambda _k$ given in Proposition \ref{conj1}. 
}}

The double zeta value $\zeta(s,t)$ in the first expression for $\ell_{2-w}$ is defined by,
\bea
\label{MZV}
\zeta(s,t) = \sum _{m, n=1}^\infty { 1 \over (m+n)^s n^t}
\eea
We note that the validity of the form of the expression given for $\ell_{2-w}$ in (\ref{L12h}) was proven for the two-loop modular graph functions functions $C_{a,b,c}$ in \cite{DHoker:2017zhq}, and may be proven by the same methods in the case of the functions $\cC_{u,v;w}$.  However, the expressions for the coefficients $\lambda_k$ can be obtained only upon assuming the validity of a conjecture.  The  coefficients in the bilinear combination of odd zeta values given in Conjecture 1.4 of \cite{DHoker:2017zhq} was obtained with the help of an identity between rational numbers given in Conjecture 6.2 of that paper, which was verified to high order by MAPLE. Applying  Conjecture 6.2 of \cite{DHoker:2017zhq} to the expression for $\ell_{2-w}$, we obtain the following expression for the coefficients $\lambda_k$,

{\prop \label{conj1} {\rm (Obtained by assuming the validity of Conjecture 6.2 of \cite{DHoker:2017zhq}.)}\\
{\sl The coefficients $\lambda _k$  in the Laurent series for $\cC_{u,v;w}$ are given as follows,
\bea
\lambda _k & = & 
(-)^w \delta_{w,v+1}{ \theta(w-z-1-\eps)\theta\left(\left[ {w-u \over 2} \right]-w + z+1 \right) \Gamma(2w-3)\over \Gamma(u) \Gamma(v) \Gamma(2w-u-v-1)}
\no\\&&
+2 (-)^v {\theta(w-k-3+\eps) \theta (k -z+1-\eps) \theta ( [{w-u \over 2} ]+z-k -1) \Gamma (2w-2k-3) \Gamma (2k +1) 
\over \Gamma (w+v-2k -2)\Gamma (w-v) \Gamma (u) \Gamma (v)\Gamma (2k -u-v+3)}
\no\\ &&
-(-)^v \sum_{m=\eps} ^{ [ {w-u \over 2} ]} {\theta(w-z-m-2+\eps) \Gamma (2w-2z-2m-1)  \over \Gamma (w-u-2m+\eps)\Gamma (w-v) \Gamma (u) \Gamma (v) \Gamma (2m+1-\eps)}
\no \\ && \hskip 0.5in  \times 
\sum_{n=2m} ^{2w-2z-3}  { \mathrm{E}_{n-2m}(0) \Gamma (2k +1) \Gamma (2z+n-1) \over \Gamma (2w-2z-n-1) \Gamma (2k -2w+2z+n+3) \Gamma (n-2m+1) }
\no \\ && + ( u \leftrightarrow v)
\eea
where $z = [(u+v)/2 ]$, $\eps= u+v\,\,\, (\mathrm{mod}\,\, 2)$, and $\mathrm{E}_n(0)$ are Euler polynomials evaluated at $0$.}}
\newline\newline
The Euler polynomials $\mathrm{E}_n(t)$ (not to be confused with the Eisenstein series $E_s(\tau)$) may be defined by the following generating function, 
\bea
{ 2\, e^{xt} \over e^x +1} = \sum _{n=0}^\infty \mathrm{E}_n(t) { x^n \over n!}
\eea
The proof of Theorem \ref{thm10} is given in Appendix \ref{sec:A}, while the proof of Proposition \ref{conj1}, assuming the validity of Conjecture 6.2 of \cite{DHoker:2017zhq}, is given in Appendix \ref{sec:B}.

\subsection{Generalized divisor sum functions}
\label{gendiv}

In the calculation of non-constant exponential contributions to $\cC_{u,v;w}$ to which we next turn, we shall encounter a generalization of the standard divisor functions $\sigma _s(N) = \sum _{0 < n |N} n^s$. We introduce these generalized divisor sums now. For $M,N \in \NN$ and $A,B,C \in \ZZ$, we  introduce the following generalized divisor function, 
\bea
\label{VABC}
V_{A,B,C}(M, N) =  \sum _{{m| M \atop m \not=0}} \sum_{{ n| N \atop n \not=0, -m}}
{  \ep(m) \ep (n)  \over m^{A-1} n^{B-1} (m+n)^C   } 
\eea
In contrast to the standard divisor sum, the summation here is carried out over both positive and negative divisors $m,n$. The function $V_{A,B,C}(M,N)$ vanishes when $A+B+C$ is an odd integer since the summand is then odd under the reversal of the signs of both $m$ and $n$.  To account for this property in a systematic manner, we introduce the function, 
\bea
\label{Indefin}
\mI_{n} = \half ( 1 + (-)^n) 
\eea
which vanishes for $n$ odd and equals 1 for $n$ even. A more explicit formula for $V_{A,B,C}(M,N)$ may be obtained by separating the contributions from positive and negative $m, n$, 
\bea
\label{expandedVABC}
V_{A,B,C}(M, N)  =   
 \sum _{0< m | M} { 2 \mI_{A+B+C} \over m^{A-1}} \Bigg ( \sum_{0< n| N}
{  1 \over n^{B-1} (m+n)^C   }  
+ \sum_{{0< n| N \atop n \not= m}}
{  (-)^B \over  n^{B-1} (m-n)^C   }  \Bigg )
\quad \eea

\subsection{The exponential part of the constant Fourier mode $\cQ_{u,v;w}^{(N)}$}

We now proceed to the calculation of the exponential part of the constant Fourier mode. To obtain this, we retain the contributions with identical powers of $q$ and $\bar q$  in $\cC_{u,v;w}$, namely all terms of the form $(q\bar q)^N$. The calculation of the exponential contributions will be carried out in Appendix \ref{sec:C}, and the contributions to $\cQ_{u,v;w}^{(N)}$ are organized in Appendix \ref{sec:CC}. The result may be stated in the form of the following theorem.

\sm

{\thm \label{thm15}
{\sl The exponential part of the constant Fourier mode of $\cC_{u,v;w}$ is given by,
\bea
\label{finalcQ}
\cQ_{u,v;w}^{(N)}(\tau_2) & = &
 \sum_{\b=1}^{w-u+v-1}   {  N^{\b-1} (  J_{u,v;w}^{(1)} (\beta,N) +J^{(3)}_{u,v;w}(\beta, N) ) \over \Gamma(\b) (4 \pi \tau_2)^{w- \b -1}}  
\no \\ &&
\hskip 0.05 in+\sum_{\beta=1} ^{u+v-1} {  N^{\beta -1} J_{u,v;w}(\beta,N) \over \Gamma (\beta) (4 \pi \tau_2)^{w-\beta-1}} + (u \leftrightarrow v)  
\eea
where the symmetrization in $u,v$ applies to both lines. The coefficients are all rational functions, and are given by,
\bea
J_{u,v;w}^{(1)}  (\beta,N) &=& 2 (-)^v \sigma _{2-2w}(N)  
\sum_{k=1}^{w+v} \mbinom{2w-k-1}{w+v-k} \mbinom{k-\b-2}{u-1} \mbinom{k-u-2}{v-1}
\no \\ && \qquad
\times[\theta (\beta -v) \theta(k-\b-u-1) + \theta(v-\b-1) \theta(k-u-v-1)]
\no\\
J_{u,v;w}^{(2)} (\beta,N)  & = &  
 \sum_{k=u+v+1}^{w+v} \,  \sum_{a=\max(1,\b-v+1)}^{\min(\beta,u)}
  \mbinom{k-\beta-2}{u-a,v+a-\beta-1,k-u-v-1} 
 \no \\ && \qquad \times 
 (-)^{a+v}  \mbinom{\beta-1}{\beta -a}   \mbinom{2w-k-1}{w+v-k}  V_{a,k-a,2w-k}(N,N)\no
\no \\
J^{(3)}_{u,v;w}(\beta, N) 
& = &
\sum_{c=\max (1,\beta-v+1)} ^{\min (w-u,\beta)} \, \sum_{k=u+v}^{w+v-c} 
   \mbinom{k+c-\beta-2}{u-1,v-\beta+c-1,k-u-v} \mbinom{2w-k-c-1}{w+v-k-c}  
\no \\ && \qquad \times
(-)^v \mbinom{\beta-1}{\beta-c}  \Big (4\, \mI_{k} \,\sigma_{1-k}(N) \sigma _{1-2w+k}(N) 
-2 \,  \sigma_{2-2w}(N) \Big )
\no\\
J_{u,v;w}(\beta, N) &=&  J^{(2)}_{u,v;w}(\beta, N) +(-)^w \sigma_{2-2w}(N) \mbinom{2w-\b-2}{u+v-\b-1}  \mbinom{u+v-2}{u-1}
\eea
The functions $V_{A,B,C}(M,N)$ and $\mI_n$ appearing above were defined in subsection \ref{gendiv}.}}

\subsection{The non-constant Fourier modes $F_{u,v;w}^{(N)}$ and $G_{u,v;w}^{(N,L)}$}

The full expressions for the non-constant Fourier modes $G_{u,v;w}^{(N,L)}$ and $F_{u,v;w}^{(N)}$ are more involved than those of $\cL_{u,v;w}$ and $\cQ^{(N)}_{u,v;w}$. They are obtained in Appendix \ref{sec:CC}, with the final result summarized in the following theorem.

\sm

{\thm \label{thmnonconst}
{\sl The non-constant Fourier modes $G_{u,v;w}^{(N,L)}$ and $F_{u,v;w}^{(N)}$ are given by 
\bea
G_{u,v;w}^{(N,L)}(\tau_2) &=& \sum_{k=k_-}^{k_+}\sum_{\ell=1}^{\ell_+}\sum_{\b=1}^{\b_+}
{\cW^{k,\ell,\beta}_{u,v;w}(N,L) \over (4 \pi \tau_2)^{w-\ell-\b}}
\no\\
F_{u,v;w}^{(N)}(\tau_2)&=&\sum_{k=k_-}^{k_+} \sum_{\ell=1}^{\ell_+}\left[  { \cM_{u,v;w}^{k,\ell}(N) \over (4 \pi \tau_2)^{w-\ell-1}}+\sum_{\g=0}^{\g_+} {\cH^{k,\ell,\b}_{u,v;w}(N) \over (4 \pi \tau_2)^{w-\ell-\g}}\right]
\eea
where
\bea
\label{klbg+}
k_- &=& 1 + u + v
\no\\
k_+ &=& \mathrm{max}\left(w+u, w+v \right)
\no\\
\ell_+ &=& \mathrm{max} \left(w-u, u \right)
\no\\
\beta_+ &=&  \mathrm{max} \left(w+|u-v|-1,\, u+v-1 \right)
\no\\
\g_+ &=& \mathrm{max} \left(u+v, \, 2w-u-v-\eps \right)
\eea
and $\eps = u+v \,\, (\mathrm{mod}\,\, 2)$. The coefficients $\cW^{k,\ell,\beta}_{u,v;w}(N,L)$ are given in (\ref{Wbigformula}), while the coefficients $\cH^{k,\ell,\beta}_{u,v;w}(N)$ are defined implicitly by (\ref{implicitcH}). Both are rational numbers for all integer-valued arguments and indices. The function $\cM_{u,v;w}^{k,\ell}(N)$ is shown in (\ref{FNcMresult}) and gives rational multiples of odd zeta values.
}}
\sm

This completes the calculation of the full Fourier expansion for general two-loop modular graph functions, all of which may be constructed from $\cC_{u,v;w}$. In the following subsection we illustrate these results by means of an explicit example.

\subsection{An example: $C_{2,1,1}$}
\label{sec:56}

We now illustrate our results for the familiar case of two-loop modular graph functions $C_{a,b,c}$. To utilize the results of the previous sections, we must first recast these modular graph functions in terms of the basis $\cC_{u,v;w}$. One finds for example\bea
\label{C211intermsofCuvw}
C_{2,1,1} &=& \cC_{2,2;4} - {3 \over 2}\left( \cC_{2,3;4}+\cC_{3,2;4}\right)
\eea
It is known that $C_{2,1,1}$ has the following Laurent polynomial \cite{DHoker:2016quv}, 
\bea
\label{C211L}
C_{2,1,1}^\cL = {2 y^4 \over 14175}  + {\zeta(3) \over 45}y +  {5\zeta(5)\over 12 y}
- {\zeta(3)^2 \over 4y^2}+{9 \zeta(7) \over 16 y^3}
\eea
with $y = \pi \tau_2$.  Using Mathematica, one verifies that Theorem \ref{thm10} reproduces this result, 
\bea
\cL_{2,2;4}-{3 \over 2} \left(\cL_{2,3;4} + \cL_{3,2;4} \right) = C_{2,1,1}^\cL 
\eea
The exponential part of the Fourier series of $C_{2,1,1}$ can be obtained by exploiting the following differential equation,
\bea
\label{C211Laplaceequation}
\left(\Delta- 2 \right)C_{2,1,1} = 9 E_4 - E_2^2
\eea 
as well as the known Fourier expansion (\ref{Eas}) of the non-holomorphic Eisenstein series. Let us define the coefficients of the Fourier expansion as $C_{2,1,1}^{(M,N)}$, such that,
\bea
\label{C211}
C_{2,1,1} (\tau) = \sum_{M,N=0}^\infty C_{2,1,1}^{(M,N)} (\tau_2) q^M \bar q^N
\eea

\sm

We begin by obtaining the exponential part of the constant Fourier mode, denoted by $\cQ^{(N)}$ in (\ref{Four}). The only contribution of the form $(q \bar q)^N$ to the right-hand side of (\ref{C211Laplaceequation}) comes from $E_2^2$, as can be seen from (\ref{Eas}). Using the latter formula gives the following differential equation for $C^{(N,N)}_{2,1,1}$, 
\bea
\left(\tau_2^2 \p_{\tau_2}^2- 2 \right)\left( C_{2,1,1}^{(N,N)}(\tau_2) (q \bar q)^N\right) = - 8  N^2 \sigma_{-3}(N)^2 P_2(4 \pi N \tau_2)^2 (q \bar q)^N
\eea
Restricting to solutions involving only powers of $\tau_2$, one finds
\bea
\label{C211a}
C^{(N,N)}_{2,1,1} = - 8{ \sigma_{-3}(N)^2 \over (4 \pi \tau_2)^2}
\eea
In other words, we expect from (\ref{C211intermsofCuvw}) that 
\bea
\cQ^{(N)}_{2,2;4}(\tau_2)-{3 \over 2}\left(\cQ^{(N)}_{2,3;4}(\tau_2)+\cQ^{(N)}_{3,2;4}(\tau_2)\right) = -8\,{\sigma_{-3}(N)^2 \over (4\pi \tau_2)^2}
\eea 
Using Mathematica, one may check that this is indeed the result given by Theorem \ref{thm15}.

\sm

One may now similarly proceed to obtain the non-constant contributions. For example, the Laplace equation (\ref{C211Laplaceequation}) together with appropriate boundary conditions yields the following results, 
\bea
C^{(1,0)}_{2,1,1} &=&  {y \over 45} + {1\over 3} + {11\over 12 y}  + {9 \over 8 y^2}   
-{ \zeta(3)\over 2 y^2} + {9\over 16 y^3}
\no\\
C^{(2,0)}_{2,1,1} &=& { y\over 40} + {31\over 16} + {385\over 128 y}  + {1033\over 512 y^2}   - {
 9 \zeta(3)\over 16 y^2} + {1161\over 2048 y^3}
\no\\
C^{(2,1)}_{2,1,1} &=&-{9 \over 16 y^2}
\eea
One may check that these are consistent with the results of Theorem \ref{thmnonconst} for $F^{(1)}$, $F^{(2)}$, and $G^{(1,1)}$ respectively, and that the first line is consistent with equation (5.7)  of \cite{DHoker:2015gmr}.

 \newpage

\section{The space of odd two-loop modular functions}
\setcounter{equation}{0}
\label{sec:6}

In this section, we shall study the space $\mA_w$ of odd two-loop modular graph functions of weight~$w$. They  are cuspidal functions with exponential decay at the cusp. Using holomorphic subgraph reduction and the sieve algorithm studied in \cite{DHoker:2016mwo,DHoker:2016quv,Gerken:2018zcy}, we shall exhibit a subspace of $\mA_w$ consisting of functions which may be expressed in terms of Eisenstein series, construct  two other linearly independent subspaces, and use these results to give a lower bound on the dimension of $\mA_w$. Finally, we shall show that $\mA_w$ is trivial for $w \leq 4$ and that the lower bound is saturated for $5 \leq w \leq 11$.

\subsection{Odd two-loop modular graph functions}

Odd two-loop modular graph functions may be obtained from general two-loop modular graph functions by taking their parity odd part, 
\bea
\label{Adef}
 \cA \left[ \begin{matrix} a_1 & a_2 & a_3 \cr b_1 & b_2 & b_3 \cr \end{matrix} \right]
=
\cC \left[ \begin{matrix} a_1 & a_2 & a_3 \cr b_1 & b_2 & b_3 \cr \end{matrix} \right]
- \cC \left[ \begin{matrix}  b_1 & b_2 & b_3 \cr a_1 & a_2 & a_3   \cr \end{matrix} \right]
\eea
where  the total exponents $a,b$, defined by,
\bea
a=a_1+a_2+a_3 \hskip 0.5in b=b_1+b_2+b_3
\eea 
are equal to one another and the weight $w$ is given by $w=a=b$.  We shall impose the following restrictions in order to obtain a space on which differential operators, such as the Laplace operator,  act consistently, 
\bea
\label{Apos}
\mA_w = \left \{ \cA\left [ \begin{matrix} a_1 & a_2 & a_3 \cr b_1 & b_2 & b_3 \cr \end{matrix} \right ] , 
\left \{ \begin{matrix} a_r , b_r \geq 0, ~ a_r+a_s+b_r+b_s \geq 3 \hbox{ for } r \not= s \cr
\hbox{ at most one } a_r \hbox{ and at most one } b_r \hbox{ vanishes} \end{matrix}  \right \} \right \} 
\eea
The following theorem is an immediate consequence of the structure of the Fourier series for two-loop modular graph functions given in the preceding section, and in particular follows from the reality of the Laurent polynomial $\cL_{u,v;w}$.

{\thm 
\label{prop36} {\sl All odd two-loop modular graph functions in $\mA_w$ are cuspidal functions.} }
\newline

The exponential suppression at the cusp of odd modular graph functions allows us to introduce the Petersson inner product on the space $\mA = \bigoplus _w \mA_w$, which is defined by,
\bea
\< \cA_1 | \cA_2 \>  = \int _{\cM} d\mu  \,  \cA_1(\tau)^*  \cA_2(\tau) 
\hskip 1in 
d \mu = { i \over 2 \tau_2^2} \, d \tau \wedge d \bar \tau
\eea
for $\cA_1, \cA_2 \in \mA$ and where $\cM$ is a fundamental domain $\cM=\cH/PSL(2,\ZZ)$.

\sm

Using Proposition \ref{prop31} all odd two-loop modular graph functions may be expressed as linear combinations of the following functions, for which we adopt a convenient notation,
\bea
\label{Ashort}
\cA_{u,v;w} = - \cA_{v,u;w} = \cA \left[\begin{matrix} u & 0 & w-u \\ 0 & v & w- v \end{matrix} \right]
\eea
Without loss of generality, we restrict the range of parameters to $1 \leq u < v \leq w-1$. This restriction gives a trivial upper bound on the dimension $\dim \mA_w \leq \half (w-1)(w-2)$. However, many linear relations exits between the functions of (\ref{Ashort}), and one readily shows  $\cA_{1,2;3}=\cA_{1,2;4}=\cA_{1,3;4}=\cA_{2,3;4}=0$ so that 
\bea
\dim \mA_3= \dim \mA_4=0
\eea 
To investigate the spaces $\mA_w$ with $w \geq 5$, we shall use holomorphic subgraph reduction and the sieve algorithm, which we review in the next two subsections.

\subsection{Holomorphic subgraph reduction}

A systematic {\sl sieve algorithm} was developed in \cite{DHoker:2016mwo,DHoker:2016quv} to prove  identities between modular graph forms. The key ingredient is the use of the Cauchy-Riemann operator $\nabla = 2 i \tau_2^2 \p_\tau$ to relate identities between modular graph forms to identities between holomorphic Eisenstein series, using {\sl holomorphic subgraph reduction}. 

\sm

Applying the Cauchy-Riemann operator to an odd modular graph function will produce a modular graph form. Thus, we shall need to work on a larger space, of two-loop modular graph forms  given by the Kronecker-Eisenstein sums of (\ref{4a1}). For independent total exponents $a,b \geq 2$, we define the spaces, 
\bea
\label{pos}
\cV_{(a,b)} = \left \{ \cC^+ \left [ \begin{matrix} a_1 & a_2 & a_3 \cr b_1 & b_2 & b_3 \cr \end{matrix} \right ] , 
\left \{ \begin{matrix} a_r , b_r \geq 0, ~ a_r+a_s+b_r+b_s \geq 3 \hbox{ for } r \not= s \cr
\hbox{ at most one } a_r \hbox{ and at most one } b_r \hbox{ vanishes} \end{matrix}  \right \} \right \} 
\eea
while for $a=1$ or $b=1$, we define the spaces as follows, 
\bea
\cV_{(a,1)} = \left \{ \cC^+ \left [ \begin{matrix} a & 0 \cr 1 & 0 \cr \end{matrix} \right ] , ~ a \geq 5 \right \} 
\hskip 1in 
\cV_{(1,b)} = \left \{ \cC^+ \left [ \begin{matrix} 1 & 0 \cr b & 0 \cr \end{matrix} \right ] , ~ b \geq 5 \right \} 
\eea
The restriction that at most one $a_r$ and at most one  $b_r$ is allowed to vanish provides a self-consistent truncation,  and guarantees that the Laplace operator maps $\cV_{(a,a)}$ to itself \cite{DHoker:2016mwo,DHoker:2016quv}.  The spaces $\cV_{(a,b)}$ for $a, b \geq 2$ and $a+b\geq 6$ also contain one-loop modular graph functions and products of one-loop modular graph functions as was shown in (\ref{algred}).  

\sm

The operator $\nabla$ acts as follows on an arbitrary two-loop modular graph form $\cC^+ \in \cV_{(a,b)}$,  \bea
\label{CauchyRiemannAction}
\nabla \cC^+ \left[ \begin{matrix} a_1 & a_2 & a_3 \cr b_1 & b_2 & b_3 \cr \end{matrix} \right]
= 
\sum _{i=1}^3 a_i \, \cC^+ \left[ \begin{matrix} a_i+1 & a_{i'} & a_{i''}  \cr b_i-1 & b_{i'} & b_{i''} \cr \end{matrix} \right]
\eea
where $(i, i', i'')$ is a cyclic permutation of $(1,2,3)$.
The operator $\nabla$ maps $\cV_{(a,b)}$ to forms of weight $(a+1,b-1)$, but these forms do not necessarily belong to $\cV_{(a+1,b-1)}$. Indeed, an exponent with value $-1$ can occur  in the entry $i$ of the lower row of exponents when the corresponding exponent $b_i$ vanishes. Given the definition of $\cV_{(a,b)}$, this means that the other two lower exponents $b_{i'}, b_{i''}$  are non-zero so that the momentum conservation identities of (\ref{3d3}) can be used to convert the modular form with a negative entry into modular forms whose entries are all non-negative. These modular forms, however, still do not necessarily belong to $\cV_{(a+1,b-1)}$ since two zeros may appear in the lower row. This is the point where {\sl holomorphic subgraph reduction} enters, as the Kronecker-Eisenstein sum now contains a holomorphic subgraph sum which, for the case of two-loops, may be reduced as follows, 
\bea
\label{holsub}
(-)^{a_0} \cC^+ \left[ \begin{matrix} a_1 ~ a_2 ~ a_3 \cr 0 \, ~ \, 0 \, ~\,  b \cr \end{matrix} \right]
\! & \!  = \! & \! 
\sum_{k=2}^{[a_1/2]}  \mbinom{a_0-1-2k}{a_1-2k} \, \cG_{2k} \, 
\cC^+ \left[ \begin{matrix} a-2k & 0 \cr b  & 0 \cr \end{matrix} \right] + (a_1 \leftrightarrow a_2)
\\ &&
-  \mbinom{a_0}{a_1} \cC^+ \left[ \begin{matrix} a ~\,  0 \cr b  ~  0 \cr \end{matrix} \right]
+ \mbinom{a_0-2}{a_1-1} \left \{ 
\cG_2 \cC^+ \left[ \begin{matrix} a-2 & 0 \cr b  & 0 \cr \end{matrix} \right]
+  \cC^+ \left[ \begin{matrix} a-1 & 0 \cr b-1  & 0 \cr \end{matrix} \right]
\right \} 
\no
\eea
where we have used the notation $a_0=a_1+a_2$. The form $\cG_{2k}$ is defined by $\cG_{2k} (\tau) = (\tau_2)^{2k} G_{2k}(\tau)$ for $k \geq 2$ while $\cG_2 = \tau _2^2 G_2$  is the non-holomorphic modular covariant  weight two Eisenstein series. It was shown in \cite{DHoker:2016mwo,DHoker:2016quv} that $\nabla$ applied to any form in $\cV_{(a,b)}$ produces linear combinations of modular graph forms  in which the last two terms on the right side in (\ref{holsub}) cancel, so that $\cG_2$ never appears in the range  $\nabla \cV_{(a,b)}$.

\subsection{The sieve algorithm}
\label{sec:sieve-alg}

The sieve algorithm, developed in \cite{DHoker:2016mwo}, can be stated simply and explicitly for two-loop graphs. Holomorphic subgraph reduction shows that the range  $\nabla \cV_{(a,b)}$ contains modular graph forms of weight $(a+1,b-1)$ which do not belong to $\cV_{(a+1,b-1)}$. Those modular graph forms belong to the space $\mC_{(a+1,b-1)}$ specified by the following lemma.\footnote{The space $\cV^0_{(a,b)}$ used in \cite{DHoker:2016mwo} will be denoted by $\mC_{(a,b)}$ in this paper for the sake of greater clarity.}

{\lem
\label{C0}
{\sl For $a-b \in 2 \ZZ$ and $a, b \geq 2$  the following sum is direct,
\bea
\label{decC}
\mC_{(a,b)}  =  \bigoplus _{k=2}^{ [a/2] } \left \{  \cG_{2k} \,  \cC^+ \left [ \begin{matrix} a-2k & 0 \cr b  & 0 \cr \end{matrix} \right]
\right \}
\eea}}

We shall prove the lemma by showing that the asymptotics near the cusp, given by a Laurent series up to exponential corrections, are linearly independent.  The Laurent series in $\tau_2$ may be obtained by combining the asymptotics of (\ref{Eas}) with the first equation in  (\ref{delE})  when $a \geq b$, and with the second equation of (\ref{delE}) when $a \leq b$. The result may be combined into the following expression which is valid in both cases, 
\bea
\cC^+ \left [ \begin{matrix} a & 0 \cr b  & 0 \cr \end{matrix} \right] = \nu_a \tau_2^a + \nu_b \tau_2 ^{1-b}
+ \cO(e^{-2 \pi \tau_2})
\eea
where $\nu_a , \nu_b \not=0$. The asymptotics of $\cG_{2k} = (\tau_2)^{2k} G_{2k}$  is given by $\tau_2^{2k}$ in view of (\ref{Gk}), so that the asymptotics of the term with summation index $k$ in (\ref{decC}) is proportional to $\nu_a' \tau_2^a + \nu_b' \tau_2^{1+2k-b}$ where $\nu_a', \nu_b' \not=0$. The linear independence of the terms in $\tau_2^{1+2k-b}$ then shows linear independence of the full functions, which proves the lemma.

\subsubsection{The sieve}

From the definition of $\mC_{(a,b)}$ it is clear that $\cV_{(a,b)} \cap \mC _{(a,b)} =0$, so that $\nabla$ acts as follows,
\bea
\label{map}
\nabla : \cV _{(a,b)} & \to & \cV_{(a+1, b-1)} \, \oplus \, \mC_{(a+1,b-1)}
\eea
We define an ordered set  of subspaces of $\mC_{(a,b)}$, or sieve, 
\bea
\cV_{(a,b)} \supset \cV_{(a,b)}^{(1)}  \supset \cV_{(a,b)}^{(2)} \supset \cdots \supset \cV_{(a,b)}^{(b-1)}
\eea
by the following iterative procedure. Starting with $n=1$, we define $\cV^{(1)}_{(a,b)}$ to be the maximal subspace of $\cV_{(a,b)}$ whose range under $\nabla$ has vanishing component along $\mC_{(a+1,b-1)}$, 
\bea
\nabla : \cV^{(1)}_{(a,b)} & \to & \cV_{(a+1,b-1)}
\eea
The process may be repeated to define $\cV^{(1)}_{(a+1,b-1)}$ to be the maximal subspace of $\cV_{(a+1,b-1)}$ on which the range of $\nabla$ has vanishing component along $\mC_{(a+2,b-2)}$ by, 
\bea
\nabla : \cV^{(1)}_{(a+1,b-1)} & \to & \cV_{(a+2,b-2)}
\eea
In turn the space $\cV^{(1)}_{(a+1,b-1)}$ allows us to define $\cV^{(2)}_{(a,b)}$ as the maximal subspace of $\cV^{(1)}_{(a,b)}$ on which the range of $\nabla$ is entirely inside $\cV^{(1)}_{(a+1,b-1)}$ so that the range of $\nabla^2$ on $\cV^{(2)}_{(a,b)}$ has vanishing component along $\mC_{(a+2,b-2)}$,
\bea
\nabla : \cV^{(2)}_{(a,b)} \to \cV^{(1)}_{(a+1,b-1)} 
\hskip 1in 
\nabla ^2 : \cV^{(2)}_{(a,b)} \to \cV_{(a+2,b-2)} 
\eea
For the general case, we continue the iterative process by defining  $\cV^{(n)} _{(a,b)}$ to be the maximal subspace of $\cV^{(n-1)}_{(a,b)}$ on which the range of $\nabla^n$ has zero component along $\mC_{(a+n,b-n)}$, 
\bea
\nabla^n : \cV^{(n)}_{(a,b)} & \to & \cV_{(a+n,b-n)}
\eea
The process ends when $n=b-1$ because it follows from the definition (\ref{pos}) that $\cV_{(a,1)}$ vanishes for even $a$ and  is one-dimensional for odd $a$ generated by $\nabla ^{w-1} E_w$ with  $2w-1=a$.

\subsubsection{Direct sum decomposition}

Of fundamental importance in the application of the sieve algorithm are the difference spaces $\cU_{(a,b)} ^{(n)}$ defined by the following direct sum decompositions,
\bea
\cV_{(a,b)} ~ & =  & \cU_{(a,b)} ^{(1)} \oplus \cV^{(1)}_{(a,b)} 
\no \\
\cV_{(a,b)} ^{(n-1)} & =  & \cU_{(a,b)} ^{(n)} \oplus \cV^{(n)} _{(a,b)}  \hskip 1in 2 \leq n \leq b-1
\eea
Therefore, we have,
\bea
\nabla ^k \cV_{(a,b)} ^{(n-1)} & \subset & \cV_{(a+k,b-k)} ^{(n-1-k)} \hskip 1.5in k \leq n-1
\no \\
\nabla ^{n} \cV_{(a,b)} ^{(n-1)} & \subset & \cV_{(a+n,b-n)} \oplus \nabla ^n \cU _{(a,b)} ^{(n)}
\eea
These relations do not define the elements in the spaces $\cU_{(a,b)}^{(n)}$ uniquely, but we achieve uniqueness by requiring,
\bea
\nabla ^n \cU_{(a,b)} ^{(n)} \subset \mC _{(a+n,b-n)}
\eea
Thus, the subspaces $\cU_{(a,b)}^{(n)}$ collect the obstruction to $\nabla$ mapping to $\cV_{(a,b)}$ spaces. 
For the special case of {\sl modular graph functions} with $a=b=w$  the  sieve reduces to,
\bea
\cV_{(w,w)} \supset \cV_{(w,w)}^{(1)}  \supset \cV_{(w,w)} ^{(2)} \supset \cdots \supset \cV_{(w,w)} ^{(w-1)}
\eea
The Eisenstein series $E_w$ belongs to each space in the sieve $E_w \in \cV_{(w,w)} ^{(n)}$ for all $0\leq n \leq w-1$.

\subsection{Odd modular graph functions from Eisenstein series}

An important subspace of odd two-loop modular graph functions of weight $w$  is constructed from the following  two-loop modular graph functions of weight $w=a_1+a_2=b_1+b_2$, 
\bea
\label{ACdef}
 \cA \left[ \begin{matrix} a_1 & a_2 & 0 \cr b_1 & b_2 & 0 \cr \end{matrix} \right]
=
\cC \left[ \begin{matrix} a_1 & a_2 & 0 \cr b_1 & b_2 & 0 \cr \end{matrix} \right]
- \cC \left[ \begin{matrix}  b_1 & b_2 & 0 \cr a_1 & a_2 & 0   \cr \end{matrix} \right]
\eea
Convergence of the sums which define the modular graph functions in (\ref{4a1}) requires $a_1+b_1 \geq 3$ and $a_2+b_2 \geq 3$. Using the decomposition of Corollary \ref{prop34}, these functions are linear combinations of the functions $\cA_{u,v;w}$ with the range for $u,v$ defined in (\ref{Ashort}). Applying  the algebraic reduction formulas of (\ref{algred})  to $\cA$, we find, 
\bea
\label{3b33}
\cA \left[ \begin{matrix} a_1 & a_2 & 0 \cr b_1 & b_2 & 0 \cr \end{matrix} \right]
= 
\cC \left[ \begin{matrix} a_1 & 0 \cr b_1 &  0 \cr \end{matrix} \right]
\cC \left[ \begin{matrix}  a_2 & 0 \cr  b_2 & 0 \cr \end{matrix} \right]
-
\cC \left[ \begin{matrix} b_1 & 0 \cr a_1 &  0 \cr \end{matrix} \right]
\cC \left[ \begin{matrix}  b_2 & 0 \cr  a_2 & 0 \cr \end{matrix} \right]
\eea
If $a_1+b_1$ is an odd integer, and thus $a_2+b_2$ is also odd, each  factor on the right side of (\ref{3b33}) vanishes and the resulting $\cA$ is zero. Assuming now that $a_1+b_1$ is even and, without loss of generality, that $a_1 \geq b_1$, we define the following combinations, 
\bea
a_1+ b_1=2 w_1 
& \hskip 0.8in &
a_1-b_1=2k
\no \\
a_2+ b_2=2 w_2 
&&
b_2-a_2=2k
\eea
where $k \geq 0$, and $w=w_1+w_2$ with $w_1, w_2 \geq 2$.  We use the expressions of (\ref{delE}) for one-loop modular graph functions in terms of derivatives of Eisenstein series to simplify $\cA$ as follows,
\bea
\label{APdef}
\cA \left[ \begin{matrix} a_1 & a_2 & 0 \cr b_1 & b_2 & 0 \cr \end{matrix} \right]
= { \Gamma (w_1) \Gamma (w_2) \over \Gamma (w_1+k) \Gamma (w_2+k) } \, \cP _k (w_1, w_2)
\eea
where  the function $\cP$ is given as follows,
\bea
\label{Pdef}
\cP _k (w_1, w_2) =\tau_2^{-2k}\left( \nabla ^k E_{w_1} \, \DDb ^k E_{w_2}
- \nabla ^k E_{w_2} \, \DDb ^k E_{w_1} \right)
\eea
Manifestly, the function $\cP_k(w_1,w_2) $ vanishes for $k=0$  and is odd under swapping $w_1$ and $w_2$. Thus we shall assume $k \geq 1$ and $w_1 > w_2$. We have $\cA \in \mA_w$ whenever none of the exponents $a_1, a_2, b_1, b_2$ vanishes or, in this case, we have $\cP_k (w_1, w_2) \in \mA_w$ whenever $k+1 \leq w_2<w_1$.  We define the corresponding subspace as follows, 
\bea
\label{A1def}
\mA_w^{(1)} = \Big \{ \cP_k(w_1,w_2)  \hbox{  with  } w=w_1+w_2 \hbox{ and } 1 \leq k < w_2 < w_1 \Big \}
\eea
Note that $\cP_{w_2} (w_1, w_2)$ does not belong to $\mA_w$ but instead belongs to $\mC_{(w,w)}$ defined in (\ref{decC}).

\subsection{The spaces $\mA_w^{(2)}$ and $\mA_w^{(3)}$}

We now introduce two further  subspaces $\mA_w^{(2)}, \mA_w^{(3)}\subset \mA_w$,  defined as follows,
\bea
\mA_w^{(2)} &=& 
\left\{\cA_{u,u+1;w} \hbox{ ~ with ~ }  1\leq u \leq \left[{w-3 \over2 }\right]\right\} 
\no\\
\mA_w^{(3)} &=& 
 \left\{\cA_{u,u+1;w} \hbox{ ~ with ~ }  \left[{w-1\over 2}\right]\leq u \leq w-5 \right\} 
\eea
The following subsections will rely on the following theorem, proven in Appendix \ref{app:indepproof}.
{\thm
\label{lem32}
{\sl The spaces $\mA_w^{(i)}$, $i=1,2,3$ are linearly independent.}  }

\subsection{Bounds on the dimension of $\mA_w$}

The construction of the subspaces $\mA_w^{(1)}$, $\mA_w^{(2)}$, and $\mA_w^{(3)}$, and Theorem \ref{lem32} allow us to put a lower bound on $\dim \mA_w$. Together with the upper bound given after (\ref{Ashort}), we have, 

{\thm
{\sl The dimension of the space $\mA_w$ for $w\geq 5$ is bounded by 
\bea
\sum_{i=1}^3 \mathrm{dim}\, \mA_w^{(i)} \leq\,\mathrm{dim}\,\mA_w \,\leq \half (w-1)(w-2)
\eea
where
\bea
\label{dimA1}
\dim \mA_w^{(1)}  = \sum _{w_2 =2} ^{[ { w-1 \over 2} ]} (w_2-1)= \half \left [ { w-1 \over 2} \right ] \left [ { w-3 \over 2} \right ] 
\eea
and
\bea
\mathrm{dim} \,\mA_w^{(2)} = \left[ {w - 3 \over 2}\right]  \hspace{0.5 in}\mathrm{dim} \mA_w^{(3)} = \left\{{ \begin{matrix} w-5 -  \left[ {w - 3 \over 2}\right]& w \geq 8 \\ 0 & \mathrm{otherwise} \\\end{matrix}}\right.
\eea
}}
\newline
Explicit computation shows that this lower bound is saturated for $w \leq 11$, while for $w\geq12$ there are additional contributions. We find the following values for $\mathrm{dim}\,\mA_w$ for $w \leq 12$,
\begin{align}
 \dim \mA_5 &= 2 & \dim \mA_6 &=2 & \dim \mA_7 &= 5
\no\\
\dim \mA_8 & =6 & \dim \mA_9 &= 10 & \dim \mA_{10} & =11
\no\\
\dim \mA_{11} &= 16 & \dim \mA_{12} & =18
\end{align}
In the next subsection we illustrate this explicit computation for the simplest case of $\mA_5$.

\subsection{Constructing a basis for $\mathrm{dim}\,\mA_5$}

To make the discussion of this section more concrete, we illustrate here the construction of a basis for $\mA_5$ using holomorphic subgraph reduction and  the sieve algorithm.  By definition (\ref{Apos}), the space $\mA_5$ is generated by the following six functions, 
\bea
\label{list51}
\cA_{u,v;5}  \hskip 1in 1 \leq u < v \leq 4
\eea
The unique weight five $\cP$-function $\cP_1(2,3)$ is a linear combination of the above functions.  We shall now use the sieve algorithm to construct all linear relations between the functions $\cA_{u,v;w}$. To this end we apply  $\nabla$ repeatedly until we produce a modular form  $\cG_{2k}= \tau_2^{2k} G_{2k}$ proportional to the holomorphic Eisenstein series $G_{2k}$ for $k\geq 2$. Applying $\nabla$ once, we obtain, 
\bea
\label{nab51}
\nabla \cA_{1,2;5},  \,  \nabla  \cA_{2,3;5} & \in & \cV_{(6,4)}  
\eea
as well as 
\bea
\label{nab52}
\nabla \cA_{1,3;5} 
-  3 \,\cG_4 \, \cC^+ \left[ \begin{smallmatrix}2 & 0 \\ 4 & 0  \end{smallmatrix}\right] & \in & \cV_{(6,4)} 
\no\\
\nabla \cA_{1,4;5} 
-3\, \cG_4 \, \cC ^+ \left[ \begin{smallmatrix}2 & 0 \\ 4 & 0  \end{smallmatrix}\right] & \in & \cV_{(6,4)}
\no\\
\nabla \cA_{2,4;5}
-3 \,\cG_4 \, \cC^+ \left[ \begin{smallmatrix}2 & 0 \\ 4 & 0  \end{smallmatrix}\right] & \in & \cV_{(6,4)}
\no\\
\nabla \cA_{3,4;5}
+ 3\, \cG_4  \,\cC^+ \left[ \begin{smallmatrix}2 & 0 \\ 4 & 0  \end{smallmatrix}\right] & \in & \cV_{(6,4)}
\eea
Retaining $\cA_{1,4;5}$ as a first basis function of $\mA_5$, we take linear combinations $\cA_{1,3;5} - \cA_{1,4;5}$, $\cA_{2,4;5} - \cA_{1,4;5} $, and $\cA_{3,4;5} + \cA_{1,4;5}$ whose image under $\nabla$ is in $\cV_{(6,4)}$ in view of (\ref{nab52}). Applying $\nabla ^2$ to these functions, and $\nabla$ to the forms in  (\ref{nab51}), we obtain, 
\bea
\label{nab54}
\nabla^2(\cA_{1,3;5} - \cA_{1,4;5}) + 6 \, \cG_4  E_3 & \in & \cV_{(7,3)}
\no \\
\nabla ^2 (\cA_{2,4;5} - \cA_{1,4;5}) + 6 \, \cG_4  E_3 & \in & \cV_{(7,3)}
\no \\
\nabla ^2 (\cA_{3,4;5} + \cA_{1,4;5}) + 6 \, \cG_4  E_3 & \in & \cV_{(7,3)}
\no \\
\nabla^2 \cA_{1,2;5} + 6\, \cG_4  E_3 & \in & \cV_{(7,3)}
\no \\
\nabla^2 \cA_{2,3;5}  +18\, \cG_4 E_3 & \in & \cV_{(7,3)}
\eea
Retaining $ \cA_{1,2;5}$ as a second basis vector of $\mA_5$ (which is independent of $\cA_{1,4;5}$ by the sieve algorithm), we  take linear combinations of all the other double derivative expressions to eliminate the $\cG_4 E_3$ term, and to obtain combinations which belong to $\cV_{(7,3)}$. The further application of $\nabla$ produces the following results for $k=3,4$, 
\bea
\label{nab55}
\nabla^k (\cA_{1,3;5} - \cA_{1,4;5} - \cA_{1,2;5} )  & \in & \cV_{(5+k,5-k)}
\no \\
\nabla ^k (\cA_{2,4;5} - \cA_{1,4;5}- \cA_{1,2;5})  & \in & \cV_{(5+k,5-k)}
\no \\
\nabla ^k (\cA_{3,4;5} + \cA_{1,4;5} - \cA_{1,2;5})  & \in & \cV_{(5+k,5-k)}
\no \\
\nabla^k ( \cA_{2,3;5}  -3 \cA_{1,2;5})  & \in & \cV_{(5+k,5-k)}
\eea
By inspection of its definition in (\ref{pos}), the space $\cV_{(9,1)}$, which enters above for $k=4$, is one-dimensional and generated by $\nabla ^4 E_5$. However, $E_5$ is even and thus the left side of each line in (\ref{nab55}) must vanish for $k=4$.  Each equation above then states that $\nabla ^4 \cA=0$ for the corresponding  $\cA$.  Lemma 1 of \cite{DHoker:2016mwo}  implies in each case that $\cA$ must be constant. Since there are no constant odd functions, $\cA$ must vanish identically, and we obtain the  identities, 
\bea
\cA_{1,3;5} = \cA_{1,2;5} + \cA_{1,4;5}
& \hskip 0.5in &
\cA_{2,3;5} = 3\cA_{1,2;5}
\no \\
\cA_{2,4;5} =  \cA_{1,2;5} + \cA_{1,4;5} 
&&
\cA_{3,4;5} = \cA_{1,2;5}- \cA_{1,4;5} 
\eea
One may use similar means to show that $\cP_1(2,3) = 6 \cA_{1,4;5}$. The differential equation satisfied by $\cA_{1,2;5}$ in the fourth line of (\ref{nab54}) and by  $\cA_{1,4;5}$ in the second line of (\ref{nab52})  imply the linear independence of these functions. 
Therefore we conclude that $\dim \mA_5=2$, and we may take $\cA_{1,2;5}$ and $\cA_{1,4;5}$ as a basis.

\subsection{Speculations on a basis for $w \geq 12$}

For weight 12, the dimensions of $\mA_{12}^{(1)}$, $\mA_{12}^{(2)}$, $\mA_{12}^{(3)}$ are respectively 10, 4, and 3, adding up to 17, one fewer than the result of explicit calculation.  The missing subspace may be generated, for example,  by the function $\cA_{4,7;12}$. For weights $w >12$, no systematic pattern could yet be proven. 
However, preliminary checks done up to weight 20 using Mathematica  indicate that the following space may be a candidate for the missing subspace for higher weights,
\bea
\label{3termthings}
\mA_{w}^{(4)} =  \left\{\cA\left[\begin{matrix}u & 0 & w-u \cr 0 & v & w-v \end{matrix} \right]\, \Big | \, 1 \leq \g \leq \g_{{\rm max}},\,\, \left[{w+1 \over2}\right] \leq v \leq w-6 \right\}\,\,\,\,
\eea
where $\gamma$ and $\gamma _{{\rm max}}$ are defined as follows,
\bea
u = \left[{w+1 \over 2}\right]-3 + \g \hspace{0.5 in} \g_{{\rm max}}(v,w) = \mathrm{min}\left\{w-v-5,\,v -\left[{w+1\over2}\right]+1\right\}
\eea
Functions in this set would be expected to produce holomorphic Eisenstein series either at order $\nabla^{u}$ or $\nabla^{w-v}$, and it can be shown that only the latter occurs. However, a complete proof of linear independence has remained elusive. 

\sm

Assuming linear independence, the dimension of the space $\mA_{w}^{(4)}$ is given by,
\bea
\mathrm{dim} \,\mA_w^{(4)} = \sum_{n=6}^{[w/2]} \mathrm{min}\left\{n-5 ,\, \left[{w \over2}\right] - n+1 \right\} =\sum_{n=2}^{\left[{(w-8)\over2}\right]} \left[{n\over2}\right] 
\eea
which would lead to the following full dimension of $\mA_w$,
\bea
\mathrm{dim}\, \mA_w = \sum_{i=1}^4\mathrm{dim} \,\mA_w^{(i)}
\eea
giving up to weight $w \leq 20$ the following speculated predictions,
\begin{align}
\dim \mA_{13} &=24    & \dim \mA_{14}&=26  &  \dim \mA_{15} &=33 
\no\\
\dim \mA_{16}&=36 &  \dim \mA_{17} &=44 & \dim \mA_{18}&=47
\no\\
\dim \mA_{19} &=56 & \dim \mA_{20} &=60
\end{align}
It would be interesting to establish a complete formula for the dimension at all weights.

\newpage

\section{Inner product of modular graph functions}
\setcounter{equation}{0}
\label{sec:7}

The Petersson inner product $\< f|g\>$ of modular functions $f,g$ is defined  by \cite{Petersson},
\bea
\< f | g \> = \int _{\cM} d\mu \,  f(\tau)^* g(\tau) \hskip 1in d\mu = {i \over 2 \tau_2^2} \, d \tau \wedge d \bar \tau
\eea 
where $\cM$ is the fundamental domain for $PSL(2,\ZZ)$ in the upper half $\tau$-plane $\cH$. The Petersson inner product defines the space $L^2(\cM)$ of modular functions which are square integrable on $\cM$. The Petersson inner product may also be defined for modular forms, but we shall not consider this case here. Clearly, the inner product vanishes when $f$ is an even modular function and $g$ is an odd modular function. Therefore, it suffices to consider the inner product between modular functions which are both even  or both odd. 

\sm

In this section, we shall investigate  the Petersson inner product involving one-loop and two-loop modular graph functions, and evaluate the product by applying the Rankin-Selberg-Zagier method \cite{Rankin, Selberg, Zagier}
 to the Poincar\'e and Fourier series obtained in preceding sections (see also \cite{Pioline:2014bra} for a useful reference in the physics literature). It will be convenient to consider the cases of even and odd modular graph functions separately. Of considerable use will be the action of the Laplace-Beltrami operator  on modular graph functions, which we shall summarize in the next subsection.

\subsection{The Laplace operator acting on $\cC_{u,v;w}$}

The Eisenstein series are eigenfunctions of the Laplace operator $\Delta = 4 \tau_2^2 \p_{\bar \tau} \p_\tau$ on $\cH$, 
\bea
\Delta E_s = s(s-1) E_s
\eea
The action of the Laplace operator on two-loop modular graph functions was studied extensively in \cite{DHoker:2015gmr} for a special class of even functions. Here we shall need a generalization to the functions $\cC_{u,v;w}$ which include both even and odd functions and  is given by,
\bea
\label{DelC}
\Delta \cC_{u,v;w} &=&  \left ( w^2+2 uv - uw-vw-w  \right ) \cC_{u,v;w} + uv \, \cC_{u+1,v+1;w} 
\no \\ && 
+u (2v -w)\, \cC_{u+1,v;w}  +v (2u-w) \, \cC_{u,v+1;w}
\no \\ && 
+u  (v-w)\, \cC_{u+1,v-1;w} +v (u-w)\, \cC_{u-1,v+1;w}
\eea
The indices $u,v$ are integer-valued, but we shall allow $w\in\CC$, subject to the convergence conditions given earlier.

\sm

The Laplace operator $\Delta$ is invariant under parity $\tau \to - \bar \tau$ and (\ref{DelC}) is invariant under interchange of $u$ and $v$ for fixed $w$. Therefore, the symmetric and anti-symmetric parts of $\cC_{u,v;w}$, respectively denoted $\cS_{u,v;w} \in \mS_w$ and $\cA_{u,v;w} \in \mA_w$, separately obey  (\ref{DelC}).  

\sm

The odd two-loop modular graph functions $\cP_k(s,t)$ for $k \in \NN$ and $s,t \in \CC$ defined in (\ref{APdef}) and expressed in terms of bilinears in derivatives of Eisenstein series in (\ref{Pdef}), 
\bea
\cP_k(s,t) (\tau) = \tau_2^{-2k} \Big ( \nabla ^k E_s \bar \nabla ^k E_t - \nabla ^k E_t  \bar \nabla ^k E_s \Big )
\eea
satisfy a particularly simple system of Laplace eigenvalue equations given by,
\bea
\label{DeltaP}
\Delta \cP_k(s,t) & = & \cP_{k+1}(s,t) + (s^2 +t^2 -s-t-2k^2) \cP_k(s,t)
\no \\ && +
(s-k)(t-k)(s+k-1)(t+k-1) \cP_{k-1}(s,t)
\eea
The validity of this formula may be verified by direct calculation, and it will be convenient to set $\cP_0(s,t)=0$. A particularly useful property of the system (\ref{DeltaP}) is that it allows us to express $\cP_{k+1}$ in terms of a linear combination of $\Delta \cP_k$ and $\cP_\ell$ with $\ell<k$.

\sm

Restricting attention to $s,t \in \NN$, the system (\ref{DeltaP}) truncates in the following sense. For $1\leq k < t <s$, the functions $\cP_k(s,t)$ belong to the space $\mA_w^{(1)}$, but  $\cP_t (s,t)$ is no longer an element of $\mA_w^{(1)}$ because it may be reduced to a combination of holomorphic Eisenstein series. Truncating the system  to $1 \leq k <t<s$ and treating the function $\cP_t(s,t)$ as an inhomogeneous  term, we find the following system of differential equations,
\bea
\Delta \cP_k (s,t) - \sum _{\ell =1}^{t-1} M_{k\ell} \cP_\ell (s,t) = \delta _{k, t-1} \cP_t (s,t)
\eea
The entries of the matrix $M$ are given by (\ref{DeltaP}) and its spectrum  obeys the theorem below. 
{\thm
{\sl The eigenvalues of the matrix $M$ have multiplicity one and are found to be,}} 
\bea
(s+t-2n)(s+t-1-2n) \hskip 1in n = 1, \cdots, t-1
\eea
The theorem is a special case of a result which was proven in Section 5.3 of \cite{DHoker:2016quv}.

\subsection{Petersson inner product on odd modular graph functions}

Odd modular graph functions do not exist at one loop order. The space of odd two-loop modular graph functions is denoted $\mA= \bigoplus _w \mA_w$ and is the direct sum of the spaces $\mA_w$ of functions of weight $w$.  In view of Theorem \ref{prop36} all odd two-loop modular graph functions  are cuspidal functions with exponential decay at the cusp so that $\mA \subset L^2(\cM)$. The Petersson inner product between any pair of odd modular graph functions is convergent and may be calculated in terms of Poincar\'e and Fourier series by  Rankin-Selberg methods.

\sm

In this section we shall investigate  the Petersson inner product between odd two-loop modular graph functions and show how the product is evaluated concretely between  $\cA_{u,v;w}$ and $\cP_k (s,t)$. The Poincar\'e series of $\cA_{u, v;w} (\tau)$ with respect to $\Gamma _\infty \backslash PSL(2,\ZZ)$ is given by, 
\bea
\cA_{u, v;w} (\tau) = \sum _{g \in \Gamma _\infty \backslash PSL(2,\ZZ)} \Lambda ^\cA _{u,v;w} (g \tau)
\eea
where $\Lambda^\cA _{u,v;w}(\tau) = \Lambda _{u,v;w} (\tau) - \Lambda _{v,u;w}(\tau)$ and $\Lambda _{u,v;w}(\tau)$ is given in equation (\ref{Lambda}) of Theorem~\ref{thm52}. The Fourier series of $\cA_{u, v;w} (\tau)$ is given in (\ref{AF}). Using standard Rankin-Selberg, we have, 
\bea
\< \cA_{u_1,v_1;w_1} | \cA_{u_2,v_2;w_2} \> & = &
2 \sum_{N=1}^\infty \int _0 ^\infty {d \tau_2 \over \tau_2^2} \cF^{(N)} _{u_1,v_1;w_1} (\tau_2) 
\int _0^1 d\tau_1 (\bar q^N-q^N) \Lambda ^\cA _{u_2,v_2;w_2}(\tau)
\eea
For $w_2 \in \CC$, we use equation (\ref{Ups}) to express the seed function and carry out the integral over $\tau_1$ to obtain the theorem below.

{\thm
{\sl The inner product between two odd modular graph functions $\cA_{u_1,v_1;w_1} $ with Fourier coefficients $\cF$ and $ \cA_{u_2,v_2;w_2}$ with Poincar\'e series seed $\Upsilon$, is given by,
\bea
\< \cA_{u_1,v_1;w_1} | \cA_{u_2,v_2;w_2} \> 
& = & 
8 \sum _{k=1}^u  \mbinom{u_2+v_2-k-1}{v_2-1} \sum_{N=1}^\infty  \, 
\sum_{0<m|N} {(2 \pi)^k  N^{k-1}  \over \Gamma (k) m^{k-1}} 
\int _0 ^\infty  d \tau_2 \, \tau_2^{w_2-2}  
\no \\ && \quad \times 
e^{- 4 \pi N \tau_2} \, 
 \cF^{(N)} _{u_1,v_1;w_1} (\tau_2) \Upsilon_{w_2,k,u_2+v_2}(2m\tau_2) -(u \leftrightarrow v)
\eea
The integral over $\tau_2$ is absolutely convergent for sufficiently large $\Re(w_2)$. }}

\subsubsection{The inner product of $\cP$ with $\cA$}

The evaluation simplifies for the special case of the odd modular graph functions $\cP_k(s,t)$.  Consider the Petersson inner product of $\cP_k(s,t)$ with $\cA_{u,v;w}$ given by,
\bea
\<  \cP_k (s,t) | \cA_{u,v;w} \> = \int _{\cM}  d\mu \, \cP_k (s,t) (\tau)  \cA_{u,v;w} (\tau)
\eea
The inner product with functions $\cP_{k}(s,t)$ for $k >1$ may be expressed in terms of inner products involving only $\cP_1(s,t)$ with the help of the Laplace equations for both functions, and the self-adjointness  of the Laplace operator on the space of odd modular graph functions,
\bea
\< \Delta \cP_k (s,t) | \cA_{u,v;w} \> = \< \cP_k (s,t) | \Delta \cA_{u,v;w} \> 
\eea
For any $k \geq 1$ and fixed $s,t$, we abbreviate $\cP_k = \cP_k (s,t)$ and obtain the following recursion relations on the index $k$,
\bea
\< \cP_{k+1} | \cA_{u,v;w} \>   & = &  
uv \<  \cP_k | \cA_{u+1,v+1;w} \>  
\\ && 
- (s-k)(t-k)(s+k-1)(t+k-1) \< \cP_{k-1} | \cA_{u,v;w} \>
\no \\ && 
\left \{ w^2+2 uv - w (1+u+v) - s^2 - t^2 + s + t + 2k^2 \right \} \< \cP_k | \cA_{u,v;w} \>
\no \\ &&
+u (2v -w)\, \< \cP_k | \cA_{u+1,v;w} \>   +v (2u-w) \, \< \cP_k | \cA_{u,v+1;w}\>
\no \\ && 
+u  (v-w)\, \< \cP_k | \cA_{u+1,v-1;w} \>  +v (u-w)\, \<  \cP_k | \cA_{u-1,v+1;w} \> 
\no
\eea
which proves the assertion.

\sm

To evaluate the Petersson inner product $\< \cP_1(s,t)|\cA_{u,v;w}\>$ we use the expression for $\cP_1(s,t)$ in terms of Eisenstein series (\ref{Pdef}). We express the Eisenstein series $E_s$ in terms of its Poincar\'e series and express $E_t$ in terms of its Fourier series expansion (\ref{EF}).  We shall keep $s \in \CC$ but restrict $t \in \NN$. Finally, we use the fact that $\cA_{u,v;w}$ is cuspidal and unfold the Poincar\'e series using standard Rankin-Selberg, 
\bea
\<\cP_1(s,t) | \cA_{u,v;w}\> & = &{ 16 \, s \, \zeta (2s) \over \pi ^{s-1} \Gamma (t)} 
\sum _{k=0}^{t-1} { \Gamma (t+k) \over (4 \pi )^k \Gamma (t-k) k!} 
\sum _{N=1}^ \infty N^{t - k} \sigma _{1-2t}(N)  
\no \\ &&
\times \int _0^\infty d \tau _2 \, \tau_2^{s-k-1}  \int _0^1 d \tau_1 \left ( \bar q^N - q^N  \right ) \cA_{u,v;w}(\tau)
\eea
The Fourier series decomposition of $\cA_{u,v;w}$, given in (\ref{AF}), then produces the theorem below.

{\thm 
{\sl The inner product of $\cP_1(s,t)$ with the odd modular graph function $\cA_{u,v;w}$  whose Fourier coefficients are $\cF^{(N)}_{u,v;w}- \cF^{(N)}_{v,u;w}$ is given as follows, 
\bea
\<\cP_1(s,t) | \cA_{u,v;w}\> & = & { 32 \, s \, \zeta (2s) \over \pi ^{s-1} \Gamma (t)} 
\sum _{k=0}^{t-1} { \Gamma (t+k) \over (4 \pi )^k \Gamma (t-k) k!} 
\sum _{N=1}^ \infty N^{t - k} \sigma _{1-2t}(N)  
\no \\ &&
\times \int _0^\infty d \tau _2 \, \tau_2^{s-k-1}  e^{-4 \pi N \tau_2}
\left( \cF^{(N)} _{u,v;w} (\tau_2) -\cF^{(N)} _{v,u;w} (\tau_2) \right)
\eea
The integral over $\tau_2$ is absolutely  convergent  for sufficiently large $\Re(s)$.}}

\sm

 Explicit formulas may be obtained in terms of the coefficients $\cW_{u,v;w}^{k,\ell,\b}(N,L)$, $\cM_{u,v;w}^{k,\ell}(N)$, and $\cH_{u,v;w}^{k,\ell,\beta}(N)$ of the Laurent polynomials which define $\cF^{(N)}_{u,v;w}$  using Theorem \ref{thmnonconst}. The  integral over $\tau_2$ of a term proportional to $\tau_2^\nu $ in these Laurent polynomials  is given by,
\bea
\int _0^\infty d \tau _2 \, \tau_2^{s-k+\nu -1}  e^{-4 \pi N \tau_2}= { \Gamma (s-k+\nu) \over (4 \pi N)^{s-k+\nu}}
\eea
which shows that the sum over $N$ will converge absolutely for $\Re(s)$ sufficiently large.

\subsection{Petersson inner product on even modular graph functions}

Even modular graph functions at one-loop order consist of non-holomorphic Eisenstein series~$E_s$. The space of even modular graph functions at two-loop order will be denoted $\mS= \bigoplus_w \mS_w$ and is the direct sum of the spaces $\mS_w$ of functions of weight $w$. Even modular graph functions, either at one or two loops, generically grow as positive powers of $\tau_2$ near the cusp, and are not in $L^2(\cM)$.  Zagier has extended the use of the Rankin-Selberg method to functions of power-growth near the cusp in terms of ``renormalized" Petersson integrals~\cite{Zagier}. For applications to string theory, however, it is more useful to consider the regularized integrals  defined on the cut-off domain 
\bea
\label{ML}
\cM_L = \cM \cap \{ \tau_2 <L\}
\eea

\subsubsection{Producing even cuspidal functions from $E_s$ and $\cC_{u,v;w}$}

We begin by defining the operator $\Delta_k$ in terms of the Laplace-Beltrami operator $\Delta$ mapping the space of modular functions into itself by, 
\bea
\label{Deltak}
\Delta_k = \prod _{\ell=1}^k (\Delta - \ell(\ell-1))
\hskip 1in 
\Delta =  \tau_2^2 \left ( \p_{\tau_1}^2 + \p_{\tau_2}^2 \right )
\eea
An alternative expression for $\Delta _k$ may be obtained in terms of the first order operators,
\bea
\DD = + 2 i \tau_2^2 \p_\tau & \hskip 1in & \DD : (0,n) \to (0,n-2)
\no \\
\DDb = - 2 i \tau_2^2 \p_{\bar \tau} && \DDb :(n,0) \to (n-2,0)
\eea
where $(m,n)$ is modular weight.  The operator $\Delta_k$   satisfies, 
\bea
\label{Deligne}
\Delta _k  = \DDb ^k \tau_2^{-2k} \, \DD ^k  = \DD ^k  \tau_2^{-2k} \, \DDb ^k 
\eea
We have $\Delta _1 = \Delta$, and its action on a monomial $\tau_2^s$ is given by,
\bea
\label{lambdak}
\Delta _k \tau_2^s =  \lambda _k (s) \tau_2^s 
\hskip 1in 
\lambda _k (s) = \prod _{\ell=1-k}^k (s-\ell) 
\eea
The following lemma follows from the observation that the operator $\Delta_w$ annihilates an arbitrary  Laurent polynomial in $\tau_2$ of degrees $(w,1-w)$.

{\lem
\label{Cusp}
{\sl The modular functions $\Delta_w (E_{w_1} E_{w_2}) $ with $w_1+w_2\leq w$, and $\Delta _w \cC_{u,v;w_1}$ for $w_1 \leq w$ are cuspidal functions.}}

\sm

Lemma \ref{Cusp} is a special case of a much more general lemma which we include below.

{\lem
\label{Cuspgen}
{\sl An arbitrary genus-one modular graph function $\cC$ of weight $w$ has a Laurent polynomial of degree $(w,1-w)$ and an associated cuspidal function $\Delta _w \cC$ whose integral is given in terms of the coefficient $\mc_1$ of $\tau_2$ in its Laurent polynomial by the following formula,
\bea
\label{cuspc1}
\int _\cM d \mu \, \Delta _w \cC  = (-)^w \, w ! \, (w-1)! \, \mc_1
\eea}}
The statement that a modular graph function of weight $w$ has a Laurent polynomial of degree $(w,1-w)$ was proven long ago in \cite{DHoker:2015wxz}, so that we have,
\bea
\cC(\tau) = \sum _{k=1-w}^w \mc_k \tau_2^k + \cO(e^{-2 \pi \tau_2})
\eea
In view of (\ref{lambdak}) we have $\Delta _w \tau_2^k=0$ for all $1-w \leq k \leq w$. It follows that  $\Delta _w \cC$ is a cuspidal function with exponential decay at the cusp. The  integral of the cuspidal function $\Delta _w \cC$ reduces to the contribution localized at the cusp only. Expressing the integral as a limit of the integral over the cutoff domain $\cM_L$ defined in (\ref{ML}), we see that it reduces to a contribution from the boundary of $\cM_L$, 
\bea
\label{intcusp}
\int _\cM d \mu \, \Delta _w \cC 
= \lim _{L \to \infty} \int _{\cM_L} d \mu \, \Delta _w \cC 
= \lim _{L \to \infty} \int _0^1 d \tau_1 \Big (\p_{\tau_2}  \Delta _w ' \cC \Big )  \Big |_{\tau_2=L}
\eea
where the operator $\Delta '_w$ is defined  by,
\bea
\Delta_w'  = \prod _{\ell=2}^w \Big ( \Delta - \ell(\ell-1) \Big )
\eea
We have $\Delta _w ' \tau_2^k=0$ for all $k \not = 0,1$ in the range $1-w \leq k \leq w$. For the remaining terms, $k=0,1$, we see that $\Delta _w' 1$ is a constant, which does not contribute to (\ref{intcusp}), and  $\Delta _w ' \tau_2 =  (-)^{w-1} \, w! \, (w-1)! \, \tau_2$ with the help of which (\ref{cuspc1}) is readily established.

\subsubsection{Calculation of the regularized inner product of $E_s$ with $\cC_{u,v;w}$}

Consider a two-loop modular graph function  $\cC_{u,v;w}(\tau)$ of weight $w$. Its Laurent polynomial has degree $(w, 1-w)$. We wish to evaluate the integral,
\bea
\mI_{u,v;w} (s,L) = \int _{\cM_L } d \mu  E_s (\tau)  \cC_{u,v;w}(\tau)
\eea
by relating it to the following integral, 
\bea
I_{u,v;w}(s) = \int _\cM d \mu  E_s (\tau) \Delta_w \cC_{u,v;w}(\tau)
\eea
In view of Lemma \ref{Cusp}, the function $\Delta _w \cC_{u,v;w}$ is cuspidal and the integral $I_{u,v;w}(s)$ is therefore  absolutely convergent for sufficiently large $\Re(s)$. The integral  $I_{u,v;w}(s)$ is readily evaluated by standard Rankin-Selberg in terms of the constant Fourier mode of $\cC_{u,v;w}$, whose Laurent polynomial part cancels out, and we are thus left with, 
\bea
I_{u,v;w} (s) = {2 \zeta(2s)\over \pi^s} \sum_{N=1}^\infty \int _0 ^\infty d\tau_2 \, \tau_2^{s-2}  \Delta_w \Big ( \cQ_{u,v;w}^{(N)} (\tau_2) e^{-4 \pi N \tau_2} \Big )
\eea
The integral is convergent for sufficiently large $\Re(s)$, and we may use the self-adjointness of $\Delta _w$ to move $\Delta _w$ onto $\tau_2^{s-2}$ to obtain the following final expression, 
\bea
I_{u,v;w} (s) = {2 \zeta(2s) \lambda _w(s-2) \over \pi^s} \sum_{N=1}^\infty \int _0 ^\infty d\tau_2 \, \tau_2^{s-2}  \cQ_{u,v;w}^{(N)} (\tau_2) e^{-4 \pi N \tau_2} 
\eea 
The expression for the Laurent polynomial coefficient functions $\cQ_{u,v;w}^{(N)} (\tau_2)$ of the exponential part of the constant Fourier mode  was given in (\ref{finalcQ}) of Theorem \ref{thm15}. 

\sm

To work out the relation between the integrals $\mI_{u,v;w}(s,L)$ and $I_{u,v;w}(s)$ we start from,
\bea
\lambda _w(s) \, \mI_{u,v;w} (s,L) = \int _{\cM_L } d\mu \,  (\Delta _w E_s (\tau))  \cC_{u,v;w}(\tau)
\eea
We use formula (\ref{Deligne}) and carry out successive integrations by parts making sure to collect all the contributions from the boundary of $\cM_L$ at $\tau_2=L$, and we derive the following relation for two arbitrary modular functions $f,g$,
\bea
\label{fgdelwrel}
&&
\int _{\cM_L } d\mu  (\Delta _w f )  g - \int _{\cM_L } d \mu  f (\Delta _w g )  
 \\ && \hskip 0.1in =
 \sum_{k=0}^{w-1} (-)^k \int _0^1 d \tau_1 \Big [ (\DDb^{w-k-1} \tau_2^{-2w} \DD^w f )  \DDb^k  g
 -(\DD^{w-k-1} \tau_2^{-2w} \DDb^w g )  \DD^k  f \Big ] \Big | _{\tau_2=L}
\no
\eea
Substituting $f=E_s$ and $g=\cC_{u,v;w}$, evaluating $\Delta_w$ on $E_s$, and partitioning the integration domain $\cM$ in $I_{u,v;w}$ into $\cM_L$ and its complement, gives the theorem below.

{\thm
\label{Theorem76}
{\sl The regularized inner product integral $\mI_{u,v;w} (s,L)$  is given in terms of the convergent integral $I_{u,v;w}(s)$ by, 
\bea
\label{JmJrel}
\lambda _w(s) \mI_{u,v;w} (s,L)  & = &
I_{u,v;w} (s) - \int _L^\infty { d \tau_2 \over \tau_2^2} \int _0 ^1 d \tau_1 E_s (\Delta _w \cC_{u,v;w} ) 
\no \\ && 
+ \sum_{k=0}^{w-1} (-)^k \int _0^1 d \tau_1 \Big [ (\DDb^{w-k-1} \tau_2^{-2w} \DD^w E_s )  \DDb^k  \cC_{u,v;w}
\no \\ && \hskip 1.3in 
 -(\DD^{w-k-1} \tau_2^{-2w} \DDb^w \cC_{u,v;w} )  \DD^k  E_s \Big ] \Big | _{\tau_2=L}
\eea
Up to corrections of order $\cO(e^{-4 \pi L})$ the second term on the right side may be neglected, while in the remaining sum we may replace $E_s$ and $\cC_{u,v;w}$ by their respective Laurent polynomial parts thereby resulting in a finite sum of powers of $L$. }}

\subsection{Summations over $N$}

The calculation of Petersson inner products for both odd-odd or even-even arrangements of modular graph functions involves infinite-range summations, for which we shall now provide some useful formulas. We begin by recalling the well-known results given in \cite{bateman},
\bea
\label{Ram}
\sum _{N=1}^\infty { \sigma _a(N) \over N^s}  & = & \zeta(s) \zeta (s-a)
\no \\
\sum _{N=1}^\infty { \sigma _a(N)\sigma _b(N)  \over N^s}  & = & { \zeta(s) \zeta (s-a)\zeta(s-b) \zeta (s-a-b)
\over \zeta(2s-a-b)}
\eea
We shall also need sums involving the  generalized divisor function $V_{A,B,C}(M,N)$ defined for $A,B,C \in \ZZ$ and $M,N \in \NN$  in (\ref{VABC}).  We shall assume that $A+B+C$ is even, since otherwise the sum vanishes, and we  compute by manner of example the following sum which is required in the integration of the constant Fourier mode, 
\bea
W_{A,B,C}(s) = \sum _{N=1}^\infty {1 \over N^s} \, V_{A,B,C}(N,N)
\eea
We choose $\Re(s)$ large enough so that the sum is absolutely convergent. 
Substituting the definition of $V$ into the summand, we obtain,
\bea
W_{A,B,C}(s) = \sum _{N=1}^\infty {1 \over N^s} \, \sum _{{m,n |N \atop m,n, m+n \not=0}} 
{ \ep(m) \ep(n) \over m^{A-1} n^{B-1} (m+n)^C}
\eea
Both positive and negative $m$ and $n$ contribute to the sum. Denoting by $k = \gcd(m,n) >0$ the greatest positive common divisor of $m$ and $n$, we may decompose $m,n$ as follows, 
\bea
m= k \mu \hskip 0.7in n = k \nu \hskip 0.7in \gcd(\mu, \nu)=1 \hskip 0.6in \nu \not= - \mu
\eea
Since $m$ and $n$ both divide $N$, clearly $k$ must divide $N$, and we write $N = kM$ with $M >0$. 
The summations over $k$ and $M$ are now independent of the summations over $\mu$ and $\nu$ and may be carried out in terms of the $\zeta$-function, 
\bea
W_{A,B,C}(s) = \zeta(s) \zeta(s+A+B+C-2)  \sum _{{\gcd(\mu,\nu)=1 \atop \mu,\nu,\mu+\nu\not=0}} 
{ \ep(\mu) \ep(\nu) \over \mu^{A-1} \nu^{B-1} (\mu+\nu)^C }
\eea
Rescaling $\mu \to m=k\mu$ and $\nu \to n = k \nu$ for $k \in \NN$ and then summing over $k$ allows us to recover a simplified sum in $m$ and $n$ without divisor conditions,  
\bea
W_{A,B,C}(s) = {\zeta(s) \zeta(s+A+B+C-2) \over \zeta(A+B+C-2)} 
\sum_{m \not=0 } \sum_{n \not=0,-m} 
{ \ep(m) \ep(n) \over m^{A-1} n ^{B-1} (m+n)^C }
\eea
Since the summand is invariant under $(m,n) \to (-m,-n)$ we sum over $m >0$ upon including  a factor of 2. Furthermore, for fixed $m$, the sum over $n$ is given by the function $S_{B-1,C}(m)$,  which is defined and evaluated  in equation in (\ref{Sabm}) of Appendix \ref{sec:C4},  
\bea
W_{A,B,C}(s) = {2 \zeta(s) \zeta(s+A+B+C-2) \over \zeta(A+B+C-2)} 
\sum_{m=1}^\infty { S_{B-1,C}(m) \over m^{A-1}} 
\eea
Evaluating the sums over $m$ with the help of the formulas given in Appendix \ref{sec:C4}, we find, 
\bea
(-)^{B-1} \sum_{m=1}^\infty { S_{B-1,C}(m) \over m^{A-1}} 
& = &
\sum _{j=3} ^{B-1}  (-)^j \mbinom{B+C-j-2}{C-1} \zsv(j) \zeta(A+B+C-2-j) 
\no \\ &&
+ \sum _{j=3} ^C   \mbinom{B+C-j-2}{B-2} \zsv(j) \zeta(A+B+C-2-j)
\no \\ &&
-  \mbinom{B+C-1}{B-1} \zeta(A+B+C-2)
\no \\ &&
- 2  \sum_{j=1}^C  \mbinom{B+C-j-2}{B-2} \zeta(A+B+C-2-j,j)
\eea
Note that in the first two lines above, the second $\zeta$-factor is an odd zeta-value since the first factor is non-zero only when $j$ is odd, and by assumption $A+B+C-2$ is even.

\subsection{Example: the inner product of $E_s$ with $\Delta _4 C_{2,1,1}$}

In this final subsection, we shall evaluate the Petersson inner product of $E_s$ with $\Delta _4 C_{2,1,1}$,
\bea
f(s) = \int _\cM d\mu \, E_s \, \Delta _4 C_{2,1,1}
\eea
Lemma \ref{Cusp} guarantees that $\Delta _4 C_{2,1,1}$ is a cuspidal function so that the above integral 
is absolutely convergent for sufficiently large $\Re(s)$, and may be analytically continued to a meromorphic function in $\CC$. We shall evaluate $f(s)$ in two different ways. First, by expressing $E_s$ as a Poincar\'e series for the coset $\Gamma _\infty \backslash \Gamma$ and evaluating $f(s)$ in terms of an integral over the constant Fourier mode of $\Delta _4 C_{2,1,1}$. The latter is obtained from the expression  for the constant Fourier mode of $C_{2,1,1}$ in subsection \ref{sec:56}. Second, by using the differential equation for $C_{2,1,1}$ and Zagier's integral formula for the product of three Eisenstein series \cite{Zagier}.

\subsubsection{Calculation by Poincar\'e and Fourier series}

With the help of the Poincar\'e series representation for $E_s$, we may express $f(s)$ in terms of an integral of the constant Fourier mode of $\Delta_4 C_{2,1,1}$, 
\bea
f(s) = c_s \int _0^\infty  d \tau_2 \, \tau_2 ^{s-2}  \int _0 ^1 d \tau_1 \, \Delta _4 C_{2,1,1}
\eea
where the asymptotics of the Eisenstein series is expressed as follows,
\bea
E_s (\tau) = c_s \tau_2^s + \tilde c_s \tau_2 ^{1-s} + \cO(e^{-2 \pi \tau_2})\, , 
\hskip 0.3in
c_s =  { 2 \zeta (2s) \over \pi ^s}\, , 
\hskip 0.3in 
\tilde c_s  = { 2 \Gamma (s-\half) \zeta (2s-1) \over \Gamma (s) \pi^{s-\half}}
\quad
\eea
Using the expression for the constant Fourier mode of $C_{2,1,1}$ given in (\ref{C211}) and (\ref{C211a}), and the absolute convergence of the integral for sufficiently large $\Re(s)$ to interchange the sum and integration, we find, 
\bea
f(s) = -8 c_s \sum_{N=1}^\infty  \sigma _{-3}(N)^2 \int _0 ^\infty d\tau _2 \, \tau_2^{s-2} \Delta _4 \left ( { e^{-4 \pi N \tau_2} \over (4 \pi \tau_2)^2} \right )
\eea
Since the integrand falls off exponentially near the cusp, we may integrate $\Delta _4$ by parts for sufficiently large $\Re(s)$ which brings out  a factor of the function $\lambda _4(s)$ defined in (\ref{lambdak}). Carrying out the remaining integral over $\tau_2$ we have,
\bea
f(s) =  -8 \lambda_4 (s) \, c_s \, {\Gamma (s-3) \over (4 \pi )^{s-1}} \sum_{N=1}^\infty  { \sigma _{-3}(N)^2 \over N^{s-3}}
\eea
Using Ramanujan's formula on the second line of (\ref{Ram}) for $a=b=-3$ and the expression for $c_s$  gives the following result, 
\bea
\label{fLsa}
f(s) = - 32 \,  \lambda _4(s) { \Gamma (s-3) \over (2\pi)^{2s-1}} \,  \zeta(s)^2 \, \zeta (s+3) \, \zeta (s-3)
\eea
We note that $f(s)$ has a simple pole at $s=1$, whose residue gives the integral of the cuspidal function $\Delta _4 C_{2,1,1}$ by Rankin-Selberg. 

\sm

The value of the residue may be obtained from (\ref{fLsa}) and equals  $ -16 \pi  \zeta(3)/5$, which also equals $-4! \, 3! \, \mc_1$ where $\mc_1$ is the coefficient of the $\tau_2$-term in the Laurent expansion of $C_{2,1,1}$ given in (\ref{C211L}). This result is in agreement with the general formula (\ref{cuspc1}) of Theorem \ref{Cuspgen} for the integral of such cuspidal functions for the special case of $\cC=C_{2,1,1}$.

\subsubsection{Calculation by Zagier's integral for triple product of Eisenstein series}

In the second calculation, we use the differential equation for $C_{2,1,1}$ in (\ref{C211Laplaceequation}) and $\Delta E_4 = 12 E_4 $ to recast $f(s)$ as follows, 
\bea
f(s) = - \int _\cM d \mu E_s \,  \Delta (\Delta -6) (\Delta-12) E_2^2
\eea
We cannot  integrate by parts over the domain $\cM$ because the Eisenstein series have polynomial growth at the cusp. Hence we shall consider the integral over a cut-off domain $\cM_L $, introduced in (\ref{ML}), for arbitrary $L \gg 1$, 
\bea
\label{fints}
f(s) = \lim _{L \to \infty} f_L(s)
\hskip 1in 
f_L(s) =  - \int _{\cM_L}  d \mu E_s \,  \Delta (\Delta^2 -18 \Delta +72) E_2^2
\eea
We may now evaluate this integral via Theorem \ref{Theorem76}. In fact, the current example is simple enough that we may rederive the result of (\ref{JmJrel}) explicitly as follows. Integrating (\ref{fints}) by parts while keeping track of the boundary contributions at $\tau_2=L$, we find, 
\bea
\label{fLs}
f_L(s)  =  
\int _0^1 d\tau_1 \Big [ \mF \,  \p_{\tau_2} E_s  - E_s \, \p_{\tau_2} \mF \Big ] _{\tau_2=L} 
- { \lambda _4(s) \over (s+1)(s-2)} 
 \int _{\cM_L}  d \mu \, E_s \,  E_2^2
\eea
where $\mF$ is given by,
\bea
\mF = \Big ( (\Delta -6)(\Delta -12) +s(s-1)(\Delta -18) +s^2(s-1)^2 \Big ) E_2^2
\eea
The power-behaved terms in the combination $\mF$ are  $\tau_2^4, \tau_2$ and $\tau_2^{-2}$, while $E_s$ has the power-behaved terms $\tau_2^s $ and $\tau_2^{1-s}$. Thus, the  power-behaved terms  in the first term in (\ref{fLs}) are given by a linear combination of $L^{s+3}, L^s, L^{s-3}, L^{4-s}, L^{1-s}$, and $L^{-2-s}$. 

\sm

The evaluation of the integral over $\cM_L$ in (\ref{fLs}) was given in \cite{Zagier}, and takes the form, 
\bea
\label{EsE2E2}
\int _{\cM_L} d \mu \, E_s \, E_2^2 & = & 
c_s \, c_2^2 \, { L^{s+3} \over s+3} + 2 \, c_s \, c_2 \, \tilde c_2 \, {L^s \over s} 
+ c_s \, \tilde c_s^2 \, { L^{s-3} \over s-3}
+ (s \to 1-s) 
\\ &&
+ { 32 \, \Gamma (s-3) (s+1)(s-2) \over (2 \pi)^{2s-1}} \, \zeta (s+3) \, \zeta(s-3) \, \zeta(s)^2 + \cO(e^{-2\pi L})
\no \eea
where the instruction to add the contribution from $s\to 1-s$ applies only to the first line. Omitting contributions to $f_L(s)$ which are exponentially suppressed as a function of $L$, using the Laurent polynomials of $E_2$ and $E_s$ to evaluate the power-behaved terms of $\mF$, we readily establish that the first integral in (\ref{fLs}) precisely cancels all the power-behaved terms in the contribution from the integral in (\ref{EsE2E2}). The remaining term on the second line of (\ref{EsE2E2}) equals the result (\ref{fLsa}) computed using Poincar\'e and Fourier series and Rankin-Selberg.

\newpage
\appendix

\section{Proof of Theorem \ref{thm10}}
\setcounter{equation}{0}
\label{sec:A}

To evaluate the Laurent polynomial $\cL_{u,v;w}$ and thereby prove Theorem \ref{thm10}, we decompose $\cL_{u,v;w}$ into contributions  $\cL^{(k)}_{u,v;w}$  from $\cC_{u,v;w}^{(k)}$ for $k=0,1,2,3,4$, so that,
\bea
\cL_{u,v;w} = \sum _{k=0}^4 \cL^{(k)}_{u,v;w}
\eea
To obtain $\cL^{(k)}_{u,v;w}$, we decompose the functions $\f_k$ and $\Phi_{a;b}$ entering $\cC^{(k)}_{u,v;w}$ with the help of the decomposition formulas (\ref{decomp1}) and (\ref{decomp2}) and retain for $\cL^{(k)}_{u,v;w}$ only the contributions proportional to $\hat \f_k$ and $\Phi_{a;b}^{(0)}$. Clearly we have,
\bea
\label{L0f}
\cL_{u,v;w} ^{(0)} = \cC^{(0)} _{u,v;w} = \ell_w (4 \pi \tau_2)^w
\eea
with $\ell_w$ given in (\ref{Lambda}), which produces the first term in (\ref{Laurent}).

\subsection{Calculation of $\cL_{u,v;w}^{(1)}+\cL_{u,v;w}^{(2)}$}

No contributions to $\cL^{(1)} _{u,v;w}+\cL^{(2)} _{u,v;w}$ arise from the terms involving $L_{u,v;k}$
since this function has only exponential contributions. The contribution from the function $K_{u,v;k}$ reduces to
\bea
K_{u,v;k}  = 2 \pi \mbinom{v+k-2}{v-1} \, \Phi _{u; v+k-1}^{(0)}  (2|m|\tau_2) \, \ep(m)^{u+v+k} + \cO(e^{ - 2 \pi \tau_2})
\eea
where $\Phi _{u; v+k-1}^{(0)}$ was given in (\ref{decomp2}). The sum over $m$ may be carried out in terms of even and odd $\zeta$-values, and we obtain, 
\bea
\cL^{(1)} _{u,v;w}+\cL^{(2)} _{u,v;w}&=&
  \sum _{k=1}^{w-u} \sum _{\alpha=0} ^{[u/2]} { 4 (-)^{v+\a}  
 \binom{2w-u-v-k-1}{w-u-k}  \binom{u+v+k-2-2\a}{v-1,k-1,u-2-2\a} \over (2 \pi)^{2 \a} (4 \pi \tau_2)^{w-1-2\a}  }
 \zeta (2 \alpha) \zeta(2w-2\a-1)
 \no\\ &&
 + \sum_{k=1}^{w-u}
 { (-)^v  \binom{2w-u-v-k-1}{w-u-k}\binom{v+k-2}{v-1}\binom{u+v+k-3}{u-1} \over (4 \pi \tau_2)^{w-2} }
 \zeta(2w-2)
+(u \leftrightarrow v)
\eea
Symmetrization  $(u \leftrightarrow v)$ applies to the entire expression. The sum over $k$ in the first line may be performed using a variant of Gauss' formula, valid for $c \in \NN$ and $a,b \in \CC$,
\bea
\sum_{k=1}^{c-1} { \Gamma (a-k) \Gamma (b+k) \over \Gamma (c-k) \Gamma (k)} =
{\Gamma (b+1) \Gamma (a+1-c) \Gamma (a+b) \over \Gamma (c-1) \Gamma (a+b+2-c) }
\eea
and we obtain, 
\bea
\cL^{(1)} _{u,v;w}+\cL^{(2)} _{u,v;w} &=&
4   \sum _{\alpha=0} ^{[u/2]} {  (-)^{v+\a}  \zeta (2 \alpha) \zeta(2w-2\a-1)
\binom{2w-2\alpha-2}{w-u-1}  \binom{u+v-1-2\a}{v-1} \over (2 \pi)^{2 \a} (4 \pi \tau_2)^{w-1-2\a}  }
 \no\\
&&+{(-)^v \over (4 \pi \tau_2)^{w-2}} \zeta(2w-2)  \mbinom{u+v-2}{u-1}\mbinom{2w-3}{w-u-1}
+(u \leftrightarrow v)
\eea
where the addition of $(u \leftrightarrow v)$ again applies to the entire preceding expression. The first term contributes to $\ell_{2k-w+1}$ in (\ref{L12f}) while the second contributes to $\ell_{2-w}$ in (\ref{L12g}).

\subsection{Calculation of $\cL_{u,v;w}^{(3)}$}

The contribution $\cL_{u,v;w}^{(3)}$ is given by,
\bea
\cL_{u,v;w}^{(3)}  =  
2\pi {\tau_2^w \over \pi ^w}   (-)^{u+v+w} \mbinom{u+v-2}{v-1}  \sum _{m\not=0} 
 \Phi _{2w-u-v; u+v-1}^{(0)}  (2 |m| \tau_2) 
\eea
Carrying out the sum over $m$ in terms of even and odd zeta values gives,
\bea
\label{L3f}
\cL_{u,v;w}^{(3)}  &=&  
 4 \mbinom{u+v-2}{v-1} 
\sum _{\alpha=0} ^{[(2w-u-v)/2]} 
{  (-)^{w+\a}  \zeta (2 \alpha)  \zeta (2w-2\a-1) \binom{2w-2\a-2}{u+v-2} \over (2 \pi)^{2\a} (4\pi  \tau_2)^{w-1-2\alpha} }
\no\\ &&
+(-)^w \mbinom{u+v-2}{v-1}\mbinom{2w-3}{u+v-2}{\zeta(2w-2) \over (4 \pi \tau_2)^{w-2}}
\eea
The first term contributes to $\ell_{2k-w+1}$ in (\ref{L12f}) while the second contributes to $\ell_{2-w}$ in (\ref{L12g}).

\subsection{Calculation of $\cL_{u,v;w}^{(4)}$}

The contribution $\cL^{(4)} _{u,v;w}$ arises from the fourth term in $\Omega$ of (\ref{Omega}), as well as from the remaining terms with $a=b=c=1$. Collecting these terms up to exponential contributions, 
\bea
\Omega_{u,v;w} & = & 
 \sum_{k=u+v+1}^{w+v} 
 { (-)^v\pi^2 \binom{2w-k-1}{w+v-k}   \binom{k-3}{u-1,v-1,k-u-v-1} 
 ( \ep(m_1) - \ep (m_2))( \ep(m_1) + \ep (m_3))   \over 
 (-m_2)^{k-2}  m_3^{2w-k} (2 \tau_2)^{2w-2}} 
\\ &&
+ \sum_{k=u+v+1}^{w+u} { (-)^u\pi^2 \binom{2w-k-1}{w+u-k} \binom{k-3}{u-1,v-1,k-u-v-1} 
 ( \ep(m_2) - \ep (m_1))( \ep(m_2) + \ep (m_3))  \over 
(- m_1)^{k-2} m_3^{2w-k}  (2 \tau_2)^{2w-2} } \quad
\no
\eea
Summing now over $m_1, m_2, m_3$ to obtain $\cL^{(4)}_{u,v;w}$, we use the symmetry under the simultaneous sign reversal of  $m_1, m_2, m_3$ to set $m_1>0$ upon including an overall factor of 2. In the first sum, we then need $m_3>0$ and set $-m_2=m_1+m_3$, while in the second sum $m_1$ and $m_2$ are interchanged. The sums over $m_1$ and $m_3$ in the first term and over $m_2$ and $m_3$ in the second term give the multiple zeta values defined in  (\ref{MZV}), 
 \bea
\label{L4f}
\cL^{(4)}_{u,v;w} = 
2 (-)^v \sum_{k=u+v+1}^{w+v} 
 { \binom{2w-k-1}{w+v-k}   \binom{k-3}{u-1,v-1,k-u-v-1} 
  \over   (4 \pi \tau_2)^{w-2}} \zeta (k-2, 2w-k) 
 + ( u \leftrightarrow v) 
\eea
This contribution accounts for the entire multiple zeta value part of $\ell_{2-w}$ in (\ref{L12g}). 

\newpage

\section{Proof of Proposition \ref{conj1}}
\setcounter{equation}{0}
\label{sec:B}

We start from the expression obtained for $\ell_{2-w}$ in Theorem \ref{thm10}, and  use a key result of \cite{DHoker:2017zhq} that expresses the odd-odd MZVs in terms of even-even MZVs plus a function $\cT$ which is a bilinear combination of odd zeta values with rational coefficients. For $N \geq 1$, we have,
\bea
\label{oddodd}
\zeta (2M-1,2N+1) & = & 
 - \cT(M,N) +  \zeta(2M-1) \zeta(2N+1)-\zeta(2M+2N)
\\ &&
+ \sum _{n=0}^{N-1} { \mathrm{E}_{2n+1}(0) \Gamma (2M+2n) \over (2n+1)! \Gamma(2M-1)} \zeta(2M+2n,2N-2n)
\no \\ &&
- \sum _{n=0}^{N-1}  { \mathrm{E}_{2n+1}(0)  \, \Gamma (2M+2n) \over (2n+1)! \, \Gamma (2M-1)}  
\Big ( \zeta(2M+2n) \zeta(2N-2n) - \zeta (2M+2N) \Big)
\no
\eea
Here $\mathrm{E}_n(0)$ is the Euler polynomial evaluated at zero argument. For $N \geq 1$, the function $\cT(M,N)$ evaluates as follows, 
\bea
\label{L2d}
\cT(M,N)  & = &  
\sum _{\alpha =1} ^{M+N-2}  \zeta (2\alpha+1 ) \zeta (2M+2N-2\alpha-1)
\sum _{n=0}^{2N-1} \half \mathrm{E}_n(0) \mbinom{2\alpha } { 2N-n} \mbinom{2M+n-2}{n}
\no \\ &&
- \sum _{\alpha =1} ^{M+N-1}  \zeta (2\alpha ) \zeta (2M+2N-2\alpha)
\sum _{n=0}^{2N-1} \half \mathrm{E}_n(0) \mbinom{2 \alpha -1} { 2N-n} \mbinom{2M+n-2}{n}
\quad
\eea
Eliminating all odd-odd MZVs from $\ell_{2-w}$ using MAPLE for $w \leq 18$ and the above formulas, we find that $\ell_{2-w}$ is given by a bilinear sum of odd zeta values, with integer coefficients $\lambda _k$,
\bea
\label{etakdef}
\ell_{2-w} = \sum _{k=1}^{w-3} \lambda _k \, \zeta (2k+1) \zeta (2w-2k-3)
\eea
This result proceeds via the cancellation of the MZVs $\zeta(u+v+k-2,2w-u-v-k)$ for each value of $k$ for which $u+v+k$ is even,  obtained by eliminating all odd-odd MZVs using (\ref{oddodd}). The cancellation takes place as if the even-even MZVs were independent variables, even though they do obey the Euler relation, 
\bea
\zeta(s,t)+\zeta(t,s) = \zeta (s) \zeta (t) -\zeta (s+t)
\eea
and are thus not necessarily independent. Below we shall exhibit the formulas required for the cancellation separately for $u+v$ even or odd.

\sm

For $u+v= 2z$ even, the contributions with even-even MZVs are obtained from the even $k$ contributions in (\ref{L12f}) and are given by, 
 \bea
X^e _{u,v;w} &=&
 2 (-)^v \sum_{k=1}^{{ [{ w-u \over 2}]}}  \mbinom{2w-u-v-2k-1}{w-u-2k}   \mbinom{u+v+2k-3}{u-1,v-1,2k-1} 
 \no \\ && \hskip 0.8in \times 
 \zeta (2z+2k-2, 2w-2z-2k) 
 + ( u \leftrightarrow v) \quad
\eea
while the contributions from eliminating odd-odd MZVs are given by,
\bea
X^\cT _{u,v;w} & = & 
2(-)^v \sum_{k=1}^{w-z-1} \sum _{\ell=1}^{2k-1} \theta \left ( 2 \left [ { w-u \over 2} \right ] -2k+\ell+1 \right )
\zeta(2z+2k-2,2w-2z-2k)
\no \\ && \qquad \times 
{\binom{2w-u-v-2k+\ell-1}{w-u-2k+\ell} \binom{u+v+2k-\ell-3}{u-1,v-1,2k-\ell-1} \mathrm{E}_\ell(0) \Gamma (u+v+2k-2) 
\over \Gamma (\ell+1) \, \Gamma (u+v+2k-\ell-2)}  + ( u \leftrightarrow v)
\eea 
Recognizing the contribution $X^e_{u,v;w}$ as being the same as the $\ell=0$ part of $X^\cT$ would be, we see that the cancellation term-by-term in $k$ of the coefficients of the even-even MZVs is equivalent to the cancellation of the following sums, 
\bea
X^e _{u,v;w} + X^\cT _{u,v;w}  = 2 \sum _{k=1} ^{w-z-1} Y_{u,v;w,k} { \Gamma (u+v+2k-2) \over \Gamma (u) \Gamma (v)} \zeta (2z+2k-2, 2w-2z-2k)
\eea
with $Y_{u,v;w,k}$ given by,
\bea
Y_{u,v;w,k} = (-)^v \sum_{\ell=0}^{2k-1} { \theta \left ( 2 \left [ { w-u \over 2} \right ] -2k+\ell+1 \right ) \mathrm{E}_\ell(0) 
\Gamma (2w-u-v-2k+\ell) \over \Gamma (w-u-2k+\ell+1) \Gamma (w-v) \Gamma (2k-\ell) \Gamma (\ell+1)} 
+ ( u \leftrightarrow v)
\quad
\eea
The constraint arising from the Heaviside $\theta$-function is automatically satisfied in view of the first $\Gamma$-function in the denominator. We have checked using MAPLE that $Y_{u,v;w,k}$  vanishes for all $w\leq 52$ with $1 \leq u \leq w-2$, $2\leq v \leq u$, and $1 \leq k \leq  w-z-1$.  An analytical proof of the vanishing of $Y_{u,v;w,l}$ is an open problem, as was the case with its analogue  in \cite{DHoker:2017zhq}. Hopefully,  such proof may be achieved using the algorithmic techniques of \cite{Blumlein:2018cms}.

\sm

It remains to analyze the case of $N=0$, for which we have 
\bea
\label{Neq0case}
\zeta(2M-1,1) =\half (2M-1) \zeta(2M) - \half \sum_{n=1}^{2M-3}\zeta(n+1)\zeta(2M-1-n)
\eea
This contains only bilinears in odd zeta values and terms which belong to $\pi^{2w-2} \QQ$. The sum of all terms proportional to  $\pi^{2w-2} \QQ$ in $\ell_{2-w}$  are known to cancel by the generalization of Theorem 1.2 of \cite{DHoker:2017zhq}, and they were verified to vanish by MAPLE.  The remaining terms are then all bilinear in odd zeta values and arise from the second term on the first line of (\ref{oddodd}), the first line of contributions to $\cT$ in (\ref{L2d}), and from the term on the second line of (\ref{Neq0case}). The coefficients $\lambda_k$ are then found to be given by the formula in Proposition \ref{conj1} for $u+v$ even. A corresponding calculation establishes Proposition \ref{conj1} for $u+v$ odd.

\section{Calculation of exponential contributions}
\setcounter{equation}{0}
\label{sec:C}

In this section, we shall evaluate the contributions which are exponentially suppressed near the cusp. The key formulas separating the exponential contributions in $\f_k$ and $\Phi_{a;b}$ are given in (\ref{decomp1}) and (\ref{decomp2}).  There are no exponential contributions arising from $\cC_{u,v;w}^{(0)}$. The exponential contribution arising from $\cC^{(i)}_{u,v;w}$ will be denoted by $\cE_{u,v;w}^{(i)}$ for $i=1,2,3,4$.

\subsection{Calculation of $\cE_{u,v;w}^{(1)}+\cE_{u,v;w}^{(2)} $}

The exponential contributions $\cE_{u,v;w}^{(1)}+\cE_{u,v;w}^{(2)} $ arising from the combination $\cC_{u,v;w}^{(1)}+\cC_{u,v;w}^{(2)} $ given in (\ref{C1C2})  take the following form, 
\bea
\cE_{u,v;w}^{(1)}+\cE_{u,v;w}^{(2)}  =(-)^v \sum_{k=1}^{w-u} \mbinom{2w-u-v-k-1}{w-u-k} \left(\cE_{u,v;w;k}^{(K)} 
+  \bar \cE_{v,u;w;k}^{(L)} \right) + \left(u \leftrightarrow v\right)^*
\eea
The contribution $\cE_{u,v;w;k}^{(K)}$ arises from the term involving $K_{u,v;k}$, which was evaluated in (\ref{Kuvw}), and may be decomposed as follows,
\bea
\cE_{u,v;w;k}^{(K)} =  \sum_{\ell=1}^v\mbinom{v+k-\ell-1}{v-\ell} \left[ \Xi_{u,v;k;\ell}^{(1)} + \Xi_{u,v;k;\ell}^{(2)} + \Xi_{u,v;k;\ell}^{(3)}  \right] + \left(v \leftrightarrow k\right)^*
\eea
where we have defined 
\bea
 \Xi_{u,v;k;\ell}^{(1)}&=& \delta_{\ell,1} \sum_{\b=1}^{v+k-1}{ \binom{u+v+k-\b-2}{u-1}  \over (4 \pi \tau_2)^{w- \b -1}} 
 \sum_{N=1}^\infty { N^{\b-1} \over \Gamma (\b)} q^N \bar q^N \sigma_{2-2w}(N)
\\
 \Xi_{u,v;k;\ell}^{(2)}&=& 4 \sum_{\a=0}^{\left[{u \over 2} \right]} {(-)^\a \zeta(2 \a) \binom{u+v+k-\ell-2\a-1}{u - 2 \a} \over (2\pi)^{2 \a} (4 \pi \tau_2)^{w-2\a-\ell}} \sum_{N=1}^\infty { N^{\ell-1} \over \Gamma (\ell)} \bar q^N \sigma_{1+ 2 \a - 2 w}(N)
\no\\
&\vphantom{.}&\hspace{0.8 in}+ {\binom{u+v+k-\ell-2}{u-1} \over (4 \pi \tau_2)^{w-\ell-1}}\sum_{N=1}^\infty {N^{\ell-1} \over \Gamma (\ell)} \bar q^N \sigma_{2-2w}(N)
\no\\
 \Xi_{u,v;k;\ell}^{(3)}&=& 
 \sum_{\b=1}^{v+k-\ell}  {2 \binom{u+v+k-\ell-\b-1}{u-1}\over (4 \pi \tau_2)^{w-\b-\ell} }  
 \sum_{N_1,N_2=1}^\infty {N_1^{\b-1}N_2^{\ell-1} \over \Gamma (\b) \Gamma (\ell)} q^{N_1}\bar q^{N_1+N_2} 
 \sigma_{2-2w}\Big ( \mathrm{gcd}(N_1,N_2) \Big )
\no
\eea
The contribution $\cE_{u,v;w;k}^{(L)}$ arises from the  term involving  $L_{u,v;k}$ which was evaluated in (\ref{Luvw}), and is given as follows, 
\bea
(-)^k\cE_{u,v;w;k}^{(L)}  & = &
 \sum _{j=\left [ { u+k+1 \over 2} \right ]}^{\left [ { u+v+k-1 \over 2} \right ]} {4 \,  (-)^{v+j} \binom{2j-u-1}{k-1} \zeta (2j) \over   (4 \pi \tau_2)^{w-u-v-k} (2 \pi)^{2j} } \sum _{N=1} ^\infty {N^{u+v+k-2j-1} \sigma _{1+2j-2w}(N) \over \Gamma (u+v+k-2j)}  \bar q^N 
 \no \\ &&
\hskip 0.2 in+(-)^{u+k} ( v \leftrightarrow k) \quad
\eea

\subsection{Calculation of $\cE_{u,v;w}^{(3)}$}

The exponential contribution $\cE^{(3)}_{u,v;w}$ arising from $\cC_{u,v;w}^{(3)}$ given in (\ref{C3})  takes the form, 
\bea
\cE^{(3)}_{u,v;w}=(-)^w \sum_{k=1}^u  \mbinom{u+v-k-1}{u-k} \left[  \Pi_{u,v;k}^{(1)} +\Pi_{u,v;k}^{(2)} +\Pi_{u,v;k}^{(3)}  \right] + \left(u \leftrightarrow v \right)^*
\eea
where we have defined 
\bea 
\Pi_{u,v;k}^{(1)} &=& \delta_{k,1} \sum_{\b=1}^{u+v-1} {\binom{2w-\b-2}{u+v-\b-1} \over  (4 \pi \tau_2)^{w-\b-1}}\sum_{N=1}^\infty { N^{\b-1} \over \Gamma(\b) } q^N \bar q^N \sigma_{2-2w}(N)
\\
\Pi_{u,v;k}^{(2)} &=&  \sum_{\a=0}^{\left[{2w-u-v \over 2} \right]} { 4 (-)^\a \zeta(2 \a) \binom{2w-2\a-k-1}{u+v-k-1} \over  (2 \pi)^{2\a} (4 \pi \tau_2)^{w-k-2\a}} \sum_{N=1}^\infty { N^{k-1} \over \Gamma (k)} q^N \sigma_{1+2 \a - 2w}(N) 
\no\\ &\vphantom{.}&\hspace{0.8 in} 
+ {\binom{2w-k-2}{2w-u-v-1} \over  (4 \pi \tau_2)^{w-k-1}}\sum_{N=1}^\infty {N^{k-1} \over \Gamma (k)} q^N \sigma_{2-2w}(N)
\no\\
\Pi_{u,v;k}^{(3)} &=& 
\sum_{\b=1}^{u+v-k} {2 \binom{2w-k-\b-1}{u+v-k-\b} \over   (4 \pi \tau_2)^{w-k-\b}} 
\sum_{N_1,N_2=1}^\infty { N_1^{\b-1}N_2^{k-1} \over \Gamma(k) \Gamma(\b)} q^{N_1+N_2} \bar q^{N_1} \sigma_{2-2w} \Big ( \mathrm{gcd}(N_1,N_2) \Big )
\no
\eea

\subsection{Calculation of $\cE_{u,v;w}^{(4)}$}

Recall that $\cC_{u,v;w}^{(4)}$ is given by, 
\bea
\cC_{u,v;w}^{(4)} = \sum _{m_r \not=0}  { \tau_2^w \over \pi^w} \, \delta _{m_1+m_2+m_3,0} \,
\Omega_{u,v;w} \Big (m_1\tau, m_2 \bar \tau, m_3\tau, m_3 \bar \tau \Big ) 
\eea
where $\Omega$ was given in (\ref{Omega}).  The sum over the $m$-variables is carried out by partitioning the contributions according to the decomposition of the $\f$-functions. The terms with two Kronecker $\delta$-symbols have already been counted towards the Laurent series. 

\sm

The contributions with a single Kronecker $\delta$ require the following sums,  
\begin{align}
\label{Sig1}
\Sigma ^{(2)}_1 (B,C; a,b) & =  
\sum _{m_r \not=0} { (-)^B \delta _{m,0} \over 2m_2^B m_3^C } 
 \Big ( \delta _{a,1} \ep(m_1) \ep(m_2)^b  \hat \f_b (\bar q ^{|m_2|}) 
 + \delta _{b,1} \ep(m_2) \ep(m_1)^a \hat \f_a(q^{|m_1|}) \Big )
\no \\
\Sigma ^{(3)}_1 (B,C;b,c) & =  
 \sum _{m_r \not=0} { (-)^B \delta _{m,0} \over 2m_2^B m_3^C } 
\Big ( \delta_{b,1} \ep(m_2) \ep(m_3)^c \hat \f_c (q ^{|m_3|})  + \delta _{c,1} \ep(m_3) \ep(m_2)^b \hat \f_b(\bar q^{|m_2|})  \Big ) 
\no \\
\Sigma ^{(4)}_1 (B,C; c,a) & =  
\sum _{m_r \not=0} { (-)^B \delta _{m,0} \over 2m_2^B m_3^C } 
\Big (  \delta _{c,1} \ep(m_3) \ep(m_1)^a  \hat \f_a(q^{|m_1|}) +\delta _{a,1} \ep(m_1) \ep(m_3)^c \hat \f_c (q ^{|m_3|})  \Big ) 
\qquad
\end{align}
These contributions are all harmonic in $q$. Each function vanishes by $m_r \to -m_r$ symmetry of the summation unless the sum of all four of its  arguments is even, so for example in $\Sigma ^{(2)}_1(B,C;a,b)$ one must have $B+C+a+b$ even. 

\sm

The contributions without Kronecker $\delta$ require the following sums, 
\bea
\label{Sig2}
\Sigma ^{(2)}_0 (B,C; a,b) & = & \sum _{m_r \not=0} 
{ \delta _{m,0} \over (-m_2)^B m_3^C } \, \ep(m_1)^a \ep(m_2)^b 
\hat  \f_a(q^{|m_1|}) \hat \f_b (\bar q ^{|m_2|}) 
\no \\
\Sigma ^{(3)}_0 (B,C;b,c) & = & \sum _{m_r\not=0} 
{ \delta _{m,0} \over (-m_2)^B m_3^C } \,  \ep(m_2)^b \ep(m_3)^c
\hat \f_b(\bar q^{|m_2|}) \hat \f_c (q ^{|m_3|})
\no \\
\Sigma ^{(4)}_0 (B,C; c,a) & = & \sum _{m_r\not=0} 
{ \delta _{m,0} \over (-m_2)^B m_3^C } \,  \ep(m_3)^c  \ep(m_1)^a
 \hat \f_c (q ^{|m_3|}) \hat \f_a(q^{|m_1|})
\eea
In terms of these functions, we have,
\bea
\cE_{u,v;w}^{(4)} =   
   \sum_{k=u+v+1}^{w+v}  \mbinom{2w-k-1}{w+v-k}  
 \sum_{i=2,3,4} \, \sum_{j=0,1} (-)^v \cR_{u,v;w,k} ^{(i,j)}
+ (u \leftrightarrow v)^* 
\eea
where we have the following expressions for $j=0,1$,
\bea
\cR_{u,v;w,k } ^{(2,j)} & = & 
 \sum _{a=1}^u \sum _{b=1}^v 
{ (-)^{b} \, \binom{k-a-b-1}{u-a,v-b,k-u-v-1} 
 \over  (4 \pi \tau_2)^{w-a-b} }  \,
 \Sigma ^{(2)}_j  (k-a-b, 2w-k; a,b)
\no \\ 
\cR_{u,v;w,k }^{(3,j)}  & = &
\sum _{b=1}^v \sum _{c=1}^{k-u-v} 
{ (-)^{b} \,   \binom{k-b-c-1}{u-1,v-b,k-u-v-c} 
 \over  (4 \pi \tau_2)^{w-b-c}  } \,
 \Sigma ^{(3)}_j (k-b-c, 2w-k; b,c)
\no \\ 
\cR_{u,v;w,k }^{(4,j)} & = &
 \sum _{c=1}^{k-u-v} \sum _{a=1}^u 
{   \binom{k-c-a-1}{u-a,v-1,k-u-v-c}  
  \over  ( 4 \pi \tau_2)^{w-c-a} } \,
  \Sigma ^{(4)}_j (k-c-a, 2w-k; a,c)
\quad
\eea

\subsection{Harmonic sums and single-valued zeta-values}
\label{sec:C4}

The functions $\Sigma ^{(i)}_1$ for $i=2,3,4$ may be evaluated in terms of harmonic sums, sums over harmonic sums, and single-valued zeta-values, which we shall briefly discuss first.  The basic harmonic sum $H_n(m)$ satisfies the following relations,
\bea
H_n(m) = \sum_{\ell =1}^m { 1 \over \ell^n} \hskip 0.8in H_n(m) = H_n(m-1) + { 1 \over m^n}
\eea
as well as $ H_n(0)=0$ and $H_n (\infty) = \zeta (n)$. The double zeta-value $\zeta(k,n)$ may be expressed in terms of the following sum over harmonic sums, 
\bea
\sum_{m=1}^\infty { H_{n} (m-1) \over m^k} = \zeta (k,n)
\eea
The single-valued $\zeta$-values were introduced in \cite{Brown:2013gia}, and may be defined as follows,
\bea
\label{zsv}
\zsv(n) = \left \{ \begin{matrix} 2 \zeta (n) & n \in 2 \NN +1 \cr 0 & \hbox{otherwise} \cr \end{matrix} \right.
\eea
As we shall see below, they arise very naturally in the exponential part of $\cC_{u,v;w}$. Though we will not need them for our current work, we note in passing that extensions of the single-valued projection to MZVs of higher depth exist as well \cite{Brown:2013gia}. 

\sm

We may apply these definitions to evaluating the ubiquitous sum defined by, 
\bea
\label{Sabm}
S_{a,b}(m) = \sum _{n \not =0,-m} { \ep (n) \over n ^a (m+n)^b} 
\eea
Partial fraction decomposition in $n$ and summation over $n$, using the sums,
\bea
\sum _{n \not= 0, -m} { \ep(n) \over n^j} & = & \zsv(j) + { (-)^j \ep(m) \over m^j}  \hskip 0.6in j \geq 2
\no \\
\sum _{n \not= 0, -m} \left ( { \ep (n) \over m+n} - { \ep (n) \over n} \right ) 
& = & - 2 H_1(m-1)
\eea
gives the following result,
\bea
\label{Sab}
(-)^a S_{a,b}(m)  & = &
\sum _{j=3} ^a { (-)^j \binom{a+b-j-1}{b-1} \zsv(j) \over m^{a+b-j} } 
+ \sum _{j=3} ^b {  \binom{a+b-j-1}{a-1} \zsv(j) \over m^{a+b-j} } 
\no \\ &&
+ {  \binom{a+b}{a} \over m^{a+b}} 
- 2  \sum_{j=1}^b { \binom{a+b-j-1}{a-1} H_j(m) \over m^{a+b-j}}
\eea

\subsection{Calculation of $\Sigma _1^{(i)}$ for $i=2,3,4$}
 
To compute the sum of the term proportional to $\delta_{a,1}$ in the summand of $\Sigma _1^{(2)}(B,C;a,b)$, given in (\ref{Sig1}), we use $m_r \to -m_r$ symmetry to choose $m=m_2>0$, so that the term becomes, 
\bea
 \delta _{a,1} (-)^{B+C} \sum _{m_2 =1}^\infty \, \sum_{m_1\not= 0, -m_2} { \ep(m_1)   \hat \f_b (\bar q ^{m_2})  \over m_2^B (m_1+m_2)^C } 
\eea
The sum over $m_1$ may be carried out in terms of the function $S_{0,C}(m_2)$ defined in (\ref{Sab}),
and upon setting $N=pm$ and eliminating the sum over $p$ in favor of a sum over $N$, we find, 
\bea
 \delta _{a,1}   (-)^{B+C} 
\sum _{N =1}^\infty  { N^{b-1} \over \Gamma (b)}   \bar q^N \sum_{0< m|N } { S_{0,C}(m)  \over m^{B+b-1} } 
\eea
For given $N$, the sum  is over a finite number of values of $m$. For each value of $m$, the contribution is a sum of a rational number and a rational number times an odd zeta-value.   To compute the sum of the term proportional to $\delta_{b,1}$ in the summand of $\Sigma _1^{(2)}(B,C;a,b)$, we use $m_r \to -m_r$ symmetry to choose $m=m_1>0$, so that this term becomes,
\bea 
 \delta _{b,1} (-)^{B+C}  \sum _{m=1}^\infty \sum _{ m_2\not=0,-m}  {\ep(m_2)  \hat \f_a (q^{m}) \over m_2^B (m+m_2)^C }   
\eea
The sum over $m_2$ may be carried out in terms of $S_{B,C}(m)$ and putting all together, we find,
\bea
\label{S12}
\Sigma _1^{(2)}(B,C;a,b) & = & 
 \delta _{a,1} (-)^{B+C} 
\sum _{N =1}^\infty  {N^{b-1}  \over \Gamma (b)}  \bar q^N \sum_{0< m|N } { S_{0,C}(m)  \over m^{B+b-1} } 
\no \\ && +
\delta _{b,1}   (-)^{B+C}  \sum_{N=1}^\infty  { N^{a-1} \over \Gamma (a)} q^N \sum _{0<m|N} {S_{B,C}(m)   \over m^{a-1}}   
\eea
The computation of $\Sigma _1^{(3)}(B,C;b,c)$ gives, 
\bea
\label{S13}
\Sigma _1^{(3)}(B,C;b,c) & = &
\delta _{b,1}    \sum_{N=1}^\infty { N^{c-1} \over \Gamma (c)} q^N 
\Big (  (-)^B  \zsv(B) \sigma_{1-C-c}(N)  +  \sigma_{1-B-C-c}(N)  \Big )
\\ && + 
 \delta _{c,1}   (-)^B \sum _{N=1}^\infty {N^{b-1} \over \Gamma (b)} \bar q^N 
 \Big ( \zsv(C) \sigma_{1-B-b} (N)  +  (-)^C \sigma_{1-B-C-b}(N)  \Big )
\no
\eea
while the computation of $\Sigma _1^{(4)}(B,C;c,a)$ gives,  
  \bea
 \label{S14}
 \Sigma _1^{(4)}(B,C;c,a) = \sum_{N=1}^\infty q^N \left ( 
 \delta _{a,1}   { N^{c-1} \over \Gamma (c)}   \sum _{0<m|N}   { S_{0,B}(m) \over m^{C+c-1} } 
+  \delta _{c,1}  {N^{a-1} \over \Gamma (a)}   \sum _{0<m|N}   { S_{C,B}(m) \over m^{a-1} }
\right )
 \eea

\subsection{Calculation of $\Sigma _0^{(i)}$ for $i=2,3,4$}

Calculating the functions $\Sigma _0^{(i)}$ for $i=2,3,4$ given in (\ref{Sig2})  involves the generalized divisor functions $V_{A,B,C}(M,N)$ introduced in subsection \ref{gendiv} and which we recall here for convenience, 
\bea
\label{VABC2}
V_{A,B,C}(M, N) =  \sum _{{m| M \atop m \not=0}} \sum_{{ n| N \atop n \not=0, -m}}
{  \ep(m) \ep (n)  \over m^{A-1} n^{B-1} (m+n)^C   }  
\eea
Recall that $V_{A,B,C}(M,N)=0$ for odd $A+B+C$. To account for this property in a systematic manner, we introduced the function, 
\bea
\mI_{n} = \half ( 1 + (-)^n) 
\eea
which vanishes for $n$ odd and equals 1 for $n$ even. A more explicit formula for $V_{A,B,C}(M,N)$ is given in (\ref{expandedVABC}).

\subsubsection{Calculation of $\Sigma _0^{(2)}(B,C;a.b)$}

To calculate  $\Sigma _0^{(2)}(B,C;a.b)$, we choose independent summation variables $m_1, m_2$ and expand the $\hat \f$-functions using (\ref{decomp2}), 
\bea
\Sigma _0^{(2)}(B,C;a.b)=
\sum_{p_1,p_2=1}^\infty 
\sum _{m_2 \not=0}  \sum_{m_1 \not=0, -m_2} 
{(-)^{B+C}  \, \ep(m_1)^a \ep(m_2)^b \over m_2^B (m_1+m_2)^C }  
{ p_1^{a-1} p_2^{b-1} \over \Gamma (a) \Gamma (b)}  q^{p_1|m_1|} \bar q^{p_2 |m_2| } 
\eea
Changing variables from $(p_1, p_2)$ to $(N_1, N_2)$ with $N_1=p_1|m_1|, N_2 = p_2 |m_2|$, and we find, 
\bea
\label{S02}
\Sigma ^{(2)}_0 (B,C; a,b) = 
(-)^{B+C} \sum_{N_1,N_2=1}^\infty    {  N_1^{a-1} N_2 ^{b-1} \over \Gamma (a) \Gamma (b)} q^{N_1}   \bar q^{N_2} 
 V_{a,B+b,C}(N_1,N_2)
\eea

\subsubsection{Calculation of $\Sigma _0^{(3)}(B,C;b,c)$}

Choosing independent summation variables $m_2,m_3$ and solving for $m_1=-m_2-m_3$, 
\bea
\Sigma ^{(3)}_0 (B,C;b,c)= (-)^B T_{B,b}(\bar q) T_{C,c}(q) 
- 2 (-)^b \sum _{m_2 =1}^\infty  { \hat \f_b(\bar q^{m_2}) \hat \f_c (q ^{m_2}) \over m_2^{B+C}  } 
\eea
where $T_{B,b}(q)$ is given by,  
\bea
T_{B,b} (q) =  \sum _{m_2 \not=0} { \ep(m_2)^b \hat \f_b(q^{|m_2|}) \over m_2^B}
\eea
This quantity vanishes unless $B+b$ is even, in which case it may be simplified as follows,
\bea
T_{B,b} (q) =  2\, \mI_{b+B}  \sum_{N=1}^\infty  {N^{b-1} \over \Gamma (b)} q^N \sigma_{1-B-b}(N)
\eea
The last sum may be evaluated by expanding $\hat \f_k$ using (\ref{decomp2}) and 
changing summation  variables from $p_1, p_2$ to to $N_1=p_2m_2$ and $N_2 = p_1 m_2$. Putting all together, we find, 
\bea
\label{S03}
\Sigma ^{(3)}_0 (B,C;b,c) & = & 
\sum_{N_1,N_2 =1}^ \infty  {N_1^{c-1} N_2 ^{b-1}  \over \Gamma (b) \Gamma (c)} q^{N_1} \bar q^{N_2 }   
\Big (4(-)^B \mI_{b+B} \mI_{c+C}  \sigma_{1-B-b}(N_2) \sigma _{1-C-c}(N_1) 
\no \\ && \hskip 1.5in 
- 2(-)^b \sigma_{2-B-C-b-c}(\gcd(N_1,N_2)) \Big )
\eea

\subsubsection{Calculation of $\Sigma _0^{(4)}(B,C;c,a)$}

Choosing independent variables $m_1,m_3$ and expanding the $\hat \f$-functions,
\bea
\Sigma _0^{(4)}(B,C;c,a) = 
\sum_{p_1,p_2=1}^\infty  \sum _{m_1\not=0} \sum_{m_3\not=0,-m_1}
{ p_1^{a-1} p_2^{c-1} \over \Gamma (a) \Gamma (c)}  { \ep(m_1)^a \ep(m_3)^c \over (m_1+m_3)^B m_3^C } \, 
q^{p_1|m_1|+p_2|m_3|}
\eea
Change variables from $p_1, p_2$ to $N_1=p_1|m_1|$ and $N_2=p_2|m_3|$, 
\bea
\label{S04}
\Sigma ^{(4)}_0 (B,C; c,a) = 
\sum_{N_1,N_2=1}^\infty   {N_1^{a-1} N_2^{c-1} \over \Gamma (a) \Gamma (c)}  \, q^{N_1+N_2}
V_{a,C+c,B}(N_1,N_2)
\eea
where $V_{A,B,C}$ was defined in (\ref{VABC2}), and takes  rational values.

\subsection{Calculation of $\cR^{(i,1)}_{u,v;w,k}$ for $i=2,3,4$}

For $i=2$ we use $\Sigma _1^{(2)}(B,C;a,b)$ given by (\ref{S12}) with  $B=k-a-b$,  $C=2w-k$ and thus $B+C=2w-a-b$. Substituting this result in gives, 
\bea
\cR_{u,v;w,k } ^{(2,1)} & =  &
-  \sum _{b=1}^v 
{ \binom{k-b-2}{u-1,v-b,k-u-v-1} \over  (4 \pi \tau_2)^{w-b-1} }  
\sum _{N =1}^\infty    { N^{b-1} \over \Gamma (b)}   \bar q^N \sum_{0< m|N } { S_{0,2w-k}(m)  \over m^{k-2} } 
 \\ && 
+ \sum _{a=1}^u 
{ (-)^{a} \, \binom{k-a-2}{u-a,v-1,k-u-v-1} \over  (4 \pi \tau_2)^{w-a-1} } 
  \sum_{N=1}^\infty   {N^{a-1} \over \Gamma (a)}   q^N \sum _{0<m|N} {S_{k-a-1,2w-k}(m)   \over m^{a-1}}   
\no
\eea
For $i=3$ we use $\Sigma _1^{(3)}(B,C;b,c)$ given by (\ref{S12}) with  $B=k-b-c$,  $C=2w-k$ and thus $B+C=2w-b-c$. Substituting this result in gives, 
\bea
\cR_{u,v;w,k }^{(3,1)}  
& = &
 \sum _{c=1}^{k-u-v} {   \binom{k-c-2}{u-1,v-1,k-u-v-c}  \over  (4 \pi \tau_2)^{w-c-1 } } 
  \sum_{N=1}^\infty  {N^{c-1} \over \Gamma (c)}    q^N 
\Big ( \zsv(k-c-1) \sigma_{1-2w+k-c}(N) \, - \sigma_{2-2w}(N)  \Big )
\no \\ &&
+ \sum _{b=1}^v  {  \binom{k-b-2}{u-1,v-b,k-u-v-1}  \over  (4 \pi \tau_2)^{w-b-1}  } 
 \sum _{N=1}^\infty  {N^{b-1} \over \Gamma (b)}   \bar q^N 
 \Big (  \zsv(2w-k) \sigma_{2-k} (N)  -  \sigma_{2-2w}(N)  \Big )
\eea
For  $i=4$ we use $\Sigma _1^{(4)}(B,C;c,a)$ given by (\ref{S12}) with  $B=k-c-a$,  $C=2w-k$ and thus $B+C=2w-c-a$. Substituting this result in gives,
 \bea
\cR_{u,v;w,k }^{(4,1)} & = &
 \sum _{c=1}^{k-u-v}  { \binom{k-c-2}{u-1,v-1,k-u-v-c}   \over  ( 4 \pi \tau_2)^{w-c-1} } 
  \sum_{N=1}^\infty   { N^{c-1} \over \Gamma (c)}  q^N  \sum _{0<m|N}   { S_{0,k-c-1}(m) \over m^{2w-k+c-1} } 
\no \\  &&
+  \sum _{a=1}^u { \binom{k-a-2}{u-a,v-1,k-u-v-1}   \over  ( 4 \pi \tau_2)^{w-a-1} } 
  \sum_{N=1}^\infty  { N^{a-1} \over \Gamma (a)}  q^N  \sum _{0<m|N}   { S_{2w-k,k-a-1}(m) \over m^{a-1} }
\eea

\subsection{Calculation of $\cR^{(i,0)}_{u,v;w,k}$ for $i=2,3,4$}

For $i=2$ we use $\Sigma _1^{(2)}(B,C;a,b)$ given by (\ref{S12}) with  $B=k-a-b$,  $C=2w-k$ and thus $B+C=2w-a-b$. Substituting this result in gives, 
\bea
\cR_{u,v;w,k } ^{(2,0)} = 
 \sum _{a=1}^u \sum _{b=1}^v 
{ (-)^{a} \, \binom{k-a-b-1}{u-a,v-b,k-u-v-1}  \over  (4 \pi \tau_2)^{w-a-b} } 
\sum_{N_1,N_2=1}^\infty    {  N_1^{a-1} N_2 ^{b-1} \over \Gamma (a) \Gamma (b)} q^{N_1}   \bar q^{N_2} 
V_{a,k-a,2w-k}(N_1, N_2)
\qquad 
\eea
For $i=3$ we use $\Sigma _1^{(3)}(B,C;b,c)$ given by (\ref{S12}) with  $B=k-b-c$,  $C=2w-k$ and thus $B+C=2w-b-c$. Substituting this result in gives, 
\bea
\cR_{u,v;w,k }^{(3,0)}  & = &
\sum _{b=1}^v \sum _{c=1}^{k-u-v} 
{   \binom{k-b-c-1}{u-1,v-b,k-u-v-c}  \over  (4 \pi \tau_2)^{w-b-c}  } 
\sum_{N_1,N_2 =1}^ \infty  {N_1^{c-1} N_2 ^{b-1}  \over \Gamma (b) \Gamma (c)} q^{N_1} \bar q^{N_2 }   
 \\ && \quad \times
\Big (4 \, \mI_{k-c} \, \sigma_{1-k+c}(N_2) \sigma _{1-2w+k-c}(N_1) 
-2 \sigma_{2-2w}(\gcd(N_1,N_2)) \Big )
\no
\eea
For $i=4$ we use $\Sigma _1^{(4)}(B,C;c,a)$ given by (\ref{S12}) with  $B=k-c-a$,  $C=2w-k$ and thus $B+C=2w-c-a$. Substituting this result in gives,
\bea
\cR_{u,v;w,k }^{(4,0)} =
 \sum _{c=1}^{k-u-v} \sum _{a=1}^u { \binom{k-c-a-1}{u-a,v-1,k-u-v-c}   \over  ( 4 \pi \tau_2)^{w-c-a} } 
\sum_{N_1,N_2=1}^\infty   {N_1^{a-1} N_2^{c-1} \over \Gamma (a) \Gamma (c)}  \, q^{N_1+N_2}
V_{a,2w-k+c, k-c-a}(N_1, N_2)
\quad\no\\
\eea

\section{Fourier coefficients of exponential contributions}
\setcounter{equation}{0}
\label{sec:CC}

In this appendix, we shall use the results of Section \ref{sec:4} and Appendix \ref{sec:C} to compute explicitly the exponential contributions to the two-loop modular graph functions $\cC_{u,v;w}$ considered in this paper. We begin with the calculation of the exponential contributions to the constant Fourier mode in subsection \ref{sec:D1} and then proceed to evaluating the non-constant Fourier modes in subsections \ref{sec:D2} and \ref{sec:D3}.

\subsection{Calculation of $\cQ_{u,v;w}^{(N)}(\tau_2)$}
\label{sec:D1}

We begin by obtaining the exponential part of the constant Fourier mode $\cQ_{u,v;w}^{(N)}(\tau_2)$. There is no contribution to $\cQ_{u,v;w}^{(N)}(\tau_2)$ from $\cC^{(0)}_{u,v;w}$. The contribution from $\cC^{(1)} _{u,v;w}+ \cC^{(2)} _{u,v;w}$ is,
\bea
&&
(-)^v \sum_{k=u+v+1}^{w+v}  \sum_{\b=1}^{k-u-1}{ 2 \binom{2w-k-1}{w+v-k} \binom{k-\b-2}{u-1} \binom{k-u-2}{v-1} \over \Gamma(\b) (4 \pi \tau_2)^{w- \b -1}} \sum_{N=1}^\infty N^{\b-1}q^N \bar q^N \sigma_{2-2w}(N)  
\no \\
&&
+ (-)^u \sum_{k=u+v+1}^{w+u}  \sum_{\b=1}^{k-v-1}{ 2 \binom{2w-k-1}{w+u-k} \binom{k-\b-2}{v-1} \binom{k-v-2}{u-1} \over \Gamma(\b) (4 \pi \tau_2)^{w- \b -1}} \sum_{N=1}^\infty N^{\b-1}q^N \bar q^N \sigma_{2-2w}(N)  
\eea
while the contribution from $\cC^{(3)}_{u,v;w}$ is given by, 
\bea
2 (-)^w  \sum_{\b=1}^{u+v-1} {\binom{2w-\b-2}{u+v-\b-1}  \binom{u+v-2}{u-1}   \over \Gamma(\b) (4 \pi \tau_2)^{w-\b-1}}\sum_{N=1}^\infty N^{\b-1} q^N \bar q^N \sigma_{2-2w}(N)  
\eea
The contribution from $\cC^{(4)}_{u,v;w}$ arises only from $\cR_{u,v;w,k}^{(i,0)}$ for $i=2,3$ and is given by, 
\bea
&& \sum_{k=u+v+1}^{w+v} \sum _{a=1}^u \sum _{b=1}^v 
{ (-)^{a+v} \, \binom{k-a-b-1}{u-a,v-b,k-u-v-1}  \binom{2w-k-1}{w+v-k}   \over  (4 \pi \tau_2)^{w-a-b} } 
\sum_{N=1}^\infty    {  N^{a+b-2}  \over \Gamma (a) \Gamma (b)} q^{N}   \bar q^{N} 
V_{a,k-a,2w-k}(N,N)
\no \\ && 
+ \sum_{k=u+v+1}^{w+v}  \sum _{b=1}^v \sum _{c=1}^{k-u-v} 
{  (-)^v \binom{k-b-c-1}{u-1,v-b,k-u-v-c} \binom{2w-k-1}{w+v-k}  \over  (4 \pi \tau_2)^{w-b-c}  } 
\sum_{N =1}^ \infty  {N^{b+c-2}   \over \Gamma (b) \Gamma (c)} q^{N} \bar q^{N}   
\no \\ && \hskip 1in  \times
\Big (4 \, \mI_{k-c} \, \sigma_{1-k+c}(N) \sigma _{1-2w+k-c}(N)  -2   \sigma_{2-2w}(N) \Big )
\no \\ && 
+ (u \leftrightarrow v)
\eea
Thus $\cQ_{u,v;w}^{(N)}(\tau_2)$, defined in (\ref{Four}), is given by, 
\bea
\label{Qlongform}
\cQ_{u,v;w}^{(N)}(\tau_2) & = &
 \sigma_{2-2w}(N) \sum_{k=1+u+v}^{w+v}  \sum_{\b=1}^{v+k-1} N^{\b-1}  { 2 (-)^v \binom{2w-k-1}{w+v-k} \binom{k-\b-2}{u-1} \binom{k-u-2}{v-1} \over \Gamma(\b) (4 \pi \tau_2)^{w- \b -1}}    
 \no \\ &&
+   \sigma_{2-2w}(N)  \sum_{\b=1}^{u+v-1} N^{\b-1}   {(-)^w \binom{2w-\b-2}{u+v-\b-1}  \binom{u+v-2}{u-1}   \over \Gamma(\b) (4 \pi \tau_2)^{w-\b-1}} 
\no \\ &&
+\sum_{k=1+u+v}^{w+v} \sum _{a=1}^u \sum _{b=1}^v 
{ (-)^{a+v} \, \binom{k-a-b-1}{u-a,v-b,k-u-v-1}  \binom{2w-k-1}{w+v-k}   \over  (4 \pi \tau_2)^{w-a-b} } 
   {  N^{a+b-2}  \over \Gamma (a) \Gamma (b)} 
V_{a,k-a,2w-k}(N,N)
\no \\ && + \sum_{k=1+u+v}^{w+v}  \sum _{b=1}^v \sum _{c=1}^k 
{   \binom{k-b-c-1}{u-1,v-b,k-u-v-c} \binom{2w-k-1}{w+v-k}  \over  (4 \pi \tau_2)^{w-b-c}  } 
 {N^{b+c-2}   \over \Gamma (b) \Gamma (c)}   
\no \\ && \qquad \times(-)^v
\Big (4 \, \mI_{k-c} \,\sigma_{1-k+c}(N) \sigma _{1-2w+k-c}(N) 
-2   \sigma_{2-2w}(N) \Big )
\no \\ && 
+ (u \leftrightarrow v)
\eea
We may now simplify this as follows. In the first term of (\ref{Qlongform}), the summations over $k,\beta$ may be interchanged and, introducing the function
\bea
J_{u,v;w}^{(1)}  (\beta,N) &=& 2 (-)^v \sigma _{2-2w}(N)  
\sum_{k=1}^{w+v} \mbinom{2w-k-1}{w+v-k} \mbinom{k-\b-2}{u-1} \mbinom{k-u-2}{v-1}
 \\ && \qquad
\times[\theta (\beta -v) \theta(k-\b-u-1) + \theta(v-\b-1) \theta(k-u-v-1)]
\no
\eea
the first line becomes, 
\bea
 \sum_{\b=1}^{w-u+v-1}   {  N^{\b-1}J_{u,v;w} ^{(1)} (\beta,N) \over \Gamma(\b) (4 \pi \tau_2)^{w- \b -1}}    
 \eea
 Likewise in the third term of (\ref{Qlongform}), we may change variables from $b$ to $\beta =a+b-1$, and introduce the  function, 
\bea
J_{u,v;w}^{(2)} (\beta,N)  & = &  
 \sum_{k=u+v+1}^{w+v} \,  \sum_{a=\max(1,\b-v+1)}^{\min(\beta,u)}
  \mbinom{k-\beta-2}{u-a,v+a-\beta-1,k-u-v-1}  
 \no \\ && \qquad \times 
 (-)^{a+v}  \mbinom{\beta-1}{\beta -a} \mbinom{2w-k-1}{w+v-k}   V_{a,k-a,2w-k}(N,N)
\eea
The third line then becomes, 
\bea
\sum_{\beta=1} ^{u+v-1} {  N^{\beta -1} J^{(2)}_{u,v;w}(\beta, N) \over \Gamma (\beta) (4 \pi \tau_2)^{w-\beta-1}} 
\eea
In the fourth term we change variables from $b$ to $\beta=b+c-1$, and introduce the function, 
{\small \bea
J^{(3)}_{u,v;w}(\beta, N) 
& = &
\sum_{c=\max (1,\beta-v+1)} ^{\min (w-u,\beta)} \sum_{k=u+v}^{w+v-c} 
   \mbinom{k+c-\beta-2}{u-1,v-\beta+c-1,k-u-v} \mbinom{2w-k-c-1}{w+v-k-c} 
\no \\ && \qquad \times
(-)^v \mbinom{\beta-1}{\beta-c}   \Big (4\, \mI_{k} \,\sigma_{1-k}(N) \sigma _{1-2w+k}(N) 
-2 \,  \sigma_{2-2w}(N) \Big )
\eea}
The fourth term then becomes, 
\bea
\sum_{\beta=1} ^{w-u+v-1} {  N^{\beta -1} J^{(3)}_{u,v;w}(\beta, N) \over \Gamma (\beta) (4 \pi \tau_2)^{w-\beta-1}} 
\eea
Assembling all the pieces gives the result quoted in (\ref{finalcQ}).

\subsection{Calculation of $G_{u,v;w}^{(N,L)}(\tau_2)$}
\label{sec:D2}

We now obtain the coefficients $G_{u,v;w}^{(N,L)}(\tau_2)$ in the Fourier expansion, defined in (\ref{Four2}). There are no contributions from $\cC^{(0)}_{u,v;w}$. The contribution from  $\cC^{(1)}_{u,v;w}+\cC^{(2)}_{u,v;w}$ is given by,
\bea
&\vphantom{.}&2\, \sigma_{2-2w}\left(\mathrm{gcd}(L,N) \right) \left[ (-)^v \sum_{k=1}^{w-u} \sum_{\ell=1}^k \sum_{\b=1}^{v+k-\ell} {\binom{2w-u-v-k-1}{w-u-k} \binom{v+k-\ell-1}{k-\ell}\binom{u+v+k-\ell-\b-1}{u-1} \over \Gamma(\ell)\Gamma(\b) (4 \pi \tau_2)^{w-\b-\ell}} L^{\b-1}N^{\ell-1}\right.
\no\\
&\vphantom{.}&\hspace{0.5 in}+ \left. (-)^u \sum_{k=1}^{w-v} \sum_{\ell=1}^u \sum_{\b=1}^{u+k-\ell} {\binom{2w-u-v-k-1}{w-v-k} \binom{u+k-\ell-1}{u-\ell}\binom{u+v+k-\ell-\b-1}{v-1} \over \Gamma(\ell)\Gamma(\b) (4 \pi \tau_2)^{w-\b-\ell}} L^{\b-1}N^{\ell-1}\right]
\eea
The contribution from $\cC^{(3)}_{u,v;w}$ is given by,
\bea
2  \,\sigma_{2-2w}\left(\mathrm{gcd}(L,N) \right) (-)^w \sum_{k=1}^u \sum_{\b=1}^{u+v-k} {\binom{u+v-k-1}{u-k} \binom{2w-k-\b-1}{u+v-k-\b} \over \Gamma(k) \Gamma(\b) (4 \pi \tau_2)^{w-k-\b}} L^{\b-1}N^{k-1}
\eea
The contributions from $\cC^{(4)}_{u,v;w}$ arise from $\cR_{u,v;w,k}^{(i,0)}$ for $i=2,3$ and are given respectively by, 
\bea
&\vphantom{.}& \sum_{k=1+u+v}^{w+v} \sum_{a=1}^u \sum_{b=1}^v {(-)^{a+v} \binom{2w-k-1}{w+v-k} \binom{k-a-b-1}{u-a,v-b,k-u-v-1} \over \Gamma(a) \Gamma(b) (4 \pi \tau_2)^{w-a-b}} (L+N)^{a-1}L^{b-1} V_{a,k-a,2w-k}(L+N,L)
\no\\
&\vphantom{.}&\hspace{0.1 in} + \sum_{k=1+u+v}^{w+u} \sum_{a=1}^u \sum_{b=1}^v {(-)^{b+u} \binom{2w-k-1}{w+u-k} \binom{k-a-b-1}{u-a,v-b,k-u-v-1} \over \Gamma(a) \Gamma(b) (4 \pi \tau_2)^{w-a-b}} (L+N)^{a-1}L^{b-1} V_{b,k-b,2w-k}(L,L+N)
\no\\
\eea
and
\bea
&\vphantom{.}&\sum_{k=1+u+v}^{w+v}\sum_{b=1}^v\sum_{c=1}^{k-u-v} {\binom{2w-k-1}{w+v-k} \binom{k-b-c-1}{u-1,v-b,k-u-v-c} \over \Gamma(b) \Gamma(c) (4 \pi \tau_2)^{w-b-c}} (L+N)^{c-1} L^{b-1}
\no\\
&\vphantom{.}&\hspace{0.5 in} \times(-)^v \left[ 4 \,\mI_{k-c}\, \sigma_{1-k+c}(L) \sigma_{1-2w+k-c}(L+N)- 2 \sigma_{2-2w} \left(\mathrm{gcd}(L,L+N) \right)\right]
\no\\
&\vphantom{.}&\hspace{0.05 in}+\sum_{k=1+u+v}^{w+u}\sum_{b=1}^u\sum_{c=1}^{k-u-v} {\binom{2w-k-1}{w+u-k} \binom{k-b-c-1}{v-1,u-b,k-u-v-c} \over \Gamma(b) \Gamma(c) (4 \pi \tau_2)^{w-b-c}} (L+N)^{b-1} L^{c-1} 
\no\\
&\vphantom{.}&\hspace{0.5 in}\times(-)^u\left[ 4 \,\mI_{k-c}\,  \sigma_{1-k+c}(L+N) \sigma_{1-2w+k-c}(L) - 2  \sigma_{2-2w} \left(\mathrm{gcd}(L,L+N) \right)\right]\no\\
\eea
The total result can be written in an abridged form as follows,
\bea
\label{finalGresult}
G_{u,v;w}^{(N,L)}(\tau_2) = \sum_{k=k_-}^{k_+}\sum_{\ell=1}^{\ell_+}\sum_{\b=1}^{\b_+}{\cW^{k,\ell,\beta}_{u,v;w}(N,L) \over (4 \pi \tau_2)^{w- \b-\ell}}
\eea
where
\bea
\label{upperboundoriginal}
k_-&=&1+u+v
\no\\
k_+&=&\max(w+ u,w+v) 
\no\\
 \ell_+&=&\max(w-u,u)  
\no\\
\b_+&=& \max \left(w+|u-v|-1,\,u+v-1 \right)
\eea

\sm

To express the coefficients $\cW^{k,\ell,\beta}_{u,v;w}(N,L)$ in a form as simple as possible, we shall adopt the convention whereby   binomial and trinomial coefficients vanish if any one of their lower entries is a negative integer,
\bea
\label{simplifyingconven}
\binom{A}{B_1, \cdots, B_r} =0 \quad \hbox{ if }   \quad -B_i \in  \NN
\eea
for any $i =1, \cdots, r$, or if $ - A \in \NN$. For all integer arguments/indices, the coefficients $\cW^{k,\ell,\beta}_{u,v;w}(N,L)$ are rational numbers. They are given by the following lengthy expression, 
\bea
\label{Wbigformula}
\cW^{k,\ell,\beta}_{u,v;w}(N,L)&=& {2L^{\b-1}N^{\ell-1}} \sigma_{2-2w}\left(\mathrm{gcd}(L,N) \right) 
\no\\
&\vphantom{.}&\hspace{-0.5 in}\times\left[ (-)^v {\binom{2w-k-1}{w+v-k} \binom{k-u-\ell-1}{k-u-v-\ell}\binom{k-\ell-\b-1}{k-\ell-\b-u}\over  \Gamma(\ell) \Gamma(\b)}  + (-)^u {\binom{2w-k-1}{w+u-k} \binom{k-v-\ell-1}{u-\ell}\binom{k-\ell-\b-1}{k-\ell-\b-v}\over  \Gamma(\ell) \Gamma(\b)} \right .
\no\\
&\vphantom{.}&\hspace{-0.2 in} \left. +(-)^w \delta_{k-u-v,1} {\binom{u+v-\ell-1}{u-\ell} \binom{2w-\ell-\b-1}{u+v-\ell-\b} \over \Gamma(\ell) \Gamma(\b)} \right]
\no\\
&\vphantom{.}&\hspace{-0.7 in}+ L^{\b-1}(L+N)^{\ell-1} 
 \left[{(-)^{\ell+v} \binom{2w-k-1}{w+v-k} \binom{k-\ell-\b-1}{u-\ell,v-\b,k-u-v-1} \over \Gamma(\ell) \Gamma(\b) }  V_{\ell,k-\ell,2w-k}(L+N,L)\right.
\no\\
&&\hskip 0.7in \left.+{(-)^{\b+u} \binom{2w-k-1}{w+u-k} \binom{k-\ell-\b-1}{u-\ell,v-\b,k-u-v-1} \over \Gamma(\ell) \Gamma(\b) } V_{\b,k-\b,2w-k}(L,L+N)\right]
\no\\
&\vphantom{.}&\hspace{-0.7 in}+ L^{\b-1}(L+N)^{\ell-1}
 \left[ {  \binom{2w-k-1}{w+v-k} \binom{k-\ell-\b-1}{u-1,v-\b,k-u-v-\ell} \over \Gamma(\ell) \Gamma(\b)}\cV^{k,\ell}_{u,v;w}(L,L+N)\right.
\no\\
&&\hskip 0.7in \left.+ {\binom{2w-k-1}{w+u-k} \binom{k-\ell-\b-1}{v-1,u-\ell,k-u-v-\b} \over \Gamma(\ell) \Gamma(\b)}\cV^{k,\b}_{v,u;w}(L+N,L) \right]
\eea
where  we have defined 
\bea
(-)^v \cV^{k,\ell}_{u,v;w}(L,M) = 4\, \mI_{k-\ell}\,\sigma_{1-k+\ell}(L) \sigma_{1-2w+k-\ell}(M)- 2\,\sigma_{2-2w} \left(\mathrm{gcd}(L,M) \right)
\eea
This gives the result cited in the first line of Theorem \ref{thmnonconst}.

\newpage

\subsection{Calculation of $F_{u,v;w}^{(N)}(\tau_2)$}
\label{sec:D3}

We now calculate the coefficients $F_{u,v;w}^{(N)}(\tau_2)$ in the Fourier expansion, defined in (\ref{Four2}). There are no contributions from $\cC^{(0)}_{u,v;w}$. The contribution from  $\cC^{(1)}_{u,v;w}+\cC^{(2)}_{u,v;w}$ is given by,
\bea
\label{FNuv12}
&\vphantom{.}&F_{u,v;w}^{(N)(12)}(\tau_2)=
\no\\
&\vphantom{.}&\hspace{0 in}(-)^v\sum_{k=1}^{w-u}\sum_{\ell=1}^k  { \binom{2w-u-v-k-1}{w-u-k} \binom{v+k-\ell-1}{k-\ell}\over \Gamma(\ell)} N^{\ell-1} \left [4 \sum_{\a=0}^{\left[{u \over 2} \right]} {(-)^\a \zeta(2 \a) \binom{u+v+k-\ell-2\a-1}{u-2\a} \over (2 \pi)^{2\a} (4 \pi \tau_2)^{w-2\a-\ell}} \sigma_{1+2\a-2w}(N)\right.
\no\\
&\vphantom{.}&\hspace{4 in} \left.+{\binom{u+v+k-\ell-2}{u-1} \over (4 \pi \tau_2)^{w-\ell-1}}\sigma_{2-2w}(N)\right]
\no\\
&\vphantom{.}&\hspace{-0.1 in}+(-)^u \sum_{k=1}^{w-v}\sum_{\ell=1}^u  { \binom{2w-u-v-k-1}{w-v-k} \binom{u+k-\ell-1}{u-\ell}\over \Gamma(\ell)} N^{\ell-1} \left [4 \sum_{\a=0}^{\left[{v \over 2} \right]} {(-)^\a \zeta(2 \a) \binom{u+v+k-\ell-2\a-1}{v-2\a} \over (2 \pi)^{2\a} (4 \pi \tau_2)^{w-2\a-\ell}} \sigma_{1+2\a-2w}(N)\right.
\no\\
&\vphantom{.}&\hspace{4 in} \left.+{\binom{u+v+k-\ell-2}{v-1} \over (4 \pi \tau_2)^{w-\ell-1}}\sigma_{2-2w}(N)\right]
\no\\
&\vphantom{.}&\hspace{-0.1 in}+4  \sum_{\ell=1}^{w-u} {(-)^{v+u} \over (4 \pi \tau_2)^{w-u-v-\ell}}\left[\sum_{j=\left[{1+v+\ell \over 2}\right]}^{\left[u+v+ \ell -1 \over 2 \right]} {(-)^{j+\ell} \binom{2w-u-v-\ell-1}{w-u-\ell} \binom{2j-v-1}{\ell-1} \over \Gamma(u+v+\ell-2j)} {\zeta(2j)  \over(2 \pi)^{2j} }N^{u+v+\ell-2j-1}\sigma_{1+2j-2w} (N)\right.
\no\\
&\vphantom{.}&\hspace{4 in}+(-)^{v} (u \leftrightarrow \ell)\Bigg]
\no\\
\eea
Note that $(u \leftrightarrow \ell)$ applies only to the portion in brackets on the penultimate line.

\sm 

The contribution from $\cC^{(3)}_{u,v;w}$ is 
\bea
\label{FNuv3}
&\vphantom{.}&F_{u,v;w}^{(N)(3)}(\tau_2)=
\no\\
&\vphantom{.}&(-)^w \sum_{\ell=1}^u {\binom{u+v-\ell-1}{u-\ell}  \over \Gamma(\ell)}N^{\ell-1} \left[\sum_{\a=0}^{\left[{2w-u-v \over 2} \right]} {4 (-)^\a \zeta(2 \a) \binom{2w-2\a-\ell-1}{u+v-\ell-1} \over (2\pi)^{2 \a} (4 \pi \tau_2)^{w-\ell-2\a}} \sigma_{1+2 \a-2w}(N) + {\binom{2w-\ell-2}{2w-u-v-1} \over ( 4 \pi \tau_2)^{w-\ell-1}}  \sigma_{2-2w}(N) \right]
\no\\
\eea

\sm 

The contributions from $\cC^{(4)}_{u,v;w}$ come from $\cR_{u,v;w,k}^{(i,1)}$ for $i=2,3,4$, as well as from $\cR_{u,v;w,k}^{(4,0)}$. They can be written collectively as 
\bea
F_{u,v;w}^{(N)(4)}(\tau_2)=\sum_{k=k_-}^{k_+} \sum_{\ell=1}^{\ell_+} {M_{u,v;w}^{k,\ell}(N) \over (4 \pi \tau_2)^{w-\ell-1}}+\sum_{k=k_-}^{w+v}\sum_{\ell=1}^{k-u-v}\sum_{\b=1}^u {H^{k,\ell,\b}_{u,v;w}(N) \over (4 \pi \tau_2)^{w-\ell-\b}}
\eea
where $k_-$, $k_+$, and $\ell_+$ are as defined in (\ref{upperboundoriginal}) and
\bea
M_{u,v;w}^{k,\ell}(N) &=&
{(-)^v \binom{2w-k-1}{w+v-k} \binom{k-\ell-2}{u-1,v-1,k-u-v-\ell} \over \Gamma(\ell)}N^{\ell-1} \mu^{(1) \,\, k,\ell}_{u,v;w}(N)
\no\\
&\vphantom{.}&\hspace{0.1 in}+ {(-)^v \binom{2w-k-1}{w+v-k} \binom{k-\ell-2}{u-\ell,v-1,k-u-v-1} \over \Gamma(\ell)}N^{\ell-1} \mu^{(2) \,\, k,\ell}_{u,v;w}(N)
\no\\
&\vphantom{.}&\hspace{0.1 in}+ {(-)^u \binom{2w-k-1}{w+u-k} \binom{k-\ell-2}{u-\ell,v-1,k-u-v-1} \over \Gamma(\ell)}N^{\ell-1} \mu^{(3) \,\, k,\ell}_{u,v;w}(N)
\no\\
\eea
Note that we are utilizing the convention put forth in (\ref{simplifyingconven}). The $\mu^{(i) \,\, k,\ell}_{u,v;w}(N)$ for $i=1,2,3$ are defined by,
\bea
\label{muidef}
\mu^{(1) \,\, k,\ell}_{u,v;w}(N)&=& \zeta_{sv}(k-\ell-1) \sigma_{1-2w+k-\ell}(N) - \sigma_{2-2w}(N) + \sum_{0<m|N} {S_{0,k-\ell-1}(m) \over m^{2w-k+\ell-1}}
\no\\
\mu^{(2) \,\, k,\ell}_{u,v;w}(N)&=& \sum _{0<m|N} {S_{2w-k,k-\ell-1}(m) \over m^{\ell-1}} + (-)^\ell \sum_{0<m|N} {S_{k-\ell-1,2w-k}(m) \over m^{\ell-1}}
\no\\
\mu^{(3) \,\, k,\ell}_{u,v;w}(N)&=& \zeta_{sv}(2w-k) \sigma_{2-k}(N) - \sigma_{2-2w}(N) - \sum_{0<m|N} {S_{0,2w-k}(m) \over m^{k-2}}
\eea
We have also defined 
\bea
 H^{k,\ell,\b}_{u,v;w}(N) &=& \theta(N-2){(-)^v \binom{2w-k-1}{w+v-k} \binom{k-\ell-\b-1}{u-\b,v-1,k-u-v-\ell} \over \Gamma(\ell) \Gamma(\b)} 
 \no\\
&\vphantom{.}&\hspace{0.1 in} \times \sum_{N_1=1}^{N-1} N_1^{\b-1}(N-N_1)^{\ell-1} V_{\b,2w-k+\ell,k-\ell-\b}(N_1,N-N_1)
\eea
where again we are using the convention of (\ref{simplifyingconven}).
\sm 

The coefficients $M_{u,v;w}^{k,\ell}(N)$ evaluated on integer-valued indices and arguments are linear combinations of rational numbers and rational multiples of odd zeta values. Odd zeta values naively appear in all $\mu^{(i) \,\, k,\ell}_{u,v;w}(N)$ for $i=1,2,3$. However, there are actually no odd zeta values arising from $\mu^{(3) \,\, k,\ell}_{u,v;w}(N)$, since the factor of $\zeta_{sv}(2w-k)$ appearing in (\ref{muidef}) cancels against the zeta value arising from the definition of $S_{0,2w-k}(m)$, see (\ref{Sab}). Hence we obtain contributions from only the first two lines of (\ref{muidef}), which are given by 
\bea
\label{FNcMresult}
\cM_{u,v;w}^{k,\ell}(N) &=& 
{2(-)^v \binom{2w-k-1}{w+v-k} \Gamma(k-\ell-1) \over \Gamma(v) \Gamma(\ell) \Gamma(k-v-\ell)} N^{\ell-1} 
\\
&\vphantom{.}&\hspace{0.1 in}\times\left[  \mbinom{k-v-\ell-1}{k-u-v-\ell} \sigma_{1-2w+k-\ell}(N) \zeta_{sv}(k-\ell-1) \right.  
\no\\
&\vphantom{.}&\hspace{0.3 in}\left. +{(-)^{k} \mbinom{k-v-\ell-1}{u-\ell}} 
\left\{  \sum_{j=3}^{k-\ell-1} \mbinom{2w-\ell-j-2}{2w-k-1}\zeta_{sv}(j) \sigma_{2+j-2w}(N)\right.\right.
\no\\
&\vphantom{.}&\hspace{2 in}\left.\left. -\sum_{j=3}^{2w-k} \mbinom{2w-\ell-j-2}{k-\ell-2} \zeta_{sv}(j) \sigma_{2+j-2w}(N)\right\} \right]
\no\eea
Once again, we are working with the convention of (\ref{simplifyingconven}).

\sm

Finally, we find, 
\bea
\label{FNuv4}
F_{u,v;w}^{(N)}(\tau_2)=\sum_{k=k_-}^{k_+} \sum_{\ell=1}^{\ell_+}\left[  { \cM_{u,v;w}^{k,\ell}(N) \over (4 \pi \tau_2)^{w-\ell-1}}+\sum_{\g=0}^{\g_+} {\cH^{k,\ell,\b}_{u,v;w}(N) \over (4 \pi \tau_2)^{w-\ell-\g}}\right]
\eea
where 
\bea
\g_+ = \mathrm{max}\left(u+v, \,2w -u -v-\eps \right)
\eea
 $\eps = u+v \,\, (\mathrm{mod}\,\, 2)$ and the function $\cH^{k,\ell,\g}_{u,v;w}(N)$ is obtained by combining $ H^{k,\ell,\b}_{u,v;w}(N) $ with the rational portions of $M_{u,v;w}^{k,\ell}(N)$, i.e. $M_{u,v;w}^{k,\ell}(N) -\cM_{u,v;w}^{k,\ell}(N) $, as well as with the contributions from (\ref{FNuv12}) and (\ref{FNuv3}), such that
 \bea
 \label{implicitcH}
\sum_{k=k_-}^{k_+} \sum_{\ell=1}^{\ell_+} \sum_{\g=0}^{\g_+} {\cH^{k,\ell,\b}_{u,v;w}(N) \over (4 \pi \tau_2)^{w-\ell-\g}}&=&F_{u,v;w}^{(N)(12)}(\tau_2)+F_{u,v;w}^{(N)(3)}(\tau_2)
\no\\
&\vphantom{.}&\hspace{-1.3in}+\sum_{k=k_-}^{w+v}\sum_{\ell=1}^{k-u-v}\sum_{\b=1}^u {H^{k,\ell,\b}_{u,v;w}(N) \over (4 \pi \tau_2)^{w-\ell-\b}}+\sum_{k=k_-}^{k_+} \sum_{\ell=1}^{\ell_+} { M_{u,v;w}^{k,\ell}(N) -\cM_{u,v;w}^{k,\ell}(N)  \over (4 \pi \tau_2)^{w-\ell-1}}
 \eea
 For all integer-valued indices and arguments $\cH^{k,\ell,\b}_{u,v;w}(N)$ is a rational number.

\newpage
\section{Proof of Theorem \ref{lem32}}
\setcounter{equation}{0}
\label{app:indepproof}

In this appendix we prove Theorem \ref{lem32}. To do so we proceed in two parts. We first prove the linear independence of the spaces $\mA_w^{(1)}$ and $\mA_w^{(2)}$, and then move on to the more complicated case of $\mA_w^{(3)}$. Both parts are proven by means of holomorphic subgraph reduction and the sieve algorithm.

\subsection{Independence of $\mA_w^{(1)}$ and $\mA_w^{(2)}$}

As reviewed in Section \ref{sec:sieve-alg}, the sieve algorithm involves determining the order in $\nabla$ at which the first holomorphic Eisenstein series appears. It is easy to see that for $\cP_k(w_1,w_2) \in \mA_w^{(1)}$, the first holomorphic Eisenstein series appears at order $\nabla^{w_2 - k}$, so that $\cP_k(w_1,w_2) \in \cV^{(w_2-k)}_{(w,w)}$.

On the other hand, elements of $\mA_{w}^{(2)}$ have $u+1 < w-u-1$, and hence the lowest entry in the exponent matrix of $\cA_{u,u+1;w} \in \mA_{w}^{(2)}$ is $u$. As a result, we have, 
\bea
\cA_{u,u+1;w} \in \cV^{(u)}_{(w,w)}\hspace{0.5 in} \mathrm{for \,\, all}\hspace{0.5 in}\cA_{u,u+1;w} \in \mA_{w}^{(2)}
\eea
That is, upon application of the Cauchy-Riemann operator to elements of $\mA_{w}^{(2)}$, the first possible appearance of a holomorphic Eisenstein series is at order $\nabla^u$. The holomorphic Eisenstein series in question appears through algebraic reduction identities (\ref{algred}), and one finds by explicit computation that,
\bea
\nabla^u \cA_{u,u+1;w}  = -(-1)^u {\Gamma(2u+1)\over \Gamma(u+1)} \,\cG_{2u+1}\, \cC^+ \left[\begin{matrix}w-u-1 & 0 \cr  w-u & 0\end{matrix} \right]+ \dots
\eea
where the ellipsis represents elements of $\cV_{(w+u,w-u)}$. However, since $2u+1$ is odd, this holomorphic Eisenstein series vanishes. So we in fact have the stricter statement that,
\bea
\cA_{u,u+1;w} \in \cV^{(u+1)}_{(w,w)} \subset \cV^{(u)}_{(w,w)}\hspace{0.5 in} \mathrm{for \,\, all}\hspace{0.5 in} \cA_{u,u+1;w} \in \mA_{w}^{(2)}
\eea
This means that we may proceed to $\nabla^{u+1}$, where one obtains the following holomorphic Eisenstein series via algebraic reduction,
\bea
\label{easyderiv}
\nabla^{u+1}\cA_{u,u+1;w} = (-1)^u {\Gamma(2u+2) \over \Gamma(u)}\,\cG_{2(u+1)} \,E_{w-u-1} + \dots 
\eea
The ellipsis represents elements of $\cV_{(w+u+1,w-u-1)}$.  
We are now in a position to prove the first part of the theorem. We begin by assuming a linear relation of the form
\bea
\label{A2minuslincomb}
\sum_{w_2 = 2}^{\left[{w-1\over 2}\right]}\sum_{k=1}^{w_2-1} \a(w_2, k)\cP_k(w-w_2,w_2)+ \sum_{u=1}^{\left[{w-3 \over2 }\right]} \b(u) \cA_{u,u+1;w} =0
\eea
and then apply $\nabla $ to it. Since $\mA_w \subset \cV_{(w,w)}$, we have
\bea
\label{nabmapA1}
\nabla : \mA_w & \to & \cV_{(w+1, w-1)} \oplus \mC_{(w+1,w-1)}
\eea
Applying $\nabla$ to (\ref{A2minuslincomb}) produces only elements of $\cV_{(w+1, w-1)}$, except for when it acts on $\cP_{w_2-1}(w-w_2,w_2)$, in which case it produces
\bea
\label{A1mustvanish}
\sum _{w_2=2}^{[{w-1\over 2}]}  \alpha (w_2, w_2-1) \tau_2^{2-2w_2} \nabla ^{w_2} E_{w_2} \DDb ^{w_2-1} E_{w-w_2}  \in \mC_{(w+1, w-1)}
\eea
By the sieve algorithm, all terms not in $\cV_{(w+1, w-1)}$ must cancel, so (\ref{A2minuslincomb}) requires the above to vanish. The form $\nabla ^{w_2} E_{w_2}$ is proportional to $G_{2w_2}$ by (\ref{Gk}) so that all the terms for different values of $w_2$ must be linearly independent of one another, and we must have $\alpha (w_2, w_2-1)=0$ for all $2 \leq w_2 \leq [ { w-1 \over 2}]$. Given these cancellations, the range of the sum over $k$ in (\ref{A2minuslincomb}) now effectively has upper limit $w_2-2$. By construction, for these values of $k$ and $w_2$, the image of $\nabla$ on each term in (\ref{A2minuslincomb}) belongs to $\cV_{(w+1,w-1)}$, and we may now proceed to $\nabla ^2$,
\bea
\nabla^2: \mA_w \rightarrow \cV_{(w+2,w-2)} \oplus \mC_{(w+2,w-2)}
\eea
In addition to the contributions coming from $\cP_{w_2-2}(w-w_2,w_2)$, there will now be a contribution towards $\mC_{(w+2,w-2)}$ coming from an element of $\mA_w^{(2)}$ as well. In particular, the image of $\nabla^2$ on each term of the abridged (\ref{A2minuslincomb}) is in $\cV_{(w+2,w-2)}$, except for the images of $\cP_{w_2-2}(w-w_2,w_2)$ and $\cA_{1,2;w}$, which have,
\bea
\label{A2mustvanish}
\sum_{w_2=3}^{\left[{w-1\over 2}\right]} \a(w_2, w_2-2)\, \cG_{2 w_2} \overline \nabla^{w_2-2} E_{w-w_2} + \b(1)\, \cG_4 \,E_{w-2}\, \in\, \mC_{(w+2,w-2)}
\eea
For simplicity, we have absorbed unimportant factors of Gamma functions and $\tau_2$ into $\a(w_2,w_2-2)$ and $\b(1)$. For this expression to vanish, we require $\a(w_2,w_2-2)=\b(1)=0$. Having eliminated the elements of $\mC_{(w+2,w-2)}$, one may then move on to order $\nabla^3$. Iterating this argument, one inductively shows that all coefficients must vanish, and thus that all elements of $\mA_w^{(1)}$ and $\mA_w^{(2)}$ are independent.

\subsection{Independence of $\mA_w^{(1)}$, $\mA_w^{(2)}$,  and $\mA_w^{(2)}$}

We shall now prove the linear independence of $\mA_w^{(3)}$ analogously to the case of $\mA_w^{(2)}$. The additional complication is  that the holomorphic Eisenstein series which arise do so via holomorphic subgraph reduction in addition to algebraic reduction. In particular, note that for any $\cA_{u,u+1;w}\in\mA_{w}^{(3)} $, by definition $w-u-1\leq u+1$. Restricting temporarily to the case $u \neq \left[{w-1 \over 2} \right]$, it follows that $w-u-1$ is the lowest element in the exponent matrix, and hence,
\bea
\cA_{u,u+1;w} \in \cV^{(w-u-1)}_{(w,w)}\hspace{0.5 in} \mathrm{for \,\, all}\hspace{0.5 in} \cA_{u,u+1;w} \in \mA_{w}^{(3)}
\eea
or equivalently 
\bea
\nabla^{w-u-1} \cA_{u,u+1;w} \in \cV_{(2w-u-1,u+1)}\oplus\mC_{(2w-u-1,u+1)}
\eea
The elements of $\mC_{(2w-u-1,u+1)}$  arise via holomorphic subgraph reduction, and are computed in closed form in Lemma \ref{lemma1} of  Appendix~\ref{app:E}, which yields
\bea
\label{V1deriv2}
\nabla ^{w-u-1} \cA_{u,u+1;w} =
\sum _{\ell=w-u}^{w-1} H_1(\ell) \, \cG_\ell \, \cC^+_\ell 
+ \sum_{\ell=w-u}^{2(w-u)-1} H_2(\ell) \, \cG_\ell \, \cC^+_\ell + \cdots
\eea
when $v=u+1$. $\cC^+_\ell$ is defined by,
\bea
\cC^+_\ell = \cC^+ \left[\begin{matrix} 2w-u-\ell-1 &0 \cr u+1 & 0 \cr\end{matrix} \right]
\eea
 The ellipsis in (\ref{V1deriv2})  represent elements of $\cV_{(2w-u-1,u+1)}$, while the functions $H_1(\ell)$, $H_2(\ell)$, and $ \cC^+_\ell$ are obtained from (\ref{H1H2f}) and  (\ref{Cstar}) by setting $v=u+1$,
\bea
\label{H1H2fintext}
H_1(\ell) & = & 
{ (-)^w \over \Gamma (w-u) } \,  \prod_{j=1}^{w-u-1} (\ell-j) \, \prod_{k=1}^{w-u-1} (w+k-\ell-1)  
\no \\
H_2(\ell) & = & 
{ (-)^{u+1}  \over \Gamma (u)  } \, 
\prod_{j=1}^{w-u-1} (\ell-j) \, \prod_{k=1}^{u-1} (2w-2u+k-\ell-1) 
\eea
 Importantly, we have noted that $u\leq w-5$ in the definition of $\mA_w^{(3)}$ implies that $w-u-1 \geq 4$, and then we have used (\ref{H1H2fintext}) to obtain the lower bounds on the sums in (\ref{V1deriv2}).  Each individual term in the remaining sums has non-vanishing coefficient.  

\sm

We reemphasize that these are only the contribution arising from holomorphic subgraph reduction.  As long as $u\neq\left[{w-1 \over 2}\right]$, this is in fact all of the contributions. However, for $u=\left[{w-1 \over 2}\right]$ one would expect an additional holomorphic Eisenstein series $\cG_w$ produced via algebraic reduction. For $w$ odd this term is vanishing and the distinction is unimportant, whereas for $w$ even we must take it into account. For $w \in 4 \NN$, the case of $u={w \over 2}$ is also exceptional, since (\ref{V1deriv2}) vanishes. For the moment, we take $u>\left[{w \over 2}\right]$ to avoid these complications, returning to them at the end of the proof.

\sm

 We now focus on the terms in (\ref{V1deriv2}) with the holomorphic Eisenstein series of lowest weight. 
We may begin by assuming that $w - u \in 2 \NN+1$. If this is the case, then (\ref{V1deriv2}) has a term containing a holomorphic Eisenstein series $\cG_{w-u+1}$ with coefficient 
\bea
H_1(w - u + 1)+H_2(w - u + 1) =   (-)^{u+1}(w-u){ \Gamma(w-1) \over \Gamma(u)}
\eea
which is non-vanishing. Hence the lowest weight holomorphic Eisenstein series is $\cG_{w-u+1}$. 

\sm

In the case of $w - u \in 2 \NN$, one begins by considering the case $\ell = w-u$, to which both sums contribute. However, in this case $H_1(w-u)+H_2(w-u)$ vanishes. Thus one moves on to $\ell = w-u+2$. There is always a contribution towards this term from the second sum in (\ref{V1deriv2}), since $u\leq w-5$ implies that $2(w-u)-1\geq w-u+4$. If the first sum does not contribute to this term, then the coefficient is automatically non-vanishing. If the first sum does contribute, one finds
\bea
\label{H1H2wu2}
H_1(w - u + 2)+H_2(w - u + 2) =  \half (-1)^{u+1} (w-2u) {\Gamma(w-2) \Gamma(w-u+2) \over \Gamma(u) \Gamma(w-u)}
\eea
which is non-vanishing for $u > \left[{w\over2}\right]$. Thus the lowest weight holomorphic Eisenstein series produced is  $\cG_{w-u+2}$. 

\sm

With this information, we may now begin to prove linear independence of $\mA_{w}^{(1)}$, $\mA_{w}^{(2)}$, and $\mA_{w}^{(3)}$. This is done in the same manner as before: we begin by writing  a tentative linear combination, now including the additional contributions due to elements of $\mA_{w}^{(3)}$,
\bea
\label{A3lincomb}
\sum_{w_2 = 2}^{\left[{w-1\over 2}\right]}\sum_{k=1}^{w_2-1} \a(w_2, k)\cP_k(w-w_2,w_2)+ \sum_{u=1}^{{w-5} } \b(u) \cA_{u,u+1;w}=0
\eea
 At any order $n<4$ in $\nabla$, the elements of $\mA_{w}^{(3)}$ do not give rise to elements of $\mC_{(w+n,w-n)}$, and one may proceed exactly as in the first part of the proof. On the other hand, at each order $4\leq n <\left[{w-1\over 2}\right]$ in $\nabla$ there is a single element of $\mA_{w}^{(3)}$ which is in $\cV^{(n)}_{(w,w)}$, namely $\cA_{w-n-1,w-n;w}$.\footnote{The upper bound on $n$ is equivalent to the earlier bound $u > \left[{w\over2}\right]$.} By the reasoning of the previous paragraphs, this function produces elements of $\mC_{(w+n,w-n)}$ which include terms proportional to $\cG_{n+2}$ ($n$ even) or $\cG_{n+3}$ ($n$ odd). At the same order $\nabla^n$, only $\cP_{w_2-n}(w-w_2,w_2)\in \mA^{(1)}_w$ and $\cA_{n-1,n;w}\in \mA^{(2)}_w$ contribute elements of $\mC_{(w+n,w-n)}$. Since the former requires $w_2 \geq n+1$, the lowest weight holomorphic Eisenstein series which it can produce is $\cG_{2(n+1)}$, while the latter contributes only a term containing $\cG_{2n}$. Because $n\geq 4$, these are necessarily of higher weight than $\cG_{n+2}$ or $\cG_{n+3}$, thus proving linear independence for all functions in $\cV^{(n)}_{(w,w)}$ for $n<\left[{w-1\over 2}\right]$.

 \sm

We now turn to the remaining cases of $n=\left[{w-1\over 2}\right]$ and,  if $w$ is even, $n={w\over 2}$. In fact, as mentioned before it is only when $w$ is even that exceptional phenomena occur, and hence we focus on the case of $w$ even for the remainder of the proof. We consider separately the cases of $w \notin 4 \NN$ and $w \in 4 \NN$. If $w \notin 4 \NN$, the result (\ref{V1deriv2}) for $n=\left[{w-1\over 2}\right]={w\over2}-1$ is unchanged, but the case of $n = {w \over 2}$ must be modified slightly: upon application of $\nabla^{{w \over 2}}$ to  $\cA_{{w \over 2}-1,{w \over 2};w}$, one not only produces elements of $\mC_{(3w/2,w/2)}$ via holomorphic subgraph reduction as per (\ref{V1deriv2}), but also a term via algebraic reduction. In particular, a term proportional to $\cG_w$ is produced by applying algebraic reduction to the image of $\nabla^{{w \over 2}}$ on the $\cC_{{w \over 2},{w \over 2}-1;w}$ contained in $\cA_{{w \over 2}-1,{w \over 2};w}$. This new contribution cancels the term proportional to $\cG_w$ in the second sum of (\ref{V1deriv2}), and we are left with
\bea
\label{exception1}
\nabla ^{w\over 2} \cA_{{w \over 2}-1,{w \over 2};w} &=&
\sum _{\ell=w/2+1}^{w-2} \left(H_1(\ell)+H_2(\ell) \right)  \cG_\ell \, \cC^+_\ell + \dots
\eea

\sm

When $w \in 4 \NN$, one must make the same modification (\ref{exception1}) to the case $n = {w \over 2}$. But there is also a more subtle modification in the case of $n= {w\over 2}-1$. Naively, we would assume that $\nabla^{{w\over2}-1}\cA_{{w \over 2}, {w \over 2}+1;w}\in \cV_{(3w/2-1,w/2+1)}\oplus \mC_{(3w/2-1,w/2+1)}$.  However, from (\ref{possiblecancellation}) we see that for these choices of $w$ and $u$ all elements of $\mC_{(3w/2-1,w/2+1)}$ in (\ref{V1deriv2}) cancel, and we must proceed to the next order in $\nabla$ to obtain the first holomorphic Eisenstein series. That is, we actually have 
\bea
\label{specialfunctdef}
\cA_{{w \over 2},{w \over 2}+1;w} \in \cV^{(w/2)}_{(w,w)}\hspace{0.45 in} \mathrm{for \,\, all}\hspace{0.45 in}\cA_{{w \over 2},{w \over 2}+1;w} \in \mA_{w}^{(3)} \,\,\,\mathrm{and}\,\,\,  w\in 4 \NN
\eea
and hence $\cA_{{w \over 2},{w \over 2}+1;w}$ contributes non-holomorphic Eisenstein series at the same order in $\nabla$ as $\cA_{{w \over 2}-1,{w \over 2};w}$. The elements of $\mC_{(3w/2,w/2)}$ appearing upon application of $\nabla^{w/2}$ to $\cA_{{w \over 2},{w \over 2}+1;w}$ are obtained in Appendix \ref{app:F}, with the final results given by (\ref{finalspecialcase}). 

\sm

To complete the proof of independence, we return to our tentative linear combination (\ref{A3lincomb}) and show that all coefficients must vanish at orders $\nabla^{{w\over2}-1}$ and $\nabla^{w\over2}$ for $w \in 2\ZZ$. If  $w \notin 4 \ZZ$, the proof proceeds as before for $\nabla^{{w\over2}-1}$, and we need special consideration only for $\nabla^{w \over 2}$. Note that there is no contribution from $\mA_w^{(1)}$ or $\mA_w^{(2)}$ at this order (since these by definition have vanishing intersection with $\cV^{(w/2)}_{(w,w)}$), but there is a non-vanishing contribution (\ref{exception1}) from $\mA_w^{(3)}$. The coefficient $\beta\left({w \over 2}-1 \right)$ in the linear combination must then vanish, completing the proof for $w \notin 4 \ZZ$. 

\sm

For $w \in 4\ZZ$, there are no contributions from elements of $\mA_{w}^{(3)}$ at order $\nabla^{{w\over2}-1}$. Instead, there are two contributions at order $\nabla^{w \over 2}$, namely from $\cA_{{w \over 2}-1,{w \over 2};w}$ and $\cA_{{w \over 2},{w \over 2}+1;w}$. As before, there are no contributions from elements of $\mA_w^{(1)}$ or $\mA_w^{(2)}$ at this order. Note that $\cA_{{w \over 2}-1,{w \over 2};w}$ produces elements of $\mC_{(3w/2,w/2)}$ as dictated by (\ref{exception1}), while $\cA_{{w \over 2},{w \over 2}+1;w}$ produces elements of $\mC_{(3w/2,w/2)}$ as dictated by (\ref{finalspecialcase}). The contributions from (\ref{exception1}) can be seen to be proportional to the first term of the first line of (\ref{finalspecialcase}). The proof is then completed by noting that (\ref{finalspecialcase}) has remaining non-zero terms.

\section{Holomorphic subgraph reduction}
\setcounter{equation}{0}
\label{app:E}

We begin with the following lemma.

{\lem
\label{lemma1}
{\sl For $1 \leq u < v \leq w-1$ and $w-v<u$, the lowest order $\nabla$-derivative on $\cA_{u,v;w}$ which yields a holomorphic subgraph is $\nabla ^{w-v}$, with the contribution from holomorphic subgraph reduction given by,
\bea
\label{nabA}
\nabla ^{w-v} \cA_{u,v;w} =
\sum _{\ell=4}^{w+u-v} H_1(\ell) \, \cG_\ell \, \cC^+_\ell 
+ \sum_{\ell=4}^{2w-u-v} H_2(\ell) \, \cG_\ell \, \cC^+_\ell + \cdots
\eea
The ellipsis represents terms in $\cV_{(2w-v,v)}$ and the functions $H_i(\ell)$ are given by,
\bea
\label{H1H2}
H_1(\ell) & = & 
{ (-)^w \, \Gamma (\ell) \, \Gamma (2w-v-\ell) \over \Gamma (w-u) \Gamma (\ell+v-w) \Gamma (w+u-v-\ell+1)}
\no \\
H_2(\ell) & = & 
{ (-)^v \, \Gamma (\ell) \, \Gamma (2w-v-\ell) \over \Gamma (u) \Gamma (\ell+v-w) \Gamma (2w-u-v-\ell+1)}
\eea
We have used the following abbreviation,
\bea
\label{Cstar}
\cC^+_\ell = \cC^+ \left[\begin{matrix} 2w-v-\ell &0 \cr v & 0 \cr\end{matrix} \right]
\eea}}

\sm

To prove this, we begin by noting that the conditions $1 \leq u < v \leq w-1$ and $w-v<u$ imply $w-v<w-u$ and $w-v<v$. Hence $w-v$ is the smallest non-zero exponent in the modular graph function $\cA_{u,v;w}$ and as such $\nabla^n \cA_{u,v;w} \in \cV_{(w+n,w-n)}$ for all $n < w-v$.  The first possible appearance of a holomorphic subgraph is at $n=w-v$.

\sm

The holomorphic subgraph reduction contribution to the derivative $\nabla ^{w-v} \cA_{u,v;w}$ arises solely from the first term in (\ref{Adef}). Focusing on this term, repeated use of (\ref{CauchyRiemannAction}) gives 
\bea
\label{Leibnitz}
\nabla ^{w-v} \cA_{u,v;w} & = &
\sum _{m=0}^{w-v} \binom{w-v}{m} {\Gamma (w+u-v-m) \Gamma (w-u+m) \over \Gamma (u) \Gamma (w-u)}
\no \\ && \hskip 0.8in 
\times \, \cC^+  \left[\begin{matrix} u+w-v-m  & 0 & w-u+m \cr -w+v+m & v & w-v-m  \cr\end{matrix} \right]
+ \cdots
\eea
where the ellipses represents terms in $\cV_{(2w-v,v)}$. The only holomorphic subgraph contribution arises from the following rearrangement using the momentum conservation identities (\ref{3d3}),
\bea
\label{appendix-momconsv}
\cC^+  \left[\begin{matrix} u+w-v-m  & 0 & w-u+m \cr -w+v+m & v & w-v-m  \cr\end{matrix} \right]
= (-)^{w+v+m} 
\cC^+  \left[\begin{matrix} u+w-v-m  & 0 & w-u+m \cr 0 & v & 0  \cr\end{matrix} \right] + \cdots
\no\\
\eea
which in turn may be evaluated using the holomorphic subgraph reduction formula of (\ref{holsub}). The second line of (\ref{holsub}) does not contribute, since the first term does not yield a holomorphic subgraph and the part in brackets cancels. Putting the pieces together, we find,
\bea
\label{lem1eq}
\nabla^{w-v}\cA_{u,v;w} = \sum_{m=0}^{w-v}\left( \sum_{\ell=4}^{w+u-v-m} h_1(m,\ell) \,\cG_\ell\,\cC^+ _\ell 
+ \sum_{\ell=4}^{w-u+m}h_2(m,\ell)\, \cG_\ell\,\cC^+ _\ell \right)+\dots
\eea
where the ellipses represents terms in $\cV_{(2w-v,v)}$ and the functions $h_i(m,\ell)$ are given by,
\bea
\label{h1andh2}
h_1(m,\ell) &=& (-1)^{w+m} \left(\begin{matrix} w-v \cr m\cr \end{matrix} \right)
{\Gamma(w+u-v-m)\Gamma(2w-v-\ell) \over \Gamma(u) \Gamma(w-u)\Gamma(w+u-v-m-\ell+1) }
\no\\
h_2(m,\ell) &=& (-1)^{w+m} \left(\begin{matrix} w-v \cr m\cr \end{matrix} \right)
{\Gamma(w-u+m)\Gamma(2w-v-\ell) \over \Gamma(u) \Gamma(w-u)\Gamma(w-u+m-\ell+1)}
\eea
Next, we interchange the summation over $m$ and $\ell$ in (\ref{lem1eq}) to obtain
\bea
\label{lem2eq}
\nabla^{w-v}\cA_{u,v;w} &=& 
\sum_{\ell=4}^{u} \left(\sum_{m=0}^{w-v} h_1(m,\ell) \right) \cG_\ell\, \cC^+_\ell 
+ \sum_{\ell=u+1}^{w+u-v} \left(\sum_{m=0}^{w+u-v-\ell}h_1(m,\ell) \right)\cG_\ell\, \cC^+_\ell
\\
&& +
\sum_{\ell=4}^{w-u} \left(\sum_{m=0}^{w-v} h_2(m,\ell)  \right)\cG_\ell \, \cC^+_\ell
+ \sum_{\ell=w-u+1}^{2w-u-v}\left(\sum_{m=\ell-w+u}^{w-v} h_2(m,\ell) \right)\cG_\ell\, \cC^+_\ell
+\cdots
\no
\eea
The sums over $m$ are proportional to hypergeometric functions evaluated at unit argument and may be evaluated using Gauss's formula, 
\bea
{}_2F_1(a,b;c;1) = {\Gamma(c) \Gamma(c-a-b) \over \Gamma(c-a) \Gamma(c-b)}
\eea
and the reflection product formula for $\Gamma$-functions, 
\bea
\label{refG}
\Gamma(1-z) \Gamma(z) = {\pi \over \sin \pi z}
\eea
with the final result being
\bea
\label{coeffniceform}
\sum_{m=0}^{w-v} h_1(m,\ell) = \sum_{m=0}^{w+u-v-\ell} h_1(m,\ell) & = & H_1(\ell) 
\no\\
\sum_{m=0}^{w-v} h_2(m,\ell) = \sum_{m=\ell-w+u}^{w-v} h_2(m,\ell) & = & H_2(\ell) 
\eea
\newline\newline
Having proven Lemma \ref{lemma1}, we have the following simple corollary, 
{\cor 
\label{H1H2cor}
For the range of parameters $u,v,w$ indicated in Lemma \ref{lemma1}, the functions $H_i(\ell) $ are polynomials in $\ell$ of respective degrees $2w-u-v-1$ and $w+u-v-1$. Their only zeros in $\ell$ within their respective summation ranges in (\ref{nabA}) are $w-v$ simple zeros at the integers $\{ 1,2, \cdots, w-v \}$.}
\newline\newline
By inspection of (\ref{H1H2}), it is clear that for the ranges of $u,v,w$ assumed in Lemma \ref{lemma1}, the functions $H_1(\ell)$ and $H_2(\ell)$ factorize as follows,
\bea
\label{H1H2f}
H_1(\ell) & = & 
{ (-)^w \over \Gamma (w-u) } \,  \prod_{j=1}^{w-v} (\ell-j) \, \prod_{k=1}^{w-u-1} (w+u-v+k-\ell) 
\no \\
H_2(\ell) & = & 
{ (-)^v  \over \Gamma (u)  } \, 
\prod_{j=1}^{w-v} (\ell-j) \, \prod_{k=1}^{u-1} (2w-u-v+k-\ell) 
\eea
Within their respective summation ranges in (\ref{nabA}), namely $4 \leq \ell \leq w+u-v$ for $H_1(\ell)$ and $4 \leq \ell  \leq 2w-u-v$ for $H_2(\ell)$, the second products in both lines of (\ref{H1H2f}) are non-vanishing. Thus the only zeros are $w-v$ simple zeros at the integers $\ell\in \{1,2,\cdots, w-v\}$, as claimed. 
\newline\newline
For the particular case of $w$ even and $u={w\over2}$, the above expressions for $H_1(\ell)$ and $H_2(\ell)$ exhibit the following relation, 
\bea
\label{possiblecancellation}
H_2(\ell) = (-)^v H_1(\ell) \hspace{0.8 in} \mathrm{for}\,\,\,w = 2u
\eea
In the case that $v=u+1\in 2 \NN +1$, all elements of $\mC_{(2w-v,v)}$ cancel from the right-hand side of (\ref{nabA}) and the sieve algorithm requires that we proceed to the next order in $\nabla$ to identify the first holomorphic subgraph. This is more involved, and is discussed in Appendix \ref{app:F}.

\newpage
\section{A special case }
\setcounter{equation}{0}
\label{app:F}

As discussed after Corollary \ref{H1H2cor}, for $w\in 4 \NN$, $u={w\over2}$, and $v=u+1$ all elements of $\mC_{(2w-v,v)}$ cancel from the right-hand side of (\ref{nabA}). In this case the sieve algorithm demands that we proceed to the next order in $\nabla$, i.e. $\nabla^{w-v+1}=\nabla^{w\over2}$, to obtain the first holomorphic Eisenstein series. In other words,  $\cA_{{w \over 2},{w \over 2}+1;w} \in \cV^{(w/2)}_{(w,w)}$ for $w\in 4 \NN$. Obtaining the elements of $\mC_{(3w/2,w/2)}$ explicitly in this case is more involved than in Appendix \ref{app:E}, but follows the same strategy.

\sm

We consider $\cA_{u,v;w}=\cC_{u,v;w}-\cC_{v,u;w}$ and focus on the first term on the right-hand side. In order to identify all elements of $\mC_{(3w/2,w/2)}$, at each order in $\nabla$ it no longer suffices to keep only the terms with the smallest lower right entry in the exponent matrix. Instead, we must also keep terms with next-to-smallest lower right entry, since these may still produce holomorphic Eisenstein series after the remaining $\nabla$ are applied. The terms with next-to-smallest lower right entry at level $\nabla^s$ arise from terms with smallest lower right entry at order $\nabla^{s-1}$. In particular, from (\ref{Leibnitz}) and (\ref{appendix-momconsv}) we see that the terms with smallest lower right entry at order  $\nabla^{s-1}$ are 
\bea
\nabla^{s-1} \cC _{u,v;w} &=&
\sum_{m=0}^{s-1} (-1)^{m+s-1} \left(\begin{matrix} s-1 \cr m \cr \end{matrix} \right) {\Gamma(u+s-1-m) \Gamma(w-u+m) \over \Gamma(u) \Gamma(w-u)}
\no\\
&\vphantom{.}&\hspace{0.5in} \times\,\cC^+ \left[\begin{matrix} u+s-1-m & 0& w-u+m \cr 0 & v & w- v-s+1 \cr\end{matrix} \right]+\dots
\eea
The terms of next-to-smallest lower right entry at order $\nabla^{s}$ are then obtained by applying $\nabla$ to the above and keeping only the terms which, via momentum conservation identities, subtract 1 from the bottom center entry of the exponent matrix, producing the modular graph form,
\bea
 - (u+s-1-m)\,\cC^+ \left[\begin{matrix} u+s-m & 0& w-u+m \cr 0 & v-1 & w- v-s+1 \cr\end{matrix} \right]+\dots
\eea
where we have neglected terms which do not have next-to-smallest lower right entry. Then in order to identify the elements of $\mC_{(w+n,w-n)}$ produced by these functions at order $\nabla^n$, we may consider only the terms for which, upon application of $\nabla$, the value of the lower right entry is reduced (i.e. we may neglect any functions which have further subtractions on the bottom center entry, since such terms cannot produce any holomorphic Eisenstein series at order $\nabla^{n}$). In particular, by the same reasoning as for (\ref{Leibnitz}), it follows that 
\bea
\nabla^{n-s} \cC^+ \left[\begin{matrix} u+s-m & 0& w-u+m \cr 0 & v-1 & w- v-s+1 \cr\end{matrix} \right] &=&
\no\\
&\vphantom{.}&\hspace{-3in} \sum_{r=0}^{n-s} (-1)^{r+n-s} \left(\begin{matrix} n-s \cr r\cr \end{matrix} \right) {\Gamma(u-m+n-r)\Gamma(w-u+m+r) \over \Gamma(u+s-m)\Gamma(w-u+m)}\no\\
&\vphantom{.}&\hspace{-2in}\times\,\cC^+ \left[\begin{matrix} u-m+n-r & 0& w-u+m+r \cr 0 & v-1 & w- v-n+1 \cr\end{matrix} \right]+\dots
\eea
are the terms which must be kept. Hence the elements of $\mC_{(w+n,w-n)}$ coming from functions which experienced the single subtraction on their bottom center entry at order $\nabla^s$ can be obtained by considering the elements of $\mC_{(w+n,w-n)}$ produced by 
\bea
\nabla^{n-s}\nabla^{s}  \cC_{u,v;w}&=&\sum_{m=0}^{s-1}\sum_{r=0}^{n-s} (-1)^{r+n+m} \left(\begin{matrix} n-s \cr r\cr \end{matrix} \right)\left(\begin{matrix} s-1 \cr m\cr \end{matrix} \right) {\Gamma(u-m+n-r)\Gamma(w-u+m+r) \over \Gamma(u)\Gamma(w-u)}
\no\\
&\vphantom{.}&\hspace{0.5in}\times\,\cC^+ \left[\begin{matrix} u-m+n-r & 0& w-u+m+r \cr 0 & v-1 & w- v-n+1 \cr\end{matrix} \right] + \dots
\eea
Finally, in order to capture all holomorphic Eisenstein series arising in this way, we must sum over all $s$ in the range $1\leq s \leq n-1$. Specifying that $n=w-v+1$, $w\in 4 \ZZ$, $u=w/2$ and $v=u+1$, the complete set of elements of $\mC_{(3w/2, w/2)}$ are contained in
\bea
\nabla^{w/2}  \cC_{{w \over 2}, {w \over 2}+1; w}=\sum_{s=1}^{w/2-1}\sum_{m=0}^{s-1}\sum_{r=0}^{w/2-s} \tilde{g}(s,m,r,\ell) \,\cC^+ \left[\begin{matrix} w-m-r & w/2+m+r& 0 \cr 0 & 0 & w/2 \cr\end{matrix} \right] + \dots\,\,
\eea 
where we have defined 
\bea
\tilde{g}(s,m,r,\ell) = (-1)^{m+r} \left(\begin{matrix} w/2-s \cr r\cr \end{matrix} \right)\left(\begin{matrix} s-1 \cr m\cr \end{matrix} \right){\Gamma(w-m-r)\Gamma(w/2+m+r) \over \Gamma(w/2)^2}
\eea
We may then use holomorphic subgraph reduction (\ref{holsub}) to obtain the final result, 
\bea
\label{firstcontr}
\nabla^{w/2}  \cC_{{w \over 2}, {w \over 2}+1; w} &=&\sum_{s=1}^{w/2-1}\sum_{m=0}^{s-1}\sum_{r=0}^{w/2-s} \left\{\sum_{\ell=4}^{w-m-r} g_1(s,m,r,\ell)\, \cG_{\ell} \,\cC^+ \left[\begin{matrix}3w/2-\ell &  0 \cr w/2 & 0 \cr\end{matrix} \right] \right.
\no\\
&\vphantom{.}&\hspace{0.3in}+\left. \sum_{\ell=4}^{w/2+m+r} g_2(s,m,r,\ell)\, \cG_{\ell}\, \cC^+ \left[\begin{matrix}3w/2-\ell &  0 \cr w/2 & 0 \cr\end{matrix} \right] \right\}+\dots
\eea
where we have defined 
\bea
g_1(s,m,r,\ell) &=&\left(\begin{matrix} 3w/2-\ell-1 \cr w-m-r-\ell\cr \end{matrix} \right) \tilde{g}(s,m,r,\ell) 
\no\\
g_2(s,m,r,\ell) &=& \left(\begin{matrix} 3w/2-\ell-1 \cr w/2+m+r-\ell\cr \end{matrix} \right) \tilde{g}(s,m,r,\ell) 
\eea

\sm

This is the contribution towards $\mC_{(3w/2,w/2)}$ at order $\nabla^{w/2}$ coming from $\cC_{{w\over 2}, {w \over 2}+1; w}$ in,
\bea
\cA_{{w\over 2}, {w \over 2}+1; w} =\cC_{{w\over 2}, {w \over 2}+1; w}-\cC_{{w\over 2}+1, {w \over 2}; w}
\eea 
When calculating the total contribution to $\mC_{(3w/2,w/2)}$  from the full odd modular graph function, we must also consider contributions coming from $\cC_{{w\over 2}+1, {w \over 2}; w}$. These include elements of $\mC_{(3w/2,w/2)}$ coming from algebraic subgraph reduction, as well as from holomorphic subgraph reduction since $u={w\over2}$. The terms arising from algebraic subgraph reduction are found to be of the form 
\bea
\nabla^{w/2}  \cC_{{w\over 2}+1, {w \over 2}; w} \Big|_{alg. red.}=  {\Gamma(w) \over \Gamma(w/2+1)} \, \cG_{w+1}\, \cC^+ \left[\begin{matrix}w/2-1 &  0 \cr w/2 & 0 \cr\end{matrix} \right]+ \dots
\eea
But since we are considering $w$ even, the holomorphic Eisenstein series in this case vanishes. Thus we have only the contribution coming from holomorphic subgraph reduction, which follows directly from (\ref{nabA}) and is 
\bea
\label{seccontr}
\nabla^{w/2}\cC_{{w\over 2}+1, {w \over 2}; w} &=& \sum _{\ell=4}^{w+1} H'_1(\ell) \, \cG_\ell \, \cC^{+'}_\ell 
+ \sum_{\ell=4}^{w-1} H'_2(\ell) \, \cG_\ell \, \cC^{+'}_\ell + \cdots
\eea
where $H'_1(\ell)$, $H'_2(\ell)$, and $\cC^{+'}_\ell$ are obtained from (\ref{H1H2}) and (\ref{Cstar}) by first interchanging $u\leftrightarrow v$, and then substituting the required values for $u$ and $v$. Adding (\ref{firstcontr}) and (\ref{seccontr}) then gives the total contribution to $\mC_{(3w/2,w/2)}$ for the special functions (\ref{specialfunctdef}), which is
\bea
\label{finalspecialcase}
\nabla^{w/2}\cA_{{w \over 2},{w \over 2}+1;w} &=& -  \sum _{\ell=w/2+2}^{w-2} \left( H'_1(\ell)+H'_2(\ell) \right) \, \cG_\ell \, \cC^{+'}_\ell -H'_1(w) \cG_w {\cC_w^+}'
\no\\
&\vphantom{.}&\hspace{0in}+\sum_{s=1}^{w/2-1}\sum_{m=0}^{s-1}\sum_{r=0}^{w/2-s} \left\{\sum_{\ell=4}^{w-m-r} g_1(s,m,r,\ell)\, \cG_{\ell} \,\cC^+ \left[\begin{matrix}3w/2-\ell &  0 \cr w/2 & 0 \cr\end{matrix} \right] \right.
\no\\
&\vphantom{.}&\hspace{1.5in}+\left. \sum_{\ell=4}^{w/2+m+r} g_2(s,m,r,\ell)\, \cG_{\ell}\, \cC^+ \left[\begin{matrix}3w/2-\ell &  0 \cr w/2 & 0 \cr\end{matrix} \right] \right\}+\dots
\no\\
\eea
for $w \in 4 \NN$. The dots represent elements of $\cV_{(3w/2,w/2)}$. Note that we have used the fact that $H_1'\left({w \over 2}\right)+H_2'\left({w \over 2}\right)=0$.

\end{document}